\RequirePackage{fix-cm}
\documentclass[smallextended]{svjour3}       
\smartqed  
\usepackage{graphicx}
\usepackage{amsmath}
\usepackage{mathrsfs}
\usepackage{amssymb}
\usepackage{subcaption}
\usepackage{hyperref}
\usepackage{graphicx}

\usepackage{pdflscape}
\begin{document}

\title{Moving Collinear Cracks in a Prestressed Dry Sandy Medium: Fracture Response under Traveling Punch Loads
}
\subtitle{Moving Cracks in Dry Sandy Medium}

\titlerunning{Moving Cracks in Dry Sandy Medium}        

\author{Diksha       \and
        Soniya Chaudhary \and Pawan Kumar Sharma.
}


\institute{
Diksha \at
Department of Mathematics and Scientific Computing,
National Institute of Technology Hamirpur,
Hamirpur, Himachal Pradesh 177005, India \\
\email{tdiksha758@gmail.com}
\and
Soniya Chaudhary \at
Department of Mathematics and Scientific Computing,
National Institute of Technology Hamirpur,
Hamirpur, Himachal Pradesh 177005, India\\
\email{soniyachaudhary18@gmail.com}
\and
Pawan Kumar Sharma \at
Department of Mathematics and Scientific Computing,
National Institute of Technology Hamirpur,
Hamirpur, Himachal Pradesh 177005, India\\
\email{sara712005@gmail.com}
}
\date{Received: date / Accepted: date}

\maketitle

\begin{abstract}
The dynamic fracture behaviour of two moving collinear Griffith cracks in an initially stressed dry sandy medium subjected to concentrated crack-face loading and moving punch pressure is investigated. A moving coordinate system transforms the transient problem into a steady-state formulation, while the effects of initial stress and sandiness are incorporated into the governing equations. Fourier integral transforms are employed to obtain the characteristic equation and the transformed traction--displacement relations. The crack-face, symmetry, and outer-surface conditions reduce the problem to coupled Cauchy-type singular integral equations. For a sufficiently thick strip, an asymptotic kernel reduction is developed, and the resulting equations are solved analytically using the finite Hilbert transform. Closed-form expressions are derived for the crack-density functions, Mode-I stress intensity factors at the inner and outer crack tips, and the crack opening displacement. Several limiting cases are recovered from the general formulation, providing analytical consistency checks. The results demonstrate the combined influence of crack speed, crack geometry, initial stress, sandiness, and moving punch loading on crack-tip intensification and crack opening. The proposed analytical framework provides useful insights for fracture assessment and the design of transportation infrastructure, geotechnical systems, underground excavations, and other engineering structures involving dry sandy media subjected to moving loads.

\end{abstract}

\section{Introduction}

\subsection{Dry Sandy Media and Engineering Applications}
Dry sandy media constitute one of the most widely encountered classes of granular geomaterials in both natural environments and engineered systems. Owing to their discrete particulate structure, these materials exhibit complex mechanical behaviour arising from interparticle interactions, frictional resistance, particle rearrangement, and stress transmission mechanisms, making their response considerably different from that of conventional homogeneous elastic solids \cite{campbell2006granular,chang1992micromechanics,huang2016overview,iwashita2020mechanics,tahmasebi2023state}. The mechanical characteristics of dry sandy media play a crucial role in determining the stability and performance of numerous civil, geotechnical, and transportation infrastructures. Typical engineering applications include highway and railway pavements, airfield pavements, shallow and deep foundations, retaining structures, embankments, underground tunnels, mining excavations, and offshore foundation systems, where sandy deposits frequently serve as the primary load-bearing medium. Consequently, understanding the mechanical response of dry sandy media is essential for the reliable design, performance assessment, and long-term durability of these engineering structures.
In many practical situations, engineering systems constructed on or within dry sandy deposits are subjected not only to static loading but also to moving, cyclic, and impact loads generated by vehicles, railway traffic, aircraft operations, construction equipment, and other dynamic sources. These loading conditions induce stress redistribution, deformation, and wave propagation within the granular medium, thereby influencing its overall mechanical behaviour \cite{muqtadir2020deformation,alam2017dispersion,tahmasebi2023state}. Repeated dynamic loading may further promote the initiation and growth of material defects, leading to progressive degradation of the load-carrying capacity and structural integrity of the surrounding medium. Therefore, developing rigorous analytical models capable of describing the mechanical and fracture response of dry sandy media under dynamic loading conditions remains an important research problem with significant implications for geotechnical engineering, transportation infrastructure, and underground construction.
\subsection{Fracture Behaviour and Failure Mechanisms in Dry Sandy Media}
The mechanical performance of dry sandy media is strongly influenced by the presence of inherent discontinuities such as pores, microcracks, weak interparticle contacts, and material heterogeneities. Under external loading, these imperfections serve as locations of stress concentration, promoting crack initiation and subsequent propagation that may eventually lead to localized or global structural failure. Unlike homogeneous elastic solids, the fracture behaviour of dry sandy media is governed not only by the applied loading conditions but also by particle rearrangement, intergranular friction, grain breakage, and energy dissipation occurring at multiple length scales. Consequently, conventional continuum mechanics alone is often insufficient to accurately characterize failure processes in granular geomaterials, making fracture mechanics an indispensable framework for understanding crack evolution and assessing the structural integrity of engineering systems involving sandy media \cite{hallett1995application,tahmasebi2023state}.
Over the past few decades, significant efforts have been devoted to investigating fracture phenomena in soils and granular materials through analytical, experimental, and computational approaches. Hallett \textit{et al.}~\cite{hallett1995application} demonstrated the applicability of linear elastic fracture mechanics to crack propagation in dry soils and highlighted the important role of the fracture process zone in controlling crack growth. Subsequently, Einav~\cite{einav2007fracture} proposed a theoretical framework for fracture propagation in brittle granular materials by incorporating particle breakage into the constitutive description, thereby establishing a link between microscale fragmentation and macroscopic mechanical response. With the rapid development of advanced imaging and numerical techniques, Druckrey and Alshibli~\cite{druckrey20163d} employed three-dimensional finite element simulations together with synchrotron X-ray imaging to investigate particle fracture mechanisms in sand, providing valuable insights into crack initiation and propagation within individual grains. More recently, image-based discrete element approaches have further improved the understanding of irregular particle breakage and its influence on the mechanical behaviour of granular assemblies \cite{amir2024new}. In addition, investigations on wave propagation in dry sandy media have demonstrated that discontinuities and interface conditions substantially modify the dynamic response of these materials, emphasizing the close relationship between fracture characteristics and stress-wave propagation \cite{rahul2026love}.
Despite these important contributions, most existing studies primarily emphasize particle-scale fracture, constitutive modelling, or static and quasi-static failure of granular materials. Comparatively less attention has been devoted to crack propagation under dynamic loading conditions, where inertia effects and stress-wave interactions play a significant role in governing fracture behaviour. These aspects have stimulated growing interest in the field of dynamic fracture mechanics.
\subsection{Dynamic Fracture of Moving Collinear Cracks}
Dynamic fracture mechanics has emerged as an important branch of fracture mechanics for analyzing crack propagation under rapidly varying loading conditions, where inertia effects and stress-wave interactions significantly influence the crack-tip fields. Unlike static fracture, the stress distribution surrounding a dynamically propagating crack continuously evolves with crack velocity, resulting in substantial variations in the stress intensity factors, crack-tip energy release rate, and crack propagation characteristics. Consequently, accurate prediction of dynamic crack behaviour is essential for assessing the integrity and reliability of engineering structures subjected to impact, seismic excitation, blast loading, and moving loads \cite{freund1998dynamic,nilsson1990dynamic,adda1999generalized}.
The pioneering work of Yoffe~\cite{yoffe1951lxxv} established the theoretical foundation for moving Griffith cracks by demonstrating the influence of crack velocity on the near-tip stress field. Since then, considerable research has been devoted to extending dynamic fracture formulations to various material systems, loading configurations, and crack geometries. Analytical investigations have shown that moving cracks exhibit strong coupling with elastic stress waves, leading to velocity-dependent stress intensity factors and significant modifications of the crack-tip stress distribution. Furthermore, generalized fracture criteria have been proposed to explain crack stability and propagation behaviour at high crack velocities, thereby advancing the theoretical understanding of dynamic fracture processes \cite{adda1999generalized,freund1998dynamic}.
In many engineering components, cracks rarely exist in isolation and often interact with neighbouring defects or discontinuities. Such crack interactions alter the local stress field and substantially influence crack initiation, propagation direction, and fracture resistance. Consequently, the behaviour of collinear Griffith cracks has attracted considerable attention in analytical fracture mechanics. Das~\cite{das2006interaction} investigated the interaction of moving interface collinear Griffith cracks under antiplane shear loading, while Li~\cite{li2016multiple} analyzed multiple collinear Griffith cracks in quasicrystalline media. Similar investigations have also been reported for functionally graded materials and thermo-mechanical environments, demonstrating that crack interaction considerably modifies the stress intensity factors and fracture characteristics of neighbouring cracks \cite{singh2020investigation,monfared2011dynamic}. Experimental studies on brittle geomaterials have further confirmed that crack interaction governs fracture evolution and failure patterns in cracked rock and granular media subjected to external loading \cite{wang2018study}.
Although considerable progress has been achieved in modelling moving and interacting cracks, most analytical studies have been restricted to conventional material systems and simplified loading configurations. In practical engineering applications, however, fracture behaviour is often further influenced by factors such as initial stresses and moving external loads, whose combined effects require careful investigation.

\subsection{Effects of Initial Stress and Moving Loading}
Besides crack interaction and crack velocity, the fracture behaviour of engineering materials is strongly influenced by the presence of initial stresses and moving external loads. These effects are frequently encountered in engineering structures and geotechnical systems, where they alter crack-tip stress fields, modify stress-wave propagation, and consequently influence crack initiation, propagation, and structural integrity. Their incorporation into fracture models is therefore essential for obtaining realistic predictions of the dynamic response of cracked solids.
Initial stresses may develop due to manufacturing processes, residual deformation, thermal gradients, geostatic conditions, or prolonged service loading. Such pre-existing stresses modify the mechanical state of a material before the application of external loads and consequently influence the evolution of fracture. Numerous analytical investigations have examined the role of initial stress in cracked isotropic, anisotropic, and functionally graded media, demonstrating that both the magnitude and orientation of the pre-stress significantly affect stress intensity factors, crack-tip fields, and fracture resistance \cite{chaudhary2025crack,das2026magnetoelastic,diksha2026dynamic}.
Moving external loads constitute another important source of dynamic excitation in many practical applications, including highways, railway tracks, airport pavements, underground excavations, and machine foundations. Unlike stationary loading, moving loads continuously generate stress waves that interact with existing cracks, producing velocity-dependent stress fields and dynamic amplification of fracture parameters. Consequently, moving loading has been widely investigated in dynamic fracture mechanics, particularly for moving Griffith cracks and interacting crack systems, where both crack velocity and loading velocity govern the resulting fracture response \cite{freund1998dynamic,yoffe1951lxxv,das2006interaction,singh2025mathematical,das2026magnetoelastic}.
The simultaneous presence of initial stress and moving loading provides a more realistic representation of service conditions encountered by engineering structures. Their coupled influence governs both stress-wave propagation and crack-tip behaviour, making these effects important considerations in the analytical modelling of dynamic fracture problems.

\subsection{Research Gap and Motivation}
Despite the substantial progress achieved in the fracture analysis of granular materials, dynamic fracture mechanics, and interacting crack systems, several important issues remain unresolved. Existing studies on dry sandy media have predominantly focused on constitutive modelling, particle-scale failure, and wave propagation, whereas analytical investigations of crack propagation remain relatively limited. Similarly, most studies on moving Griffith cracks have been developed for homogeneous, functionally graded, or anisotropic materials without considering the mechanical characteristics of dry sandy media. Furthermore, the combined influence of initial stress, moving punch loading, finite strip geometry, and the interaction of moving collinear Griffith cracks has not been systematically investigated within a unified analytical framework. Since these factors coexist in many practical engineering applications, neglecting their coupled effects may lead to inaccurate predictions of fracture behaviour and structural integrity. These limitations highlight the need for a comprehensive analytical model capable of evaluating the dynamic interaction of moving collinear Griffith cracks in an initially stressed dry sandy strip subjected to moving punch loading, which forms the primary motivation for the present study.

\subsection{Objectives and Contributions of the Present Study}
The dynamic interaction of moving collinear Griffith cracks in an initially stressed dry sandy strip subjected to moving punch loading has not been adequately addressed within a unified analytical framework. To investigate this problem, the present work develops a mathematical model that enables a systematic analysis of the associated fracture behaviour. The principal objectives and scientific contributions of this study are summarized as follows:
\begin{enumerate}
    \item To develop an analytical formulation for the dynamic interaction of two moving collinear Griffith cracks embedded in an initially stressed dry sandy strip subjected to moving punch loading.

    \item To derive closed-form expressions for the crack opening displacement and the mode-I stress intensity factors using the Fourier transform technique in conjunction with the Schmidt method.

    \item To examine the influence of key governing parameters, including crack speed, initial stress, strip thickness, crack geometry, sandiness parameter, and moving punch loading, on the fracture response of the cracked strip through a comprehensive parametric investigation.

    \item To provide new insights into the coupled effects of crack interaction, initial stress, and moving loading on dynamic fracture in dry sandy media, thereby establishing benchmark analytical results for future theoretical, numerical, and experimental studies.
\end{enumerate}

\subsection{Organization of the Manuscript}
The remainder of this paper is organized as follows. Section~\ref{sec:mathematical_framework} presents the mathematical formulation of the problem, including the physical configuration, constitutive relations and governing field equations, moving coordinate framework, and boundary conditions. Section~\ref{sec:transform_analysis} develops the analytical solution by introducing the Fourier transform formulation, deriving the transformed stress fields, and establishing the amplitude reduction equations. The singular integral formulation is presented in Section~\ref{sec:singular_integral_formulation}, where the large-thickness reduction, point loading solution, stress intensity factors, crack opening displacement, and relevant special cases are discussed. Numerical results illustrating the influence of the governing parameters on the fracture behaviour of the cracked strip are presented and discussed in Section~\ref{sec:Numerical_Results}. Finally, the major findings and concluding remarks are summarized in Section~\ref{Conclusion}.

\section{Mathematical Framework for Interacting Moving Cracks}
\label{sec:mathematical_framework}

\subsection{Physical Configuration and Modeling Assumptions}
\label{subsec:physical_configuration}

\begin{figure}
    \centering    \includegraphics[width=1\linewidth]{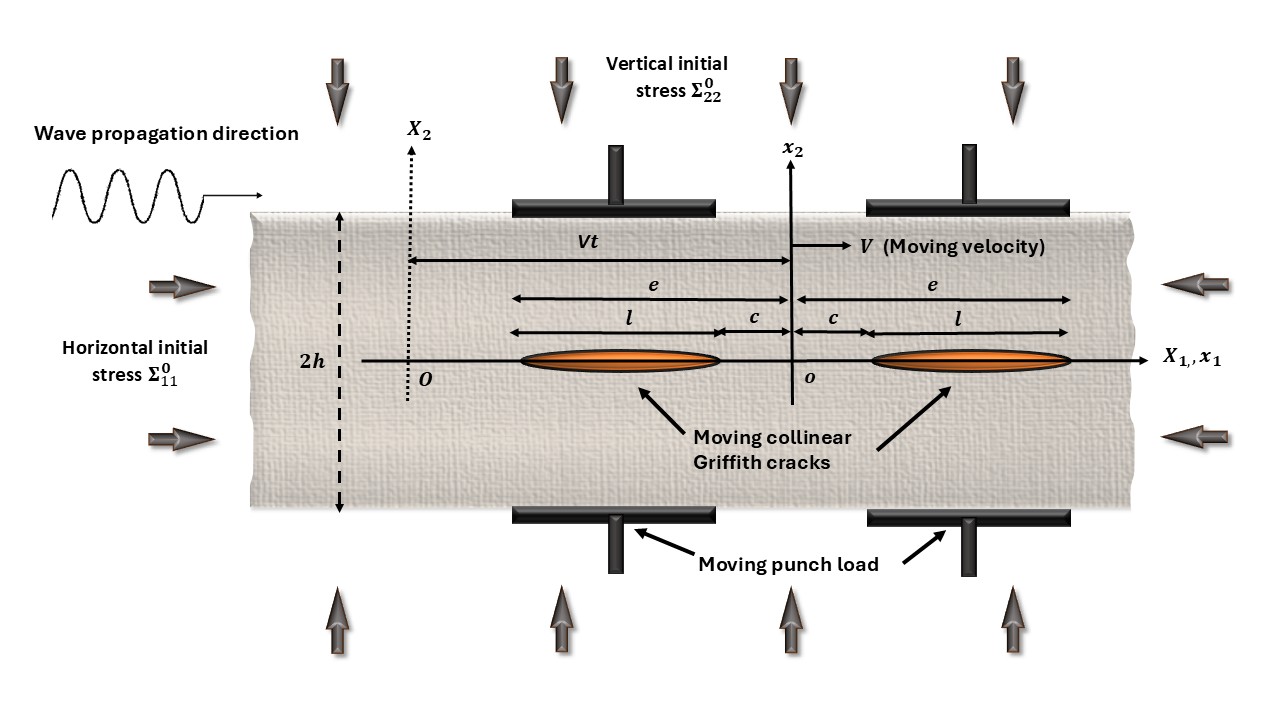}
    \caption{Schematic configuration of the problem showing two moving collinear Griffith cracks embedded in an initially stressed dry sandy elastic strip of finite thickness subjected to moving punch loading.}
    \label{fig:geometry}
\end{figure}

Consider a homogeneous dry sandy elastic strip of finite thickness $2h$
and unbounded extent in the longitudinal direction. A fixed Cartesian
coordinate system $(X_1,X_2,X_3)$ is introduced such that the strip
occupies the region
\begin{equation}
-\infty<X_1<\infty,
\qquad
-h\leq X_2\leq h,
\end{equation}
with the deformation confined to the $(X_1,X_2)$-plane.
Two finite collinear Griffith cracks are embedded symmetrically along the
mid-plane $X_2=0$. In the reference configuration, the cracks occupy the
intervals
\begin{equation}
-e\leq X_1\leq -c,
\qquad
c\leq X_1\leq e,
\qquad
0<c<e.
\label{eq:crack_intervals}
\end{equation}
Accordingly, each crack has length
\begin{equation}
\ell=e-c,
\end{equation}
whereas the intact ligament separating the two inner crack tips has
length $2c$. The crack tips located at $X_1=\pm c$ are hereafter referred
to as the \emph{inner tips}, while those at $X_1=\pm e$ are termed the
\emph{outer tips}. This distinction is retained throughout the analysis
to quantify the interaction between the neighboring cracks.
Normal punch loads are applied to the upper and lower boundaries of the
strip over regions aligned with the crack locations. The crack system and
the associated surface loading are assumed to translate together in the
positive $X_1$-direction with a uniform velocity $V$, while the crack
lengths and the ligament spacing remain unchanged. Under this steady
propagation assumption, the transient problem may be reformulated as a
stationary boundary-value problem in a coordinate system moving with the
cracks.
Prior to crack propagation, the strip is subjected to a uniform biaxial
initial stress field characterized by the normal components
$\Sigma_{11}^{0}$ and $\Sigma_{22}^{0}$ along the longitudinal and
transverse directions, respectively. The material is modeled as a dry
sandy elastic continuum, with the influence of sandiness incorporated
through a constitutive parameter introduced in the following subsection.
Body forces are neglected, and the deformation is assumed to remain
within the framework of infinitesimal plane strain.
The resulting configuration therefore combines the effects of crack
interaction, material sandiness, initial stress, finite strip thickness,
moving surface loading, and steady crack propagation. The geometry of the
problem is illustrated schematically in Fig.~\ref{fig:geometry}.

\subsection{Constitutive Description and Incremental Field Equations}
\label{subsec:constitutive_field_equations}

The mechanical response of the dry sandy strip is described through a
modified isotropic elastic constitutive law in which the resistance to
shear deformation is influenced by the degree of sandiness. The
incremental stress--strain relation is expressed as \cite{singh2025mathematical}
\begin{equation}
\sigma_{ij}
=
\lambda \varepsilon_{kk}\delta_{ij}
+
2\frac{\mu}{\chi}\varepsilon_{ij},
\qquad
\chi>1,
\label{eq:dry_sandy_constitutive}
\end{equation}
where $\sigma_{ij}$ and $\varepsilon_{ij}$ denote the incremental stress
and infinitesimal strain tensors, respectively, $\delta_{ij}$ is the
Kronecker delta, and $\lambda$ and $\mu$ are the Lam\'e elastic
constants. The dimensionless parameter $\chi$ characterizes the
sandiness of the medium. The limiting value $\chi=1$ suppresses the
sandiness effect and recovers the corresponding classical isotropic
elastic response.
For subsequent developments, the following effective elastic
coefficients are introduced:
\begin{equation}
\mathscr{E}
=
\lambda+\frac{2\mu}{\chi},
\qquad
\mathscr{L}
=
\lambda,
\qquad
\mathscr{S}
=
\frac{\mu}{\chi}.
\label{eq:effective_coefficients}
\end{equation}
These quantities represent, respectively, the effective direct,
volumetric-coupling, and shear coefficients of the dry sandy continuum.
The deformation associated with the propagating Mode-I crack system is
assumed to satisfy the plane-strain conditions
\begin{equation}
u_1=u_1(X_1,X_2,t),
\qquad
u_2=u_2(X_1,X_2,t),
\qquad
u_3=0,
\qquad
\frac{\partial}{\partial X_3}\equiv0.
\label{eq:plane_strain}
\end{equation}
Accordingly, the non-zero incremental stress components become \cite{singh2025mathematical}
\begin{align}
\sigma_{11}
&=
\mathscr{E}\frac{\partial u_1}{\partial X_1}
+
\mathscr{L}\frac{\partial u_2}{\partial X_2},
\label{eq:sigma11}
\\
\sigma_{22}
&=
\mathscr{L}\frac{\partial u_1}{\partial X_1}
+
\mathscr{E}\frac{\partial u_2}{\partial X_2},
\label{eq:sigma22}
\\
\sigma_{12}
&=
\mathscr{S}
\left(
\frac{\partial u_1}{\partial X_2}
+
\frac{\partial u_2}{\partial X_1}
\right).
\label{eq:sigma12}
\end{align}
The strip is assumed to possess a uniform biaxial initial stress state
before the onset of crack motion. The non-vanishing components of the
initial stress tensor are taken as
\begin{equation}
\Sigma_{11}^{0}\neq0,
\qquad
\Sigma_{22}^{0}\neq0,
\qquad
\Sigma_{12}^{0}=0.
\label{eq:initial_stress_state}
\end{equation}
In the absence of body forces, the incremental motion of the initially
stressed medium is governed by
\begin{equation}
\sigma_{ij,j}
+
\left(
u_{i,k}\Sigma_{kj}^{0}
\right)_{,j}
=
\rho\ddot{u}_i,
\qquad i=1,2,
\label{eq:incremental_motion}
\end{equation}
where $\rho$ denotes the mass density of the medium and repeated indices
imply summation.
Substitution of Eqs.~(\ref{eq:sigma11})--(\ref{eq:sigma12}) into
Eq.~(\ref{eq:incremental_motion}) yields the coupled field equations
\begin{align}
\left(\mathscr{E}+\Sigma_{11}^{0}\right)
\frac{\partial^2u_1}{\partial X_1^2}
+
\left(\mathscr{S}+\Sigma_{22}^{0}\right)
\frac{\partial^2u_1}{\partial X_2^2}
+
\left(\mathscr{L}+\mathscr{S}\right)
\frac{\partial^2u_2}{\partial X_1\partial X_2}
&=
\rho\frac{\partial^2u_1}{\partial t^2},
\label{eq:field_fixed_1}
\\
\left(\mathscr{S}+\Sigma_{11}^{0}\right)
\frac{\partial^2u_2}{\partial X_1^2}
+
\left(\mathscr{E}+\Sigma_{22}^{0}\right)
\frac{\partial^2u_2}{\partial X_2^2}
+
\left(\mathscr{L}+\mathscr{S}\right)
\frac{\partial^2u_1}{\partial X_1\partial X_2}
&=
\rho\frac{\partial^2u_2}{\partial t^2}.
\label{eq:field_fixed_2}
\end{align}

\subsection{Steady-State Field Equations in the Crack-Fixed Frame}
\label{subsec:moving_frame}

Since two collinear cracks propagate steadily along positive
$X_1$-direction with constant velocity $V$, a translating coordinate
system $(x_1,x_2,x_3)$ attached to the moving crack configuration is
introduced through
\begin{equation}
x_1=X_1-Vt,
\qquad
x_2=X_2,
\qquad
x_3=X_3.
\label{eq:moving_coordinates}
\end{equation}
Using Eq.~(\ref{eq:moving_coordinates})
in Eqs.~(\ref{eq:field_fixed_1}) and (\ref{eq:field_fixed_2}), the
governing equations in the crack-fixed coordinate system become
\begin{align}
\left(
\mathscr{E}+\Sigma_{11}^{0}-\rho V^2
\right)
\frac{\partial^2u_1}{\partial x_1^2}
+
\left(
\mathscr{S}+\Sigma_{22}^{0}
\right)
\frac{\partial^2u_1}{\partial x_2^2}
+
\left(
\mathscr{L}+\mathscr{S}
\right)
\frac{\partial^2u_2}{\partial x_1\partial x_2}
&=0,
\label{eq:moving_field_1}
\\
\left(
\mathscr{S}+\Sigma_{11}^{0}-\rho V^2
\right)
\frac{\partial^2u_2}{\partial x_1^2}
+
\left(
\mathscr{E}+\Sigma_{22}^{0}
\right)
\frac{\partial^2u_2}{\partial x_2^2}
+
\left(
\mathscr{L}+\mathscr{S}
\right)
\frac{\partial^2u_1}{\partial x_1\partial x_2}
&=0.
\label{eq:moving_field_2}
\end{align}

\subsection{Mechanical Boundary and Crack-Face Conditions}
\label{subsec:boundary_conditions}

In the crack-fixed coordinate system, the two collinear cracks remain
stationary along the mid-plane $x_2=0$ and occupy the intervals
\begin{equation}
-e\leq x_1\leq -c,
\qquad
c\leq x_1\leq e.
\end{equation}
The moving punch loads are likewise stationary in this frame and act
normally on the external surfaces $x_2=\pm h$ over regions corresponding
to the crack locations. Owing to the symmetry of the geometry and
loading with respect to the mid-plane, it is sufficient to formulate
the problem in the upper half-strip
\begin{equation}
0\leq x_2\leq h.
\end{equation}

At the upper boundary $x_2=h$, the surface is constrained outside the
loaded intervals, whereas a prescribed normal punch pressure acts over
the regions aligned with the cracks. The corresponding conditions are
\begin{align}
u_1(x_1,h)&=0,
&& |x_1|<c,\quad |x_1|>e,
\label{eq:upper_u1}
\\
u_2(x_1,h)&=0,
&& |x_1|<c,\quad |x_1|>e,
\label{eq:upper_u2}
\\
\sigma_{12}(x_1,h)&=0,
&& -\infty<x_1<\infty,
\label{eq:upper_shear}
\\
\sigma_{22}(x_1,h)&=-q_s(x_1),
&& c\leq |x_1|\leq e,
\label{eq:upper_normal}
\end{align}
where $q_s(x_1)$ denotes the prescribed normal pressure exerted by the
moving punch.

Along the mid-plane $x_2=0$, the displacement field vanishes over the
intact portions of the strip, while the crack faces are subjected to a
prescribed normal traction. Accordingly,
\begin{align}
u_1(x_1,0)&=0,
&& |x_1|<c,\quad |x_1|>e,
\label{eq:mid_u1}
\\
u_2(x_1,0)&=0,
&& |x_1|<c,\quad |x_1|>e,
\label{eq:mid_u2}
\\
\sigma_{12}(x_1,0)&=0,
&& -\infty<x_1<\infty,
\label{eq:mid_shear}
\\
\sigma_{22}(x_1,0)&=-q_c(x_1),
&& c\leq |x_1|\leq e,
\label{eq:crack_normal}
\end{align}
where $q_c(x_1)$ represents the normal traction prescribed on the crack
faces.

The conditions above, together with the governing equations
(\ref{eq:moving_field_1}) and (\ref{eq:moving_field_2}), define the
boundary-value problem for the steadily propagating interacting crack
system in the initially stressed dry sandy strip.

\section{Transform-Domain Analysis}
\label{sec:transform_analysis}

\subsection{Fourier Representation and Characteristic Structure}
\label{subsec:fourier_characteristic}

The symmetry of the Mode-I deformation with respect to the plane
$x_1=0$ suggests representing the longitudinal and transverse
displacement components by Fourier sine and cosine integrals,
respectively. Accordingly, the displacement field is written as
\begin{align}
u_1(x_1,x_2)
&=
\int_{0}^{\infty}
U(\zeta,x_2)\sin(\zeta x_1)\,d\zeta,
\label{eq:u1_transform}
\\
u_2(x_1,x_2)
&=
\int_{0}^{\infty}
W(\zeta,x_2)\cos(\zeta x_1)\,d\zeta,
\label{eq:u2_transform}
\end{align}
where $\zeta$ denotes the transform parameter, while
$U(\zeta,x_2)$ and $W(\zeta,x_2)$ are unknown transform-domain
amplitudes.
Substitution of Eqs.~(\ref{eq:u1_transform}) and
(\ref{eq:u2_transform}) into the moving-frame field equations
(\ref{eq:moving_field_1}) and (\ref{eq:moving_field_2}) gives
\begin{align}
\left(\mathscr{S}+\Sigma_{22}^{0}\right)
\frac{d^2U}{dx_2^2}
-
\left(\mathscr{L}+\mathscr{S}\right)
\zeta\frac{dW}{dx_2}
-
\left(
\mathscr{E}+\Sigma_{11}^{0}-\rho V^2
\right)
\zeta^2U
&=0,
\label{eq:transform_ode_1}
\\
\left(\mathscr{E}+\Sigma_{22}^{0}\right)
\frac{d^2W}{dx_2^2}
+
\left(\mathscr{L}+\mathscr{S}\right)
\zeta\frac{dU}{dx_2}
-
\left(
\mathscr{S}+\Sigma_{11}^{0}-\rho V^2
\right)
\zeta^2W
&=0.
\label{eq:transform_ode_2}
\end{align}
For compactness, the following velocity- and initial-stress-dependent
coefficients are introduced:
\begin{equation}
\begin{aligned}
\mathscr{D}_1
&=
\mathscr{E}+\Sigma_{11}^{0}-\rho V^2,
&
\mathscr{D}_2
&=
\mathscr{S}+\Sigma_{22}^{0},
\\
\mathscr{D}_3
&=
\mathscr{S}+\Sigma_{11}^{0}-\rho V^2,
&
\mathscr{D}_4
&=
\mathscr{E}+\Sigma_{22}^{0},
\\
\mathscr{D}_5
&=
\mathscr{L}+\mathscr{S}.
\end{aligned}
\label{eq:dynamic_coefficients}
\end{equation}
Hence, Eqs.~(\ref{eq:transform_ode_1}) and
(\ref{eq:transform_ode_2}) may be written in the compact form
\begin{align}
\mathscr{D}_2 U''
-
\mathscr{D}_5\zeta W'
-
\mathscr{D}_1\zeta^2U
&=0,
\label{eq:compact_ode_1}
\\
\mathscr{D}_4 W''
+
\mathscr{D}_5\zeta U'
-
\mathscr{D}_3\zeta^2W
&=0,
\label{eq:compact_ode_2}
\end{align}
where a prime denotes differentiation with respect to $x_2$.
To determine the admissible thickness-direction modes, solutions of the
form
\begin{equation}
U(\zeta,x_2)=U_0e^{r\zeta x_2},
\qquad
W(\zeta,x_2)=W_0e^{r\zeta x_2}
\label{eq:modal_trial}
\end{equation}
are considered. Substitution into
Eqs.~(\ref{eq:compact_ode_1}) and (\ref{eq:compact_ode_2}) leads to
\begin{equation}
\begin{bmatrix}
\mathscr{D}_2r^2-\mathscr{D}_1
&
-\mathscr{D}_5r
\\[1ex]
\mathscr{D}_5r
&
\mathscr{D}_4r^2-\mathscr{D}_3
\end{bmatrix}
\begin{bmatrix}
U_0\\
W_0
\end{bmatrix}
=
\begin{bmatrix}
0\\
0
\end{bmatrix}.
\label{eq:modal_system}
\end{equation}
A non-trivial modal solution exists only when the determinant of the
coefficient matrix vanishes. This condition yields the biquadratic
characteristic equation
\begin{equation}
\mathscr{B}_1r^4
-
\mathscr{B}_2r^2
+
\mathscr{B}_3
=
0,
\label{eq:characteristic_equation}
\end{equation}
where
\begin{align}
\mathscr{B}_1
&=
\mathscr{D}_2\mathscr{D}_4,
\label{eq:B1}
\\
\mathscr{B}_2
&=
\mathscr{D}_2\mathscr{D}_3
+
\mathscr{D}_1\mathscr{D}_4
-
\mathscr{D}_5^2,
\label{eq:B2}
\\
\mathscr{B}_3
&=
\mathscr{D}_1\mathscr{D}_3.
\label{eq:B3}
\end{align}
The two admissible positive roots are therefore given by
\begin{equation}
r_{1,2}
=
\left[
\frac{
\mathscr{B}_2
\pm
\sqrt{\mathscr{B}_2^2-4\mathscr{B}_1\mathscr{B}_3}
}{
2\mathscr{B}_1
}
\right]^{1/2}.
\label{eq:characteristic_roots}
\end{equation}
Only parameter ranges for which the selected roots are compatible with
the physically admissible steady-state response are considered in the
subsequent analysis.
The characteristic equation (\ref{eq:characteristic_equation}) admits the
four roots $\pm r_1$ and $\pm r_2$. Accordingly, the general solutions of
Eqs.~(\ref{eq:compact_ode_1}) and (\ref{eq:compact_ode_2}) may initially
be expressed as
\begin{align}
U(\zeta,x_2)
={}&
\Phi_1(\zeta)\cosh(r_1\zeta x_2)
+
\Phi_2(\zeta)\cosh(r_2\zeta x_2)
\nonumber\\
&+
\Psi_1(\zeta)\sinh(r_1\zeta x_2)
+
\Psi_2(\zeta)\sinh(r_2\zeta x_2),
\label{eq:U_general}
\\
W(\zeta,x_2)
={}&
\widehat{\Phi}_1(\zeta)\sinh(r_1\zeta x_2)
+
\widehat{\Phi}_2(\zeta)\sinh(r_2\zeta x_2)
\nonumber\\
&+
\widehat{\Psi}_1(\zeta)\cosh(r_1\zeta x_2)
+
\widehat{\Psi}_2(\zeta)\cosh(r_2\zeta x_2),
\label{eq:W_general_initial}
\end{align}
where $\Phi_j$, $\Psi_j$, $\widehat{\Phi}_j$, and
$\widehat{\Psi}_j$ $(j=1,2)$ are transform-domain functions.

Substitution of Eqs.~(\ref{eq:U_general}) and
(\ref{eq:W_general_initial}) into either of the coupled equations
(\ref{eq:compact_ode_1})--(\ref{eq:compact_ode_2}) establishes the
following relations between the corresponding modal amplitudes:
\begin{equation}
\widehat{\Phi}_j(\zeta)
=
\vartheta_j\Phi_j(\zeta),
\qquad
\widehat{\Psi}_j(\zeta)
=
\vartheta_j\Psi_j(\zeta),
\qquad j=1,2,
\label{eq:modal_amplitude_relations}
\end{equation}
where
\begin{equation}
\vartheta_j
=
\frac{
\mathscr{D}_2r_j^2-\mathscr{D}_1
}{
\mathscr{D}_5r_j
},
\qquad j=1,2.
\label{eq:modal_coupling}
\end{equation}
\subsection{Transform-Domain Stress Fields}
\label{subsec:transform_stress_fields}

The stress components required for enforcing the mechanical boundary
conditions are obtained by substituting the Fourier representations
(\ref{eq:u1_transform})--(\ref{eq:u2_transform}), together with the
reduced transform-domain solutions (\ref{eq:U_general}) and
(\ref{eq:W_general_initial}), into the constitutive relations
(\ref{eq:sigma12}) and (\ref{eq:sigma22}).
For compactness, the following modal traction coefficients are introduced:
\begin{align}
\Lambda_j
&=
\mathscr{S}
\left(
r_j-\vartheta_j
\right),
\label{eq:Lambda_j}
\\
\Pi_j
&=
\mathscr{L}
+
\mathscr{E}\vartheta_j r_j,
\qquad j=1,2.
\label{eq:Pi_j}
\end{align}
The coefficients $\Lambda_j$ and $\Pi_j$ characterize the contributions
of the $j$th thickness mode to the shear and normal tractions,
respectively.
The shear stress component is then obtained as
\begin{align}
\sigma_{12}(x_1,x_2)
=
\int_{0}^{\infty}
\Big[
&\Lambda_1\Phi_1(\zeta)
\sinh(r_1\zeta x_2)
+
\Lambda_2\Phi_2(\zeta)
\sinh(r_2\zeta x_2)
\nonumber\\
&+
\Lambda_1\Psi_1(\zeta)
\cosh(r_1\zeta x_2)
+
\Lambda_2\Psi_2(\zeta)
\cosh(r_2\zeta x_2)
\Big]
\zeta\sin(\zeta x_1)\,d\zeta.
\label{eq:sigma12_transform}
\end{align}
Similarly, the normal stress component takes the form
\begin{align}
\sigma_{22}(x_1,x_2)
=
\int_{0}^{\infty}
\Big[
&\Pi_1\Phi_1(\zeta)
\cosh(r_1\zeta x_2)
+
\Pi_2\Phi_2(\zeta)
\cosh(r_2\zeta x_2)
\nonumber\\
&+
\Pi_1\Psi_1(\zeta)
\sinh(r_1\zeta x_2)
+
\Pi_2\Psi_2(\zeta)
\sinh(r_2\zeta x_2)
\Big]
\zeta\cos(\zeta x_1)\,d\zeta.
\label{eq:sigma22_transform}
\end{align}
\subsection{Reduction of the Transform Amplitudes}
\label{subsec:amplitude_reduction}

Imposing the shear-free condition (\ref{eq:mid_shear})
$\sigma_{12}(x_1,0)=0$ on Eq.~(\ref{eq:sigma12_transform}) yields
\begin{equation}
\Psi_2(\zeta)
=
-\frac{\Lambda_1}{\Lambda_2}\Psi_1(\zeta).
\label{eq:Psi2_relation}
\end{equation}
The displacement condition \ref{eq:mid_u2} $u_2(x_1,0)=0$ over the intact portions of
the mid-plane is satisfied by introducing the representation
\begin{equation}
\Psi_1(\zeta)
=
\frac{1}{\zeta}
\int_c^e
\mathfrak{g}_1(s)\sin(\zeta s)\,ds.
\label{eq:Psi1_representation}
\end{equation}
Similarly, the transverse displacement condition at the upper boundary
is incorporated through
\begin{align}
&\vartheta_1\Phi_1(\zeta)\sinh(r_1\zeta h)
+
\vartheta_2\Phi_2(\zeta)\sinh(r_2\zeta h)
\nonumber\\
&\quad+
\vartheta_1\Psi_1(\zeta)\cosh(r_1\zeta h)
+
\vartheta_2\Psi_2(\zeta)\cosh(r_2\zeta h)
=
\frac{1}{\zeta}
\int_c^e
\mathfrak{g}_2(s)\sin(\zeta s)\,ds.
\label{eq:g2_representation}
\end{align}
The auxiliary density functions $\mathfrak{g}_1(s)$ and
$\mathfrak{g}_2(s)$ satisfy
\begin{equation}
\int_c^e\mathfrak{g}_1(s)\,ds=0,
\qquad
\int_c^e\mathfrak{g}_2(s)\,ds=0.
\label{eq:density_constraints}
\end{equation}
Using the upper-surface shear-free condition (\ref{eq:upper_shear})
$\sigma_{12}(x_1,h)=0$ together with
Eq.~(\ref{eq:g2_representation}), the remaining transform amplitudes are
obtained as
\begin{align}
\Phi_1(\zeta)
={}&
-\frac{1+\mathcal{F}_1(\zeta)}{\zeta}
\int_c^e
\mathfrak{g}_1(s)\sin(\zeta s)\,ds
+
\frac{\mathcal{F}_2(\zeta)}{\zeta}
\int_c^e
\mathfrak{g}_2(s)\sin(\zeta s)\,ds,
\label{eq:Phi1_final}
\\
\Phi_2(\zeta)
={}&
\frac{\Lambda_1}{\Lambda_2}
\frac{1+\mathcal{F}_3(\zeta)}{\zeta}
\int_c^e
\mathfrak{g}_1(s)\sin(\zeta s)\,ds
-
\frac{\mathcal{F}_4(\zeta)}{\zeta}
\int_c^e
\mathfrak{g}_2(s)\sin(\zeta s)\,ds,
\label{eq:Phi2_final}
\end{align}
where
\begin{align}
\mathcal{F}_1(\zeta)
&=
\frac{e^{-r_1\zeta h}}
{\sinh(r_1\zeta h)},
&
\mathcal{F}_3(\zeta)
&=
\frac{e^{-r_2\zeta h}}
{\sinh(r_2\zeta h)},
\label{eq:F13}
\\
\mathcal{F}_2(\zeta)
&=
\frac{\Lambda_2}
{\Delta\,\sinh(r_1\zeta h)},
&
\mathcal{F}_4(\zeta)
&=
\frac{\Lambda_1}
{\Delta\,\sinh(r_2\zeta h)},
\label{eq:F24}
\end{align}
with
\begin{equation*}
\Delta
=
\vartheta_1\Lambda_2
-
\vartheta_2\Lambda_1.
\label{eq:Delta}
\end{equation*}

\section{Coupled Singular Integral Formulation}
\label{sec:singular_integral_formulation}

Substitution of Eqs.~(\ref{eq:Psi1_representation}),
(\ref{eq:Psi2_relation}), (\ref{eq:Phi1_final}), and
(\ref{eq:Phi2_final}) into the normal-traction conditions (\ref{eq:upper_normal}) and (\ref{eq:crack_normal})
\begin{equation}
\sigma_{22}(x_1,0)=-q_c(x_1),
\qquad
\sigma_{22}(x_1,h)=-q_s(x_1),
\qquad
c\leq |x_1|\leq e,
\end{equation}
leads to a pair of coupled singular integral equations for the unknown
density functions $\mathfrak{g}_1(s)$ and $\mathfrak{g}_2(s)$:
\begin{align}
\int_c^e
\frac{s\,\mathfrak{g}_1(s)}
{s^2-x_1^2}\,ds
&+
\frac{1}{2}
\int_c^e
\mathcal{K}_{11}(x_1,s)\mathfrak{g}_1(s)\,ds
\nonumber\\
&+
\frac{1}{2}
\int_c^e
\mathcal{K}_{12}(x_1,s)\mathfrak{g}_2(s)\,ds
=
-\mathcal{Q}_1q_c(x_1),
\label{eq:singular_equation_1}
\\
\int_c^e
\frac{s\,\mathfrak{g}_2(s)}
{s^2-x_1^2}\,ds
&+
\frac{1}{2}
\int_c^e
\mathcal{K}_{21}(x_1,s)\mathfrak{g}_1(s)\,ds
\nonumber\\
&+
\frac{1}{2}
\int_c^e
\mathcal{K}_{22}(x_1,s)\mathfrak{g}_2(s)\,ds
=
-\mathcal{Q}_2q_s(x_1).
\label{eq:singular_equation_2}
\end{align}
The singular integrals in
Eqs.~(\ref{eq:singular_equation_1}) and
(\ref{eq:singular_equation_2}) are interpreted in the Cauchy
principal-value sense.
For compactness, introduce
\begin{equation}
\mathcal{C}
=
\Pi_2\Lambda_1-\Pi_1\Lambda_2,
\label{eq:C_coefficient}
\end{equation}
together with the normalization coefficients
\begin{equation}
\mathcal{Q}_1
=
\frac{\Lambda_2}{\mathcal{C}},
\qquad
\mathcal{Q}_2
=
-\frac{\Delta}{\mathcal{C}},
\label{eq:Q_coefficients}
\end{equation}
where $\Delta$ is defined in Eq.~(\ref{eq:Delta}).
The regular kernels appearing in the coupled system are represented by
\begin{equation}
\mathcal{K}_{ij}(x_1,s)
=
\int_{0}^{\infty}
\mathcal{N}_{ij}(\zeta)
\left[
\sin\{\zeta(s+x_1)\}
+
\sin\{\zeta(s-x_1)\}
\right]d\zeta,
\qquad i,j=1,2,
\label{eq:kernel_definition}
\end{equation}
where the corresponding transform-domain kernel multipliers are
\begin{align}
\mathcal{N}_{11}(\zeta)
=
\mathcal{N}_{22}(\zeta)
&=
\frac{1}{\mathcal{C}}
\left[
\Pi_2\Lambda_1\mathcal{F}_3(\zeta)
-
\Pi_1\Lambda_2\mathcal{F}_1(\zeta)
\right],
\label{eq:N11N22}
\\
\mathcal{N}_{12}(\zeta)
&=
\frac{\Lambda_2}{\mathcal{C}}
\left[
\Pi_1\mathcal{F}_2(\zeta)
-
\Pi_2\mathcal{F}_4(\zeta)
\right].
\label{eq:N12}
\end{align}
To express the remaining kernel multiplier, define
\begin{align}
\mathcal{F}_5(\zeta)
&=
\Pi_1
\left[
e^{-r_1\zeta h}
+
\mathcal{F}_1(\zeta)
\cosh(r_1\zeta h)
\right],
\label{eq:F5}
\\
\mathcal{F}_6(\zeta)
&=
\Pi_2
\left[
e^{-r_2\zeta h}
+
\mathcal{F}_3(\zeta)
\cosh(r_2\zeta h)
\right].
\label{eq:F6}
\end{align}
Accordingly,
\begin{equation}
\mathcal{N}_{21}(\zeta)
=
\mathcal{Q}_2
\left[
\mathcal{F}_5(\zeta)
-
\frac{\Lambda_1}{\Lambda_2}
\mathcal{F}_6(\zeta)
\right].
\label{eq:N21}
\end{equation}
The auxiliary density functions are further constrained by
\begin{equation}
\int_c^e\mathfrak{g}_1(s)\,ds=0,
\qquad
\int_c^e\mathfrak{g}_2(s)\,ds=0.
\label{eq:supplementary_conditions}
\end{equation}
Thus, Eqs.~(\ref{eq:singular_equation_1}) and
(\ref{eq:singular_equation_2}), together with
Eq.~(\ref{eq:supplementary_conditions}), provide the reduced integral
description of the interacting moving-crack problem.
\subsection{Large-Thickness Asymptotic Reduction}
\label{subsec:large_thickness_reduction}

For sufficiently large values of the strip half-thickness $h$, the
finite-thickness contributions contained in the kernel functions are
approximated by retaining the leading non-vanishing terms. Using the
series identities
\begin{equation}
\sum_{k=1}^{\infty}\frac{1}{k^2}
=
\frac{\pi^2}{6},
\qquad
\sum_{k=0}^{\infty}\frac{1}{(2k+1)^2}
=
\frac{\pi^2}{8},
\label{eq:series_identities}
\end{equation}
the regular kernels admit the asymptotic forms
\begin{align}
\mathcal{K}_{11}(x_1,s)
=
\mathcal{K}_{22}(x_1,s)
&=
-\frac{\pi^2\mathcal{P}}{6h^2}\,s
+
O\left(h^{-4}\right),
\label{eq:K11_asymptotic}
\\
\mathcal{K}_{12}(x_1,s)
&=
\frac{\pi^2\Lambda_2\mathcal{P}}
{2\Delta h^2}\,s
+
O\left(h^{-4}\right),
\label{eq:K12_asymptotic}
\\
\mathcal{K}_{21}(x_1,s)
&=
\frac{\pi^2\Delta\mathcal{P}}
{2\Lambda_2h^2}\,s
+
O\left(h^{-4}\right),
\label{eq:K21_asymptotic}
\end{align}
where
\begin{equation}
\mathcal{P}
=
\frac{1}{\mathcal{C}}
\left(
\frac{\Pi_1\Lambda_2}{r_1^2}
-
\frac{\Pi_2\Lambda_1}{r_2^2}
\right).
\label{eq:P_coefficient}
\end{equation}
Accordingly, the unknown density functions are expanded asymptotically as
\begin{equation}
\mathfrak{g}_j(s)
=
\mathfrak{g}_j^{(0)}(s)
+
\frac{1}{h^2}\mathfrak{g}_j^{(1)}(s)
+
O\left(h^{-4}\right),
\qquad j=1,2.
\label{eq:density_asymptotic_expansion}
\end{equation}
Substitution of Eqs.~(\ref{eq:K11_asymptotic})--
(\ref{eq:K21_asymptotic}) and
(\ref{eq:density_asymptotic_expansion}) into
Eqs.~(\ref{eq:singular_equation_1}) and
(\ref{eq:singular_equation_2}), followed by comparison of the
coefficients of the terms of orders $O(1)$ and $O(h^{-2})$, yields
\begin{align}
\int_c^e
\frac{2s\,\mathfrak{g}_1^{(0)}(s)}
{s^2-x_1^2}\,ds
&=
-2\mathcal{Q}_1q_c(x_1),
\label{eq:g10}
\\
\int_c^e
\frac{2s\,\mathfrak{g}_2^{(0)}(s)}
{s^2-x_1^2}\,ds
&=
-2\mathcal{Q}_2q_s(x_1),
\label{eq:g20}
\\
\int_c^e
\frac{2s\,\mathfrak{g}_1^{(1)}(s)}
{s^2-x_1^2}\,ds
&=
\frac{\pi^2\mathcal{P}}{6}
\int_c^e
s\,\mathfrak{g}_1^{(0)}(s)\,ds
\nonumber\\
&\quad
-
\frac{\pi^2\Lambda_2\mathcal{P}}
{2\Delta}
\int_c^e
s\,\mathfrak{g}_2^{(0)}(s)\,ds,
\label{eq:g11}
\\
\int_c^e
\frac{2s\,\mathfrak{g}_2^{(1)}(s)}
{s^2-x_1^2}\,ds
&=
\frac{\pi^2\mathcal{P}}{6}
\int_c^e
s\,\mathfrak{g}_2^{(0)}(s)\,ds
\nonumber\\
&\quad
-
\frac{\pi^2\Delta\mathcal{P}}
{2\Lambda_2}
\int_c^e
s\,\mathfrak{g}_1^{(0)}(s)\,ds.
\label{eq:g21}
\end{align}

\subsection{Solution for Concentrated Crack-Face Loading}
\label{subsec:point_loading_solution}

Consider a concentrated normal load of intensity $Q_c$ applied
symmetrically to the crack faces at $x_1=\pm x_0$, where
$c<x_0<e$. Owing to the symmetry of the formulation, the loading over
the positive crack interval may be represented as
\begin{equation}
q_c(x_1)
=
Q_c\,\delta(x_1-x_0),
\qquad
c<x_0<e,
\label{eq:point_loading}
\end{equation}
where $\delta(\cdot)$ denotes the Dirac delta distribution. The normal
pressure exerted by the moving punch on the outer surface is taken to be
uniform over the loaded interval, such that
\begin{equation}
q_s(x_1)=Q_s,
\qquad
c\leq x_1\leq e.
\label{eq:uniform_surface_loading}
\end{equation}
Application of the finite Hilbert transform \cite{srivastava1970xx} to
Eqs.~(\ref{eq:g10})--(\ref{eq:g21}) gives the asymptotic density
functions in the form
\begin{align}
\mathfrak{g}_1(s)
={}&
\frac{4\mathcal{Q}_1Q_c}{\pi^2}
\sqrt{\frac{s^2-c^2}{e^2-s^2}}
\sqrt{\frac{e^2-x_0^2}{x_0^2-c^2}}
\frac{x_0}{x_0^2-s^2}
+
\frac{\mathcal{D}_1}
{\sqrt{(s^2-c^2)(e^2-s^2)}}
\nonumber\\
&+
\frac{1}{h^2}
\left[
\frac{\mathcal{H}_1}{\pi}
\sqrt{\frac{s^2-c^2}{e^2-s^2}}
+
\frac{\mathcal{D}_3}
{\sqrt{(s^2-c^2)(e^2-s^2)}}
\right]
+
O\left(h^{-4}\right),
\label{eq:g1_point}
\\
\mathfrak{g}_2(s)
={}&
-\frac{2\mathcal{Q}_2Q_s}{\pi}
\sqrt{\frac{s^2-c^2}{e^2-s^2}}
+
\frac{\mathcal{D}_2}
{\sqrt{(s^2-c^2)(e^2-s^2)}}
\nonumber\\
&+
\frac{1}{h^2}
\left[
\frac{\mathcal{H}_2}{\pi}
\sqrt{\frac{s^2-c^2}{e^2-s^2}}
+
\frac{\mathcal{D}_4}
{\sqrt{(s^2-c^2)(e^2-s^2)}}
\right]
+
O\left(h^{-4}\right),
\label{eq:g2_point}
\end{align}
where $\mathcal{D}_j$ $(j=1,2,3,4)$ are integration constants
determined from the supplementary conditions
(\ref{eq:supplementary_conditions}). The coefficients
$\mathcal{H}_1$ and $\mathcal{H}_2$ account for the leading
finite-thickness correction and are given by
The coefficients $\mathcal{H}_1$ and $\mathcal{H}_2$, which characterize
the leading finite-thickness corrections, are given by
\begin{align}
\mathcal{H}_1
={}&
\frac{\pi^2\mathcal{P}}{12}
\left[
-\frac{4\mathcal{Q}_1Q_c x_0}{\pi}
\sqrt{\frac{e^2-x_0^2}{x_0^2-c^2}}
+
\pi\mathcal{D}_1
\right]
\nonumber\\
&-
\frac{\pi^2\Lambda_2\mathcal{P}}{4\Delta}
\left[
-\mathcal{Q}_2Q_s\left(e^2-c^2\right)
+
\pi\mathcal{D}_2
\right],
\label{eq:H1}
\\[1ex]
\mathcal{H}_2
={}&
\frac{\pi^2\mathcal{P}}{12}
\left[
-\mathcal{Q}_2Q_s\left(e^2-c^2\right)
+
\pi\mathcal{D}_2
\right]
\nonumber\\
&-
\frac{\pi^2\Delta\mathcal{P}}{4\Lambda_2}
\left[
-\frac{4\mathcal{Q}_1Q_c x_0}{\pi}
\sqrt{\frac{e^2-x_0^2}{x_0^2-c^2}}
+
\pi\mathcal{D}_1
\right].
\label{eq:H2}
\end{align}

\subsection{Dynamic Stress Intensity Factors}
\label{subsec:stress_intensity_factors}

The Mode-I dynamic stress intensity factors associated with the inner
and outer tips of the moving collinear cracks are defined, respectively \cite{diksha2026dynamic},
by
\begin{align}
K_I^{(c)}
&=
\lim_{x_1\to c^-}
\sqrt{2\pi(c-x_1)}\,
\sigma_{22}(x_1,0),
\label{eq:SIF_inner}
\\
K_I^{(e)}
&=
\lim_{x_1\to e^+}
\sqrt{2\pi(x_1-e)}\,
\sigma_{22}(x_1,0).
\label{eq:SIF_outer}
\end{align}
Substitution of the crack-density solution
(\ref{eq:g1_point}) into the crack-plane stress representation, followed
by evaluation of the near-tip asymptotic fields, gives the Mode-I
dynamic stress intensity factors at the inner and outer crack tips as
\begin{equation}
K_I^{(c)}
=
\frac{\pi}{2\mathcal{Q}_1}
\frac{1}
{\sqrt{c}\sqrt{e^2-c^2}}
\left(
\mathcal{D}_1
+
\frac{\mathcal{D}_3}{h^2}
\right).
\label{eq:KI_inner}
\end{equation}
At the outer crack tip, the corresponding stress intensity factor is
\begin{align}
K_I^{(e)}
={}&
\frac{\pi}{2\mathcal{Q}_1}
\frac{1}
{\sqrt{e}\sqrt{e^2-c^2}}
\Bigg[
\frac{4\mathcal{Q}_1Q_c}{\pi^2}
\frac{x_0(e^2-c^2)}
{\sqrt{x_0^2-c^2}\sqrt{e^2-x_0^2}}
\nonumber\\
&\qquad\qquad
+
\frac{\mathcal{H}_1(e^2-c^2)}
{\pi h^2}
-
\left(
\mathcal{D}_1
+
\frac{\mathcal{D}_3}{h^2}
\right)
\Bigg].
\label{eq:KI_outer}
\end{align}
\subsection{Crack Opening Displacement}
\label{subsec:crack_opening_displacement}

The crack opening displacement associated with the moving collinear
cracks is obtained from the displacement discontinuity across the crack
faces. Substitution of the density function
(\ref{eq:g1_point}) into the corresponding displacement representation
gives \cite{diksha2026dynamic}
\begin{align}
\Delta u_2(x_1,0)=&
u_2(x_1,0^+)-u_2(x_1,0^-)\nonumber\\
=&{}
\pi\mathcal{B}
\Bigg\{
\frac{4\mathcal{Q}_1Q_c}{\pi^2}
\sqrt{\frac{e^2-x_0^2}{x_0^2-c^2}}\,
x_0
\int_{x_1}^{e}
\frac{\sqrt{s^2-c^2}}
{\sqrt{e^2-s^2}\,(x_0^2-s^2)}
\,ds
\nonumber\\
&\qquad
+
\frac{\mathcal{H}_1}{\pi h^2}
\left[
eE(\phi,\kappa)
-
\frac{c^2}{e}F(\phi,\kappa)
\right]
\nonumber\\
&\qquad
+
\frac{1}{e}
\left(
\mathcal{D}_1
+
\frac{\mathcal{D}_3}{h^2}
\right)
F(\phi,\kappa)
\Bigg\},
\qquad
c\leq x_1\leq e,
\label{eq:COD_point}
\end{align}
where $\mathcal{B}$ denotes the displacement-coupling coefficient
arising from the modal representation
\begin{equation}
\mathcal{B}
=
\vartheta_1
-
\vartheta_2\frac{\Lambda_1}{\Lambda_2}.
\label{eq:B_COD}
\end{equation}.
The arguments of the incomplete
elliptic integrals are defined by
\begin{equation}
\phi
=
\sin^{-1}
\sqrt{
\frac{e^2-x_1^2}{e^2-c^2}
},
\qquad
\kappa
=
\frac{\sqrt{e^2-c^2}}{e}.
\label{eq:elliptic_parameters}
\end{equation}
Here, $F(\phi,\kappa)$ and $E(\phi,\kappa)$ denote the incomplete
elliptic integrals of the first and second kinds, respectively.

\subsection{Special Cases and Limiting Configurations}
\label{subsec:special_cases}

The general formulation admits several limiting configurations that
provide useful consistency checks and facilitate the isolation of the
effects associated with crack motion and initial stress.

\subsubsection{Stationary-crack limit}

The stationary configuration is recovered by setting the crack
propagation speed equal to zero,
\begin{equation}
V=0.
\label{eq:stationary_limit}
\end{equation}
Accordingly, the velocity-dependent effective coefficients reduce to
\begin{equation}
\mathscr{D}_j^{(\mathrm{s})}
=
\left.\mathscr{D}_j\right|_{V=0},
\qquad j=1,2,\ldots,5,
\label{eq:D_stationary}
\end{equation}
where the superscript $(\mathrm{s})$ denotes the stationary
configuration. The coefficients of the characteristic equation
therefore become
\begin{equation}
A_k^{(\mathrm{s})}
=
\left.A_k\right|_{V=0},
\qquad k=1,2,3,
\label{eq:A_stationary}
\end{equation}
and the attenuation parameters are determined from
\begin{equation}
A_1^{(\mathrm{s})}r^4
-
A_2^{(\mathrm{s})}r^2
+
A_3^{(\mathrm{s})}
=
0.
\label{eq:characteristic_stationary}
\end{equation}
Hence,
\begin{equation}
\left(r_j^{(\mathrm{s})}\right)^2
=
\frac{
A_2^{(\mathrm{s})}
+
(-1)^{j+1}
\sqrt{
\left(A_2^{(\mathrm{s})}\right)^2
-
4A_1^{(\mathrm{s})}A_3^{(\mathrm{s})}
}
}{
2A_1^{(\mathrm{s})}
},
\qquad j=1,2.
\label{eq:roots_stationary}
\end{equation}
The corresponding modal and traction coefficients follow as
\begin{align}
\vartheta_j^{(\mathrm{s})}
&=
\left.\vartheta_j\right|_{V=0},
&
\Lambda_j^{(\mathrm{s})}
&=
\left.\Lambda_j\right|_{V=0},
&
\Pi_j^{(\mathrm{s})}
&=
\left.\Pi_j\right|_{V=0},
\qquad j=1,2.
\label{eq:modal_stationary}
\end{align}
Consequently, all quantities entering the singular integral equations
are evaluated using these stationary coefficients, and the corresponding
Mode-I stress intensity factors are obtained as
\begin{equation}
K_{I,\mathrm{s}}^{(c)}
=
\left.K_I^{(c)}\right|_{V=0},
\qquad
K_{I,\mathrm{s}}^{(e)}
=
\left.K_I^{(e)}\right|_{V=0}.
\label{eq:SIF_stationary}
\end{equation}

\subsubsection{Initially unstressed-medium limit}

The fracture response in the absence of initial stress is obtained by
setting
\begin{equation}
\sigma_{11}^{0}=0,
\qquad
\sigma_{22}^{0}=0.
\label{eq:unstressed_limit}
\end{equation}
The effective coefficients then reduce to
\begin{equation}
\mathscr{D}_j^{(\mathrm{u})}
=
\left.
\mathscr{D}_j
\right|_{\sigma_{11}^{0}=\sigma_{22}^{0}=0},
\qquad j=1,2,\ldots,5,
\label{eq:D_unstressed}
\end{equation}
where the superscript $(\mathrm{u})$ identifies the initially
unstressed configuration. Accordingly,
\begin{equation}
A_k^{(\mathrm{u})}
=
\left.
A_k
\right|_{\sigma_{11}^{0}=\sigma_{22}^{0}=0},
\qquad k=1,2,3,
\label{eq:A_unstressed}
\end{equation}
and the characteristic equation becomes
\begin{equation}
A_1^{(\mathrm{u})}r^4
-
A_2^{(\mathrm{u})}r^2
+
A_3^{(\mathrm{u})}
=
0.
\label{eq:characteristic_unstressed}
\end{equation}
The corresponding roots are
\begin{equation}
\left(r_j^{(\mathrm{u})}\right)^2
=
\frac{
A_2^{(\mathrm{u})}
+
(-1)^{j+1}
\sqrt{
\left(A_2^{(\mathrm{u})}\right)^2
-
4A_1^{(\mathrm{u})}A_3^{(\mathrm{u})}
}
}{
2A_1^{(\mathrm{u})}
},
\qquad j=1,2.
\label{eq:roots_unstressed}
\end{equation}
Thus,
\begin{align}
\vartheta_j^{(\mathrm{u})}
&=
\left.
\vartheta_j
\right|_{\sigma_{11}^{0}=\sigma_{22}^{0}=0},
&
\Lambda_j^{(\mathrm{u})}
&=
\left.
\Lambda_j
\right|_{\sigma_{11}^{0}=\sigma_{22}^{0}=0},
\nonumber\\
\Pi_j^{(\mathrm{u})}
&=
\left.
\Pi_j
\right|_{\sigma_{11}^{0}=\sigma_{22}^{0}=0},
\qquad j=1,2.
\label{eq:modal_unstressed}
\end{align}
The resulting stress intensity factors are therefore
\begin{equation}
K_{I,\mathrm{u}}^{(c)}
=
\left.
K_I^{(c)}
\right|_{\sigma_{11}^{0}=\sigma_{22}^{0}=0},
\qquad
K_{I,\mathrm{u}}^{(e)}
=
\left.
K_I^{(e)}
\right|_{\sigma_{11}^{0}=\sigma_{22}^{0}=0}.
\label{eq:SIF_unstressed}
\end{equation}

\subsubsection{Punch-Free, Sandiness-Free, and Initially Unstressed Limit}
\label{subsubsec:baseline_material_limit}

A further limiting configuration is obtained by simultaneously removing
the outer-surface punch pressure, the sandiness effect, and the initial
stress. The corresponding conditions are
\begin{equation}
Q_s=0,
\qquad
\chi=0,
\qquad
\sigma_{11}^{0}=\sigma_{22}^{0}=0,
\label{eq:punch_sandiness_stress_free_limit}
\end{equation}
where $\chi$ denotes the parameter characterizing the sandiness effect
in the constitutive description of the medium. The crack propagation
speed $V$ is retained, so that the resulting formulation continues to
describe a dynamically propagating pair of collinear cracks.

Under the conditions in Eq.~(\ref{eq:punch_sandiness_stress_free_limit}),
the effective coefficients entering the governing equations reduce to
\begin{equation}
\mathscr{D}_j^{(\mathrm{b})}
=
\left.
\mathscr{D}_j
\right|_{
Q_s=0,\,
\chi=0,\,
\sigma_{11}^{0}=\sigma_{22}^{0}=0
},
\qquad
j=1,2,\ldots,5,
\label{eq:D_baseline_material}
\end{equation}
where the superscript $(\mathrm{b})$ identifies the present baseline
configuration. The associated coefficients of the characteristic
equation are therefore
\begin{equation}
A_k^{(\mathrm{b})}
=
\left.
A_k
\right|_{
\chi=0,\,
\sigma_{11}^{0}=\sigma_{22}^{0}=0
},
\qquad
k=1,2,3,
\label{eq:A_baseline_material}
\end{equation}
and the attenuation parameters satisfy
\begin{equation}
A_1^{(\mathrm{b})}r^4
-
A_2^{(\mathrm{b})}r^2
+
A_3^{(\mathrm{b})}
=
0.
\label{eq:characteristic_baseline_material}
\end{equation}
Accordingly,
\begin{equation}
\left(r_j^{(\mathrm{b})}\right)^2
=
\frac{
A_2^{(\mathrm{b})}
+
(-1)^{j+1}
\sqrt{
\left(A_2^{(\mathrm{b})}\right)^2
-
4A_1^{(\mathrm{b})}A_3^{(\mathrm{b})}
}
}{
2A_1^{(\mathrm{b})}
},
\qquad
j=1,2.
\label{eq:roots_baseline_material}
\end{equation}
The corresponding modal and traction coefficients are obtained as
\begin{align}
\vartheta_j^{(\mathrm{b})}
&=
\left.
\vartheta_j
\right|_{
\chi=0,\,
\sigma_{11}^{0}=\sigma_{22}^{0}=0
},
&
\Lambda_j^{(\mathrm{b})}
&=
\left.
\Lambda_j
\right|_{
\chi=0,\,
\sigma_{11}^{0}=\sigma_{22}^{0}=0
},
\nonumber\\
\Pi_j^{(\mathrm{b})}
&=
\left.
\Pi_j
\right|_{
\chi=0,\,
\sigma_{11}^{0}=\sigma_{22}^{0}=0
},
\qquad j=1,2.
\label{eq:modal_baseline_material}
\end{align}
Since the external punch pressure is absent, the leading-order
surface-loading contribution to the second density function vanishes.
Consequently,
\begin{equation}
\mathfrak{g}_2^{(0)}(s)
=
\frac{\mathcal{D}_2}
{\sqrt{(s^2-c^2)(e^2-s^2)}}.
\label{eq:g2_baseline_leading}
\end{equation}
The first density function retains the contribution associated with the
concentrated crack-face loading and is given by
\begin{align}
\mathfrak{g}_1^{(0)}(s)
={}&
\frac{4\mathcal{Q}_1^{(\mathrm{b})}Q_c}{\pi^2}
\sqrt{\frac{s^2-c^2}{e^2-s^2}}
\sqrt{\frac{e^2-x_0^2}{x_0^2-c^2}}
\frac{x_0}{x_0^2-s^2}
\nonumber\\
&+
\frac{\mathcal{D}_1}
{\sqrt{(s^2-c^2)(e^2-s^2)}}.
\label{eq:g1_baseline_leading}
\end{align}
For this limiting configuration, the first-order correction coefficients
reduce to
\begin{align}
\mathcal{H}_1^{(\mathrm{b})}
={}&
\frac{\pi^2\mathcal{P}^{(\mathrm{b})}}{12}
\left[
-\frac{4\mathcal{Q}_1^{(\mathrm{b})}Q_c x_0}{\pi}
\sqrt{\frac{e^2-x_0^2}{x_0^2-c^2}}
+
\pi\mathcal{D}_1
\right]
\nonumber\\
&-
\frac{\pi^3\Lambda_2^{(\mathrm{b})}
\mathcal{P}^{(\mathrm{b})}}
{4\Delta^{(\mathrm{b})}}
\mathcal{D}_2,
\label{eq:H1_baseline}
\\
\mathcal{H}_2^{(\mathrm{b})}
={}&
\frac{\pi^3\mathcal{P}^{(\mathrm{b})}}{12}
\mathcal{D}_2
\nonumber\\
&-
\frac{\pi^2\Delta^{(\mathrm{b})}
\mathcal{P}^{(\mathrm{b})}}
{4\Lambda_2^{(\mathrm{b})}}
\left[
-\frac{4\mathcal{Q}_1^{(\mathrm{b})}Q_c x_0}{\pi}
\sqrt{\frac{e^2-x_0^2}{x_0^2-c^2}}
+
\pi\mathcal{D}_1
\right].
\label{eq:H2_baseline}
\end{align}
The complete asymptotic density functions for this limiting case are
therefore
\begin{align}
\mathfrak{g}_1^{(\mathrm{b})}(s)
={}&
\mathfrak{g}_1^{(0)}(s)
+
\frac{1}{h^2}
\left[
\frac{\mathcal{H}_1^{(\mathrm{b})}}{\pi}
\sqrt{\frac{s^2-c^2}{e^2-s^2}}
+
\frac{\mathcal{D}_3}
{\sqrt{(s^2-c^2)(e^2-s^2)}}
\right]
+
O(h^{-4}),
\label{eq:g1_baseline_complete}
\\
\mathfrak{g}_2^{(\mathrm{b})}(s)
={}&
\frac{\mathcal{D}_2}
{\sqrt{(s^2-c^2)(e^2-s^2)}}
+
\frac{1}{h^2}
\left[
\frac{\mathcal{H}_2^{(\mathrm{b})}}{\pi}
\sqrt{\frac{s^2-c^2}{e^2-s^2}}
+
\frac{\mathcal{D}_4}
{\sqrt{(s^2-c^2)(e^2-s^2)}}
\right]
.\label{eq:g2_baseline_complete}
\end{align}
Thus, this limiting configuration isolates the fracture response
generated by the concentrated crack-face loading and the dynamic
interaction between the propagating collinear cracks, while excluding
the effects of the moving punch pressure, material sandiness, and
initial stress. Under these restrictions, the present formulation
reduces to the corresponding special case reported in
\cite{diksha2026dynamic}, thereby providing an analytical consistency
check for the generalized model developed herein.

\section{Numerical Results and Discussion}
\label{sec:Numerical_Results}
This section presents the numerical results illustrating the dynamic fracture behaviour of two moving collinear Griffith cracks propagating in an initially stressed dry sandy elastic strip subjected to a moving concentrated load. The analytical solutions developed in the preceding sections are employed to examine the influence of the governing material, geometrical, loading, and initial stress parameters on the normalized Mode-I stress intensity factors at the inner and outer crack tips. The material properties adopted in the present study are taken from \cite{singh2025mathematical} unless otherwise stated. In each numerical example, only the parameter under investigation is varied, while all other parameters are maintained at their respective reference values to clearly demonstrate its individual influence on the fracture response.
The stress intensity factors are presented in the normalized forms
$\frac{|K_I^{(c)}|}{Q_c\sqrt{e}}
\quad \text{and} \quad
\frac{|K_I^{(e)}|}{Q_c\sqrt{e}},$
corresponding to the inner and outer crack tips, respectively. The crack propagation speed is characterized by the normalized crack-speed ratio $V/\beta$, where $V$ denotes the crack propagation speed and
$\beta=\sqrt{\frac{S_{\mathrm{eff}}}{\rho}}
$
is the effective shear-wave velocity of the dry sandy medium, with $S_{\mathrm{eff}}=\mu/\chi$ representing the effective shear modulus.

\begin{table}[!ht]
\centering

\rotatebox{90}{%
\begin{minipage}{0.92\textheight}
\centering

\captionof{table}{Reference values of the material, geometrical, loading, and initial stress parameters used in the numerical study. In each figure, only the parameter indicated as ``Varied'' is changed, while all remaining parameters are maintained at their corresponding reference values.}
\label{tab:NumericalParameters}

\vspace{0.3cm}

\renewcommand{\arraystretch}{1.2}
\small

\begin{tabular}{lccccc}
\hline
\textbf{Parameter} &
\textbf{Fig.~\ref{fig:sandiness}} &
\textbf{Fig.~\ref{fig:prestress}} &
\textbf{Fig.~\ref{fig:cracklength}} &
\textbf{Fig.~\ref{fig:thickness}} &
\textbf{Fig.~\ref{fig:loading}} \\
\hline

Lam\'e constant, $\lambda$ (Pa)
& $2.510\times10^{10}$ & $2.510\times10^{10}$ &
$2.510\times10^{10}$ & $2.510\times10^{10}$ &
$2.510\times10^{10}$ \\

Shear modulus, $\mu$ (Pa)
& $1.987\times10^{10}$ & $1.987\times10^{10}$ &
$1.987\times10^{10}$ & $1.987\times10^{10}$ &
$1.987\times10^{10}$ \\

Density, $\rho$ (kg\,m$^{-3}$)
& 4705 & 4705 & 4705 & 4705 & 4705 \\

Sandiness parameter, $\chi$
& Varied & 1.30 & 1.30 & 1.30 & 1.30 \\

Initial horizontal stress, $\sigma_{11}^{0}$ (Pa)
& $1.0\times10^{9}$ & Varied &
$1.0\times10^{9}$ & $1.0\times10^{9}$ &
$1.0\times10^{9}$ \\

Initial vertical stress, $\sigma_{22}^{0}$ (Pa)
& $1.0\times10^{9}$ & Varied &
$1.0\times10^{9}$ & $1.0\times10^{9}$ &
$1.0\times10^{9}$ \\

Normalized inner crack coordinate, $c/e$
& 0.10 & 0.10 & Varied & 0.10 & 0.10 \\

Normalized strip half-thickness, $h/e$
& 5 & 5 & 5 & Varied & 5 \\

Normalized load position, $x_0/e$
& 0.96 & 0.96 & 0.96 & 0.96 & Varied \\

Tangential loading ratio, $Q_s/Q_c$
& 2.1 & 2.1 & 2.1 & 2.1 & Varied \\

Normal concentrated load, $Q_c$ (N)
& $1.0\times10^{8}$ &
$1.0\times10^{8}$ &
$1.0\times10^{8}$ &
$1.0\times10^{8}$ &
$1.0\times10^{8}$ \\

Outer crack-tip coordinate, $e$
& 1.40 & 1.40 & 1.40 & 1.40 & 1.40 \\

Normalized crack-speed ratio, $V/\beta$
& $0.01$--$0.94$ &
$0.01$--$0.94$ &
$0.01$--$0.94$ &
$0.01$--$0.94$ &
$0.01$--$0.94$ \\

\hline
\end{tabular}

\end{minipage}
}

\end{table}
\subsection{Variation of stress intensity factors with crack speed}
Figures~\ref{fig:sandiness}--\ref{fig:loading} collectively illustrate the variation of the normalized Mode-I stress intensity factors at the inner and outer crack tips with the normalized crack-speed ratio $V/\beta$ for different material, geometrical, loading, and initial-stress parameters. The corresponding reference values of the parameters used in these parametric studies are listed in Table~\ref{tab:NumericalParameters}.
It is evident from Figs.~\ref{fig:sandiness}--\ref{fig:loading} that, irrespective of the governing material, initial stress, geometrical, or loading parameter, all curves exhibit a common dynamic trend. At relatively low values of the normalized crack-speed ratio $V/\beta$, the normalized Mode-I stress intensity factors at both the inner and outer crack tips vary only slightly. However, as $V/\beta$ approaches unity, both stress intensity factors increase rapidly, indicating significant dynamic amplification of the crack-tip stress field.
This rapid increase is associated with the inertial effects generated by the moving crack and the reduction of the effective wave-transmission capability of the surrounding medium. Consequently, the mechanical fields become increasingly localized near the crack tips, producing higher stress concentrations. In all cases, the stress intensity factors remain finite over the investigated range but attain their maximum values close to the limiting crack speed, suggesting that unstable crack propagation is most likely to occur in this high-speed regime.
A comparison of the inner and outer crack-tip responses in Figs.~\ref{fig:sandiness}--\ref{fig:loading} further reveals that the normalized Mode-I stress intensity factor at the outer crack tip is generally larger than that at the inner crack tip. This difference is attributed to the closer proximity of the moving concentrated load to the outer crack tip, which causes greater stress concentration and makes the outer crack tip more susceptible to crack extension.

\subsubsection{Effect of the sandiness parameter}

\begin{figure*}[!t]
\centering
\begin{subfigure}{0.48\textwidth}
\includegraphics[width=\linewidth]{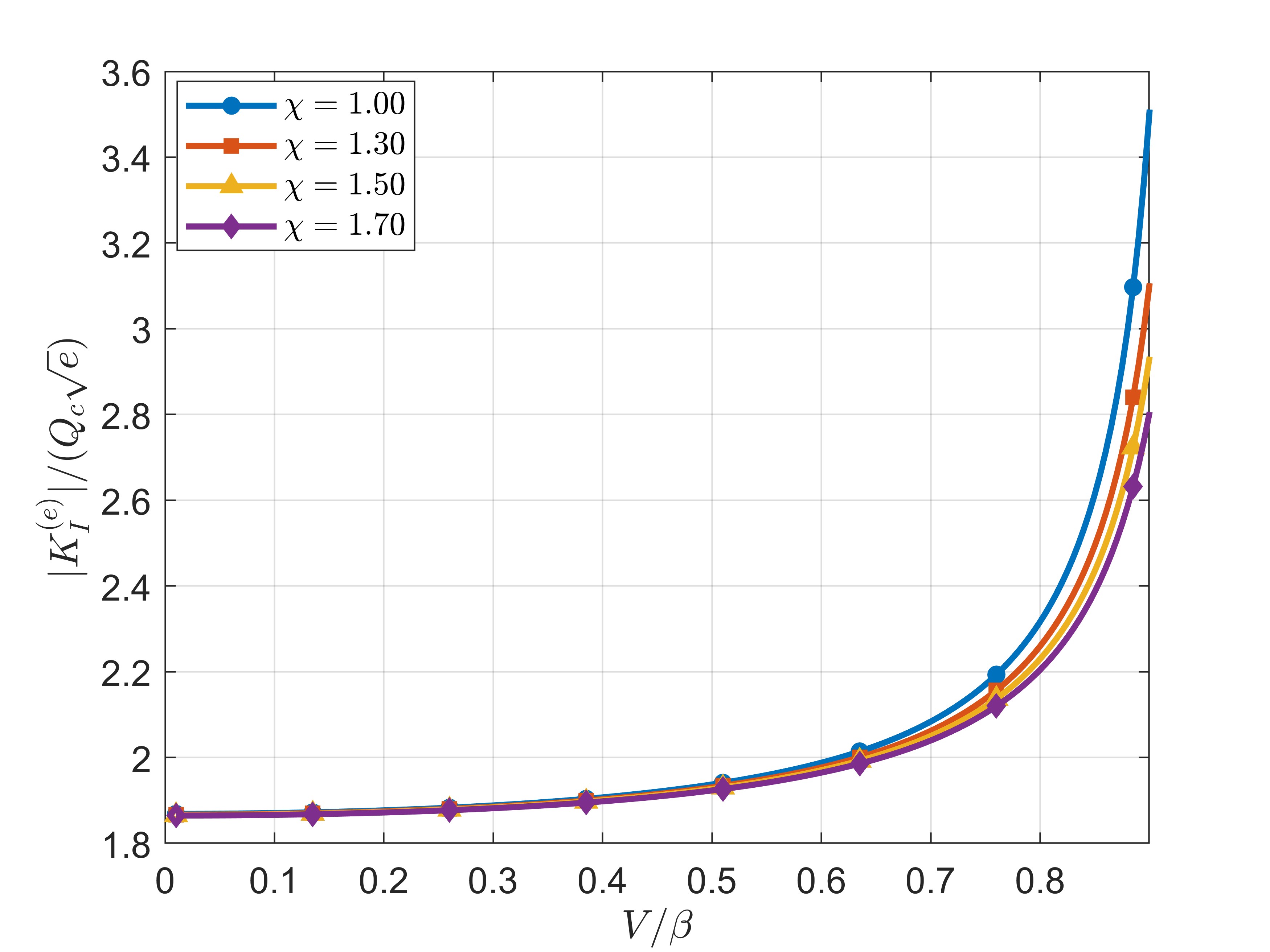}
\caption{Outer crack tip.}
\end{subfigure}
\hfill
\begin{subfigure}{0.48\textwidth}
\includegraphics[width=\linewidth]{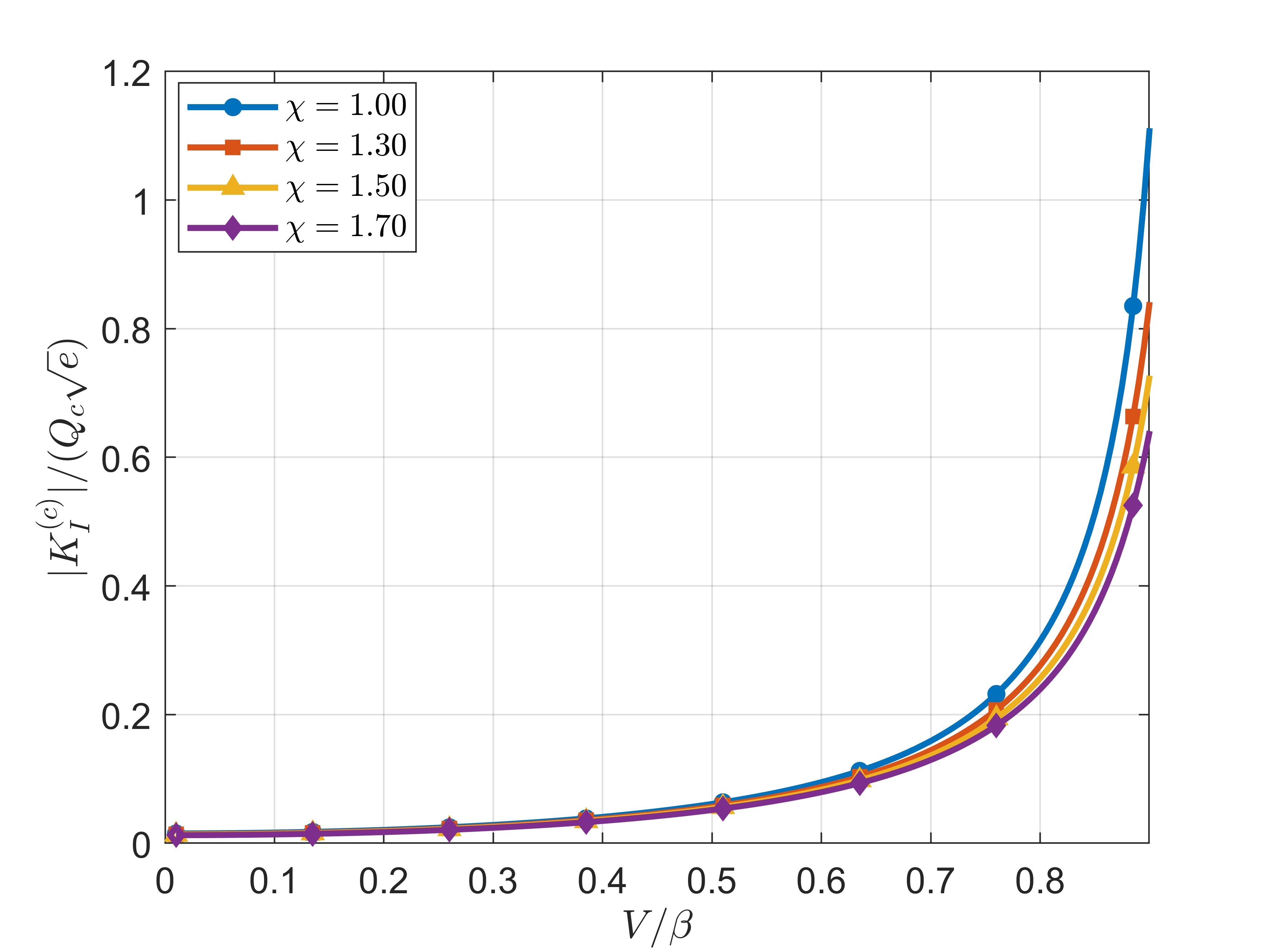}
\caption{Inner crack tip.}
\end{subfigure}
\caption{Variation of the normalized Mode-I stress intensity factor with crack-speed ratio for different values of the sandiness parameter $\chi$.}
\label{fig:sandiness}
\end{figure*}

Figure~\ref{fig:sandiness} presents the influence of the sandiness parameter $\chi$ on the normalized Mode-I stress intensity factors. It is observed that increasing the sandiness parameter reduces both the inner and outer crack-tip stress intensity factors throughout the investigated crack-speed range. The reduction is relatively small at low crack speeds but becomes increasingly significant as the crack velocity approaches the limiting wave speed.
The decrease in stress intensity factor can be attributed to the reduction in the effective shear rigidity associated with increasing sandiness. As the material becomes more compliant, a larger portion of the deformation is distributed throughout the surrounding medium rather than being concentrated near the crack tips. Consequently, the crack-tip stress concentration decreases, leading to improved resistance against dynamic crack propagation. The influence of the sandiness parameter is more pronounced at the outer crack tip because of its stronger interaction with the moving concentrated load.

\subsubsection{Effect of the initial stresses}

\begin{figure*}[!t]
\centering

\begin{subfigure}{0.48\textwidth}
\includegraphics[width=\linewidth]{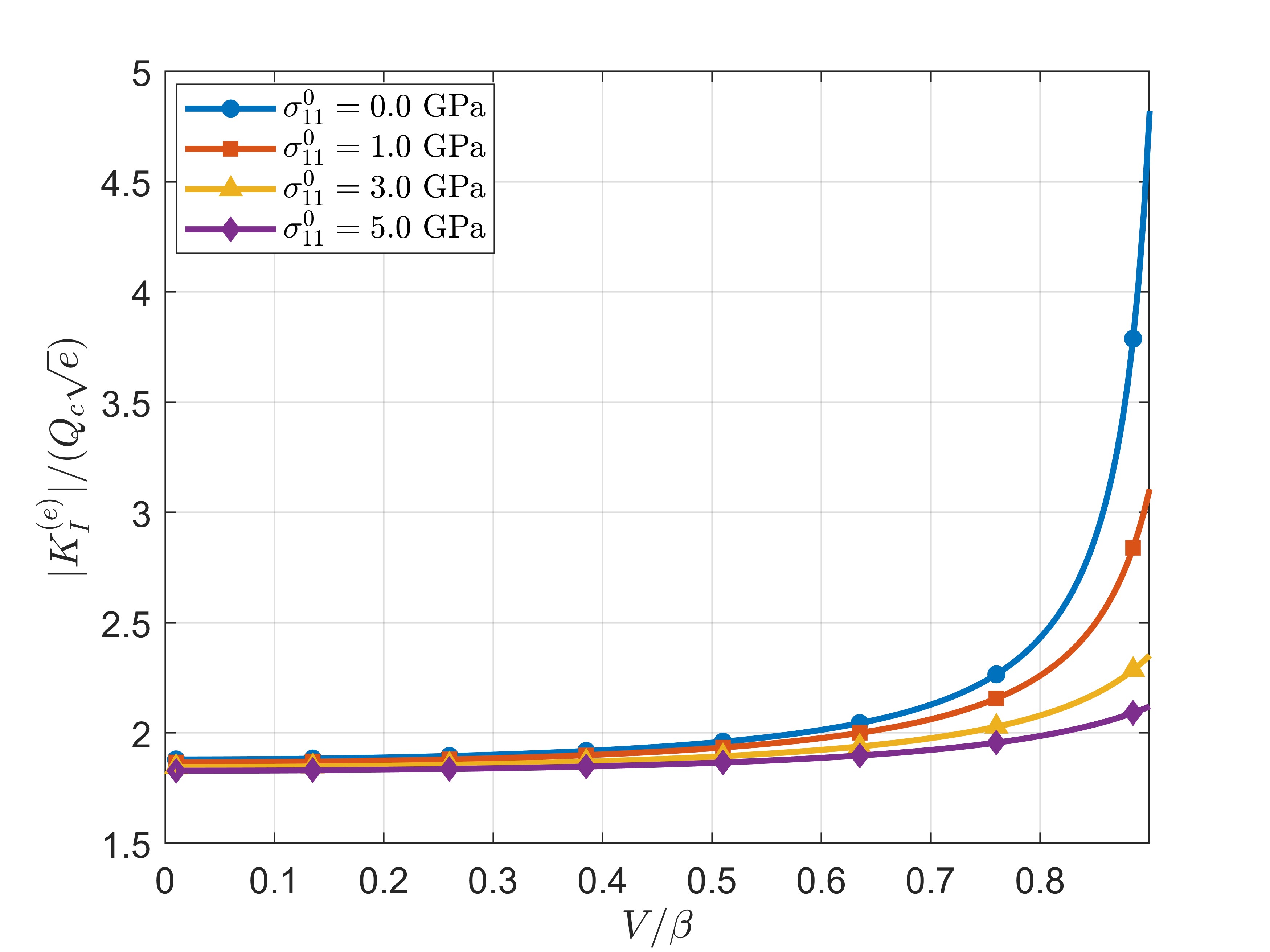}
\caption{Effect of $\sigma_{11}^{0}$.}
\label{fig:prestress(a)}
\end{subfigure}
\hfill
\begin{subfigure}{0.48\textwidth}
\includegraphics[width=\linewidth]{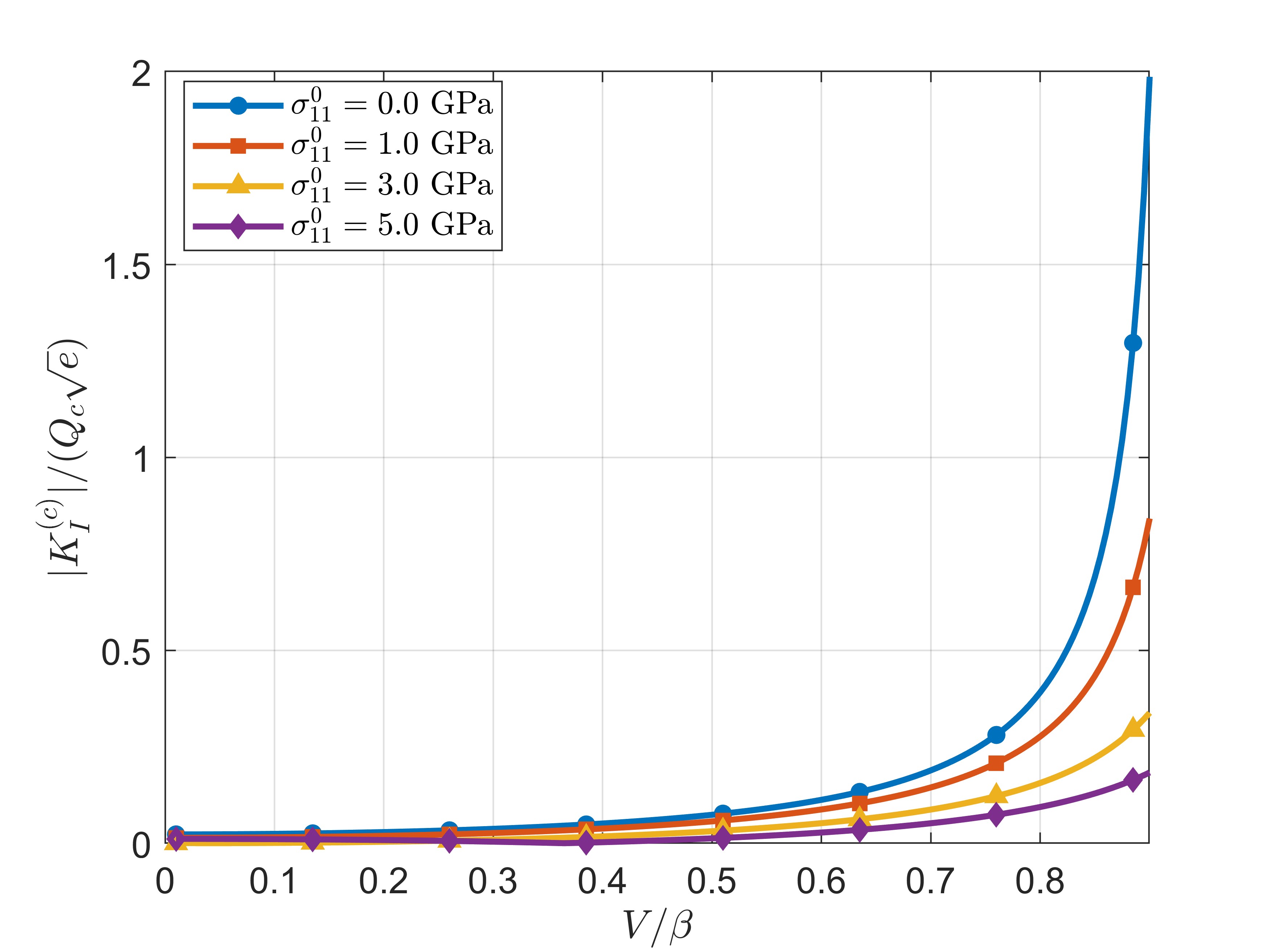}
\caption{Effect of $\sigma_{11}^{0}$.}
\label{fig:prestress(b)}
\end{subfigure}

\vspace{0.3cm}

\begin{subfigure}{0.48\textwidth}
\includegraphics[width=\linewidth]{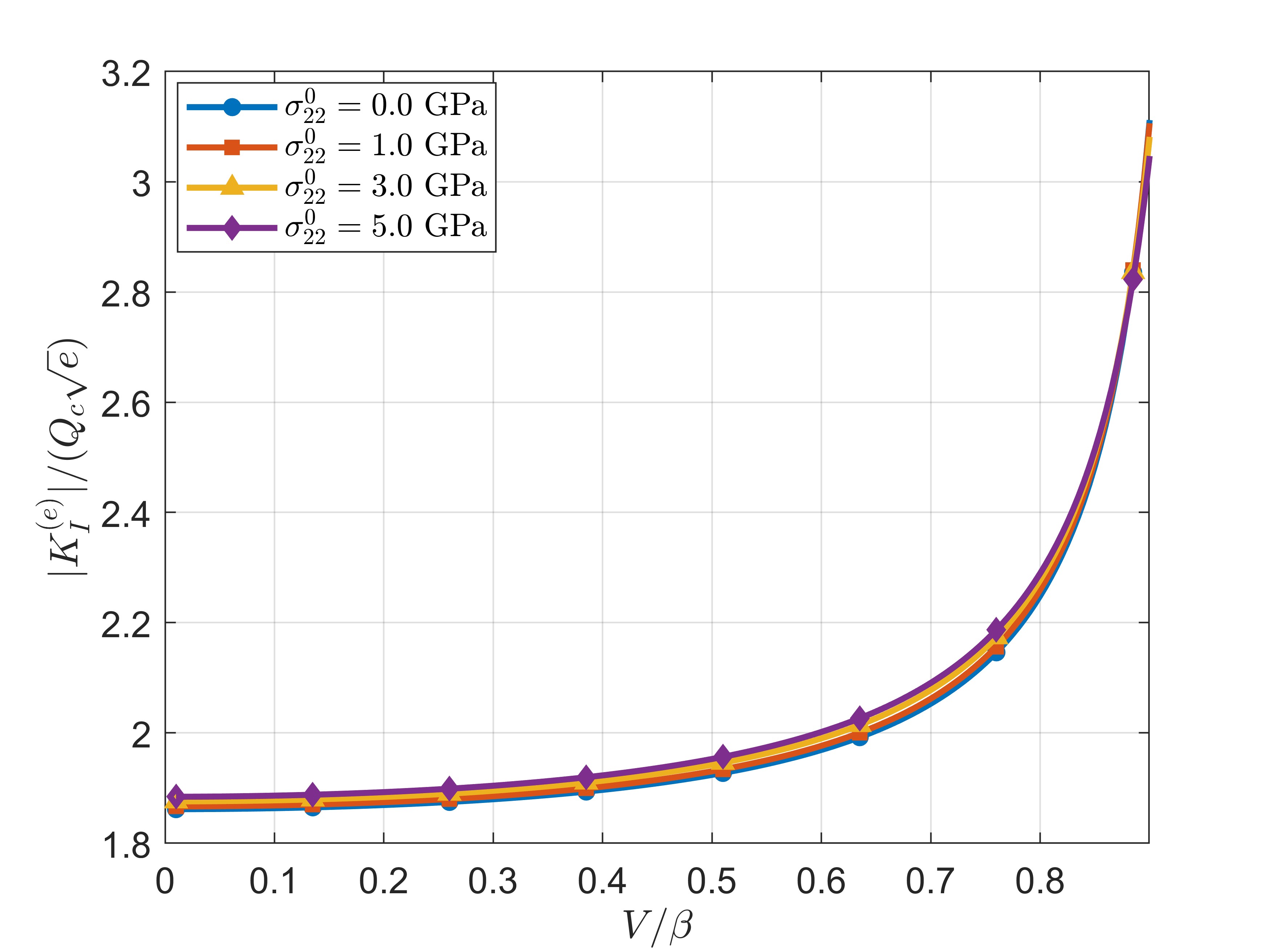}
\caption{Effect of $\sigma_{22}^{0}$.}
\label{fig:prestress(2c)}
\end{subfigure}
\hfill
\begin{subfigure}{0.48\textwidth}
\includegraphics[width=\linewidth]{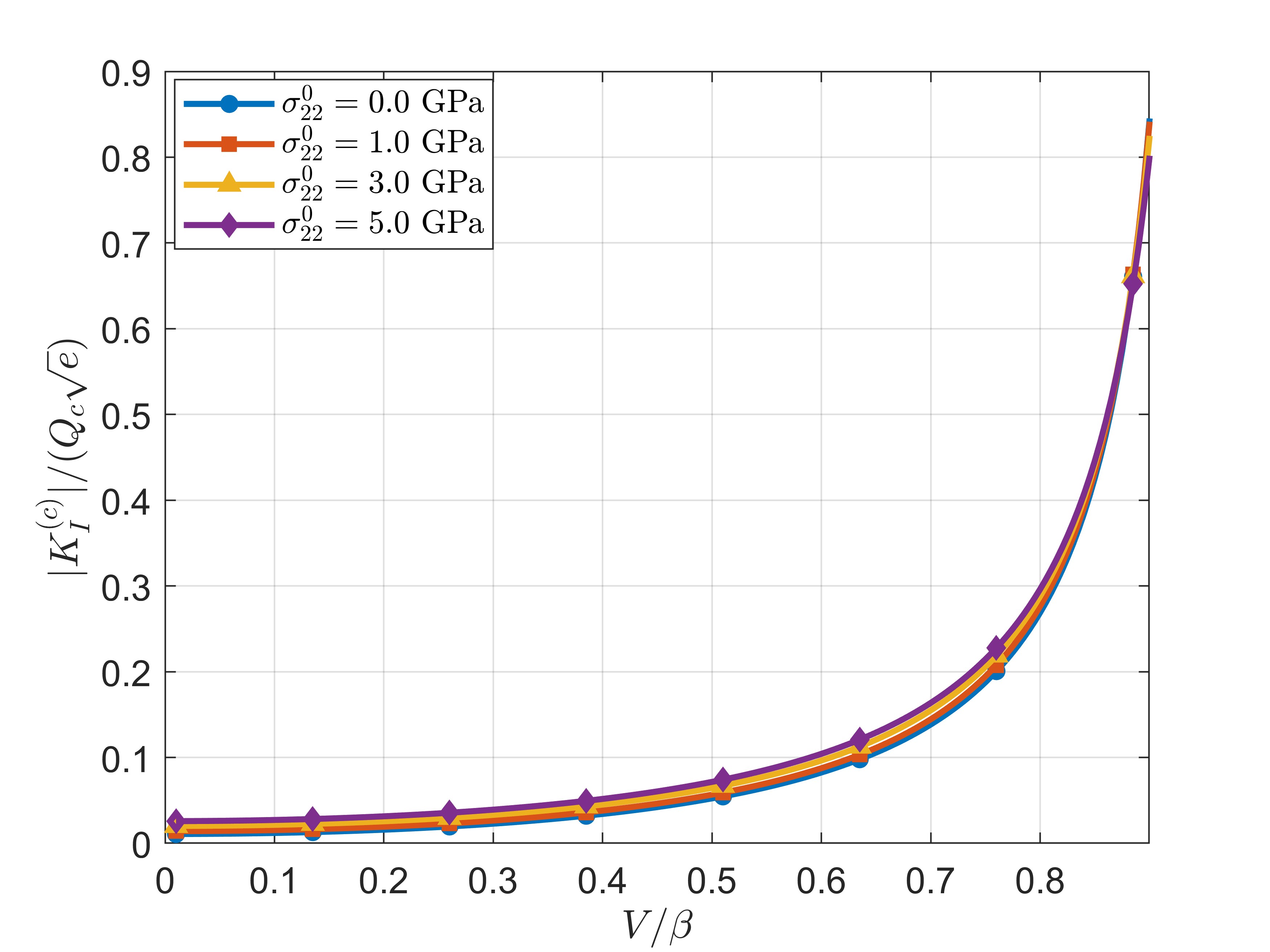}
\caption{Effect of $\sigma_{22}^{0}$.}
\label{fig:prestress(d)}
\end{subfigure}

\caption{Variation of the normalized Mode-I stress intensity factors with the crack-speed ratio for different values of the initial stresses: (a) outer crack tip for varying $\sigma_{11}^{0}$, (b) inner crack tip for varying $\sigma_{11}^{0}$, (c) outer crack tip for varying $\sigma_{22}^{0}$, and (d) inner crack tip for varying $\sigma_{22}^{0}$.}
\label{fig:prestress}
\end{figure*}

Figure~\ref{fig:prestress} illustrates the variation of the normalized Mode-I stress intensity factors with the crack-speed ratio for different values of the initial stresses acting parallel and perpendicular to the crack plane. It is evident from Figs.~\ref{fig:prestress(a)} and \ref{fig:prestress(b)} that increasing the longitudinal initial stress $\sigma_{11}^{0}$ leads to a significant reduction in both the outer and inner crack-tip stress intensity factors over the entire range of crack speeds. Although the reduction is relatively small at lower crack velocities, it becomes increasingly pronounced as the crack speed approaches the limiting shear-wave velocity. This behavior indicates that longitudinal pre-stressing effectively suppresses the dynamic stress concentration developed at the crack tips.

In contrast, Figs.~\ref{fig:prestress(2c)} and \ref{fig:prestress(d)} show that the transverse initial stress $\sigma_{22}^{0}$ exerts a comparatively weaker influence on the fracture response. A gradual increase in the stress intensity factors is observed with increasing $\sigma_{22}^{0}$; however, the overall variation remains considerably smaller than that produced by $\sigma_{11}^{0}$. This difference arises because the longitudinal initial stress directly modifies the stress field in the crack-propagation direction, whereas the transverse initial stress primarily alters the constraint normal to the crack plane.

\subsubsection{Effect of crack geometry}

\begin{figure*}[!t]
\centering

\begin{subfigure}{0.48\textwidth}
\includegraphics[width=\linewidth]{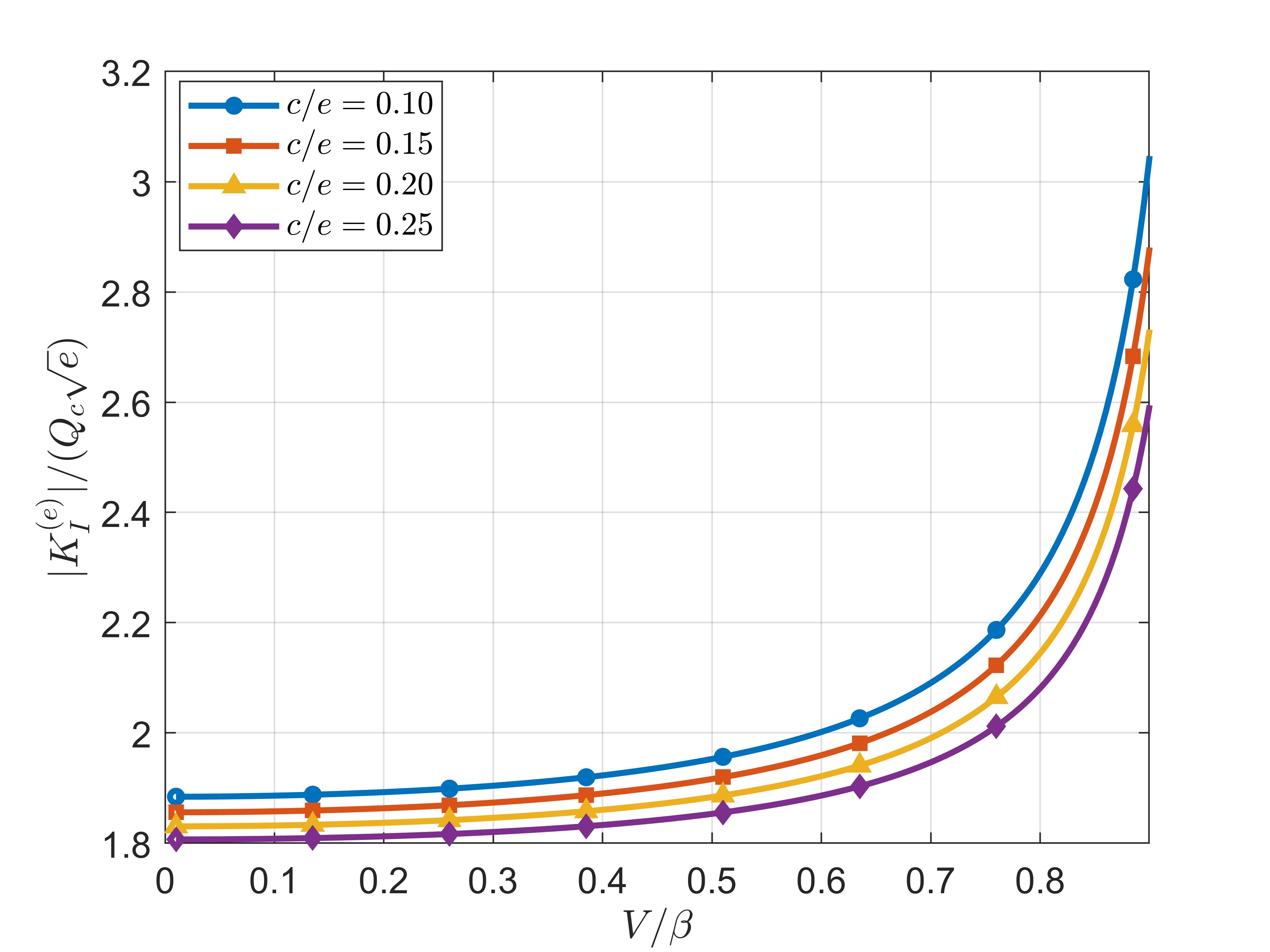}
\caption{Outer crack tip.}
\end{subfigure}
\hfill
\begin{subfigure}{0.48\textwidth}
\includegraphics[width=\linewidth]{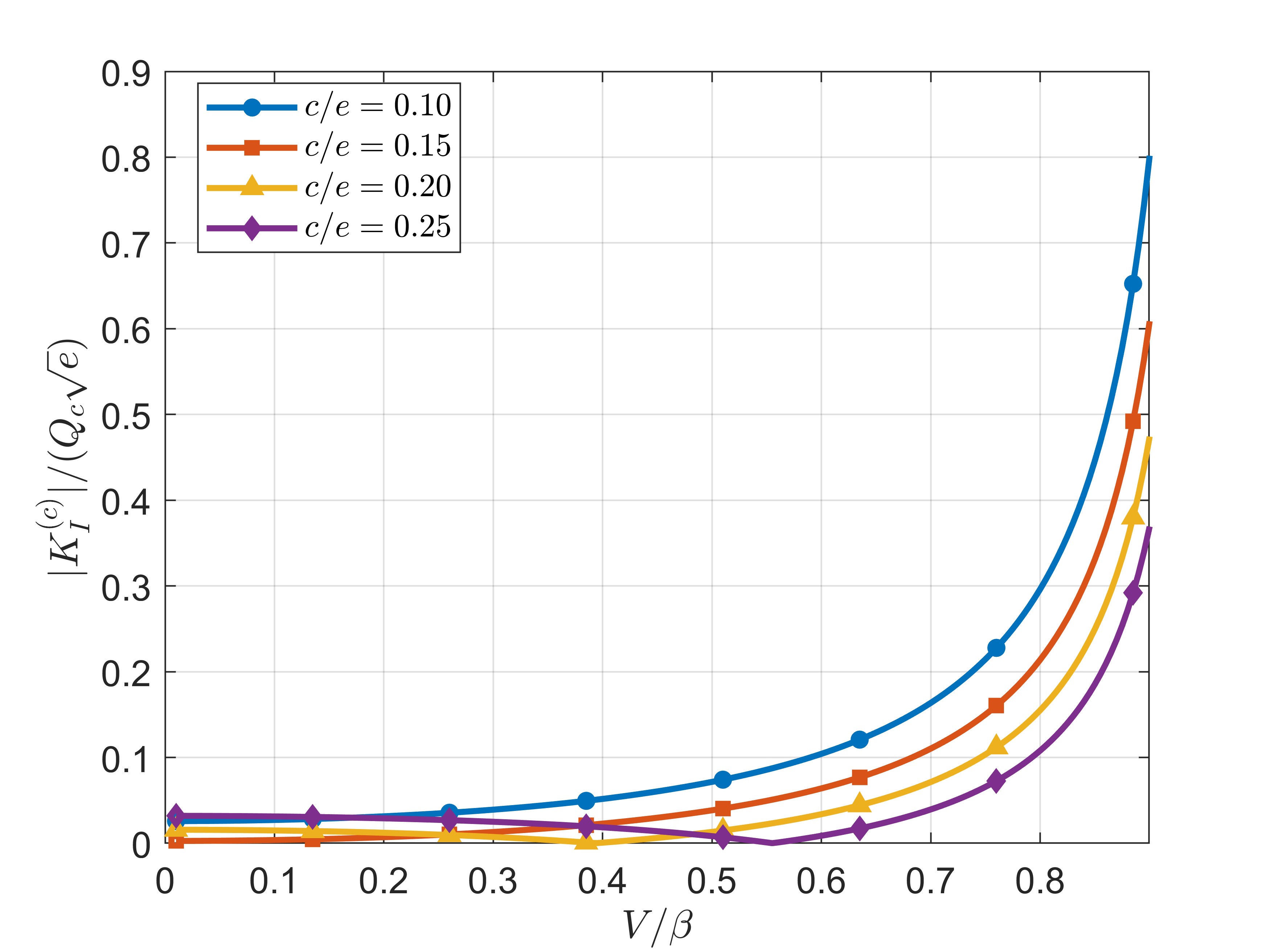}
\caption{Inner crack tip.}
\end{subfigure}

\caption{Variation of the normalized Mode-I stress intensity factors at the outer and inner crack tips with the crack-speed ratio for different values of the normalized inner crack coordinate $c/e$.}
\label{fig:cracklength}
\end{figure*}

Figure~\ref{fig:cracklength} presents the variation of the normalized Mode-I stress intensity factors with the crack-speed ratio for different values of the normalized inner crack coordinate $c/e$. It is observed that increasing $c/e$ reduces the stress intensity factors at both the outer and inner crack tips throughout the investigated crack-speed range. Although the reduction is moderate at lower crack velocities, it becomes increasingly pronounced as the crack speed approaches the limiting shear-wave velocity.
The observed reduction arises because an increase in $c/e$ effectively decreases the crack length while simultaneously increasing the ligament between the two collinear cracks. Consequently, the mutual interaction between the cracks weakens, leading to a lower concentration of stresses at the crack tips. This effect is more pronounced at the inner crack tip, where the crack interaction is strongest due to the presence of the ligament separating the two crack segments.

\subsubsection{Effect of strip thickness}

\begin{figure*}[!t]
\centering

\begin{subfigure}{0.48\textwidth}
\includegraphics[width=\linewidth]{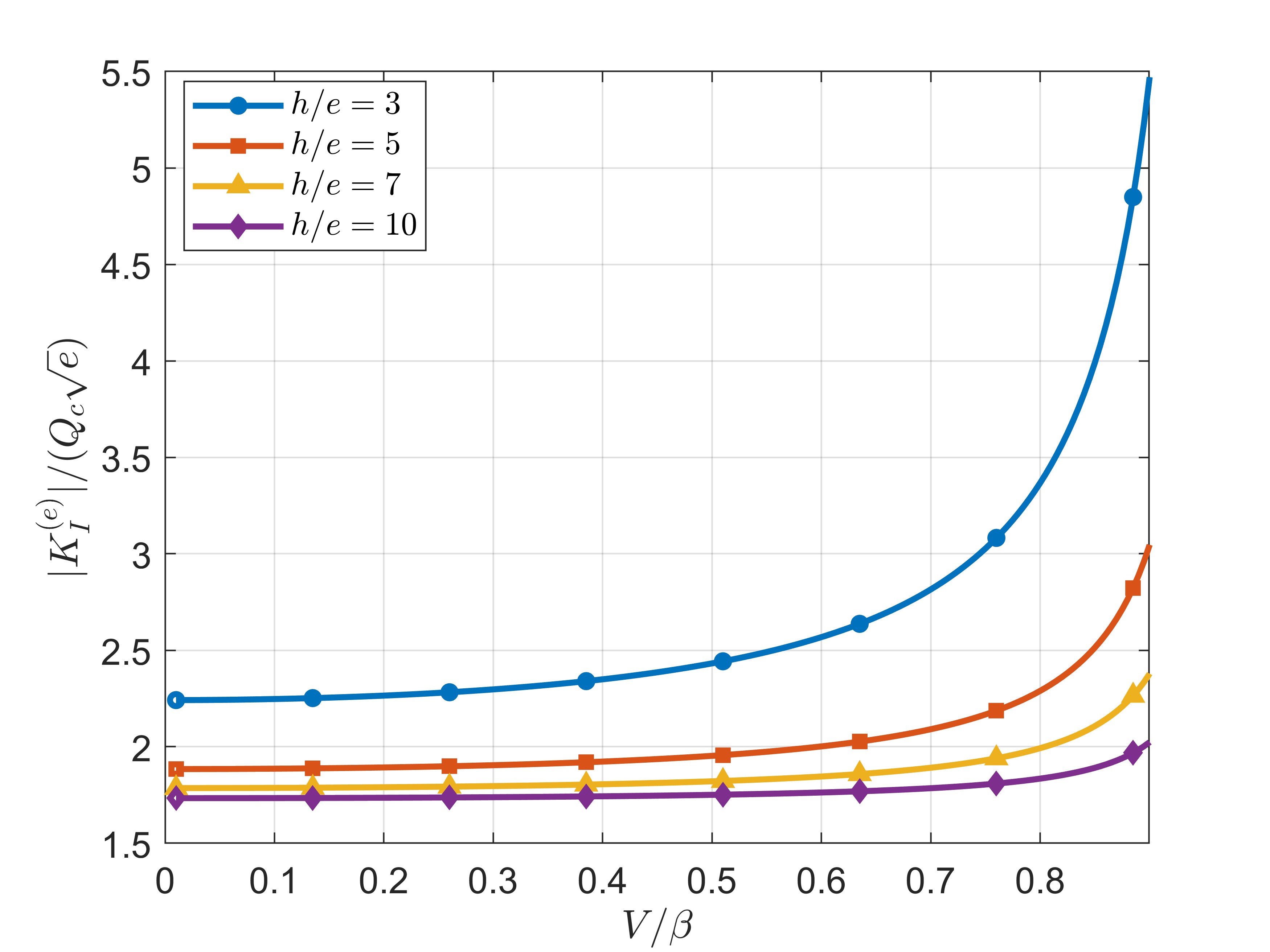}
\caption{Outer crack tip.}
\end{subfigure}
\hfill
\begin{subfigure}{0.48\textwidth}
\includegraphics[width=\linewidth]{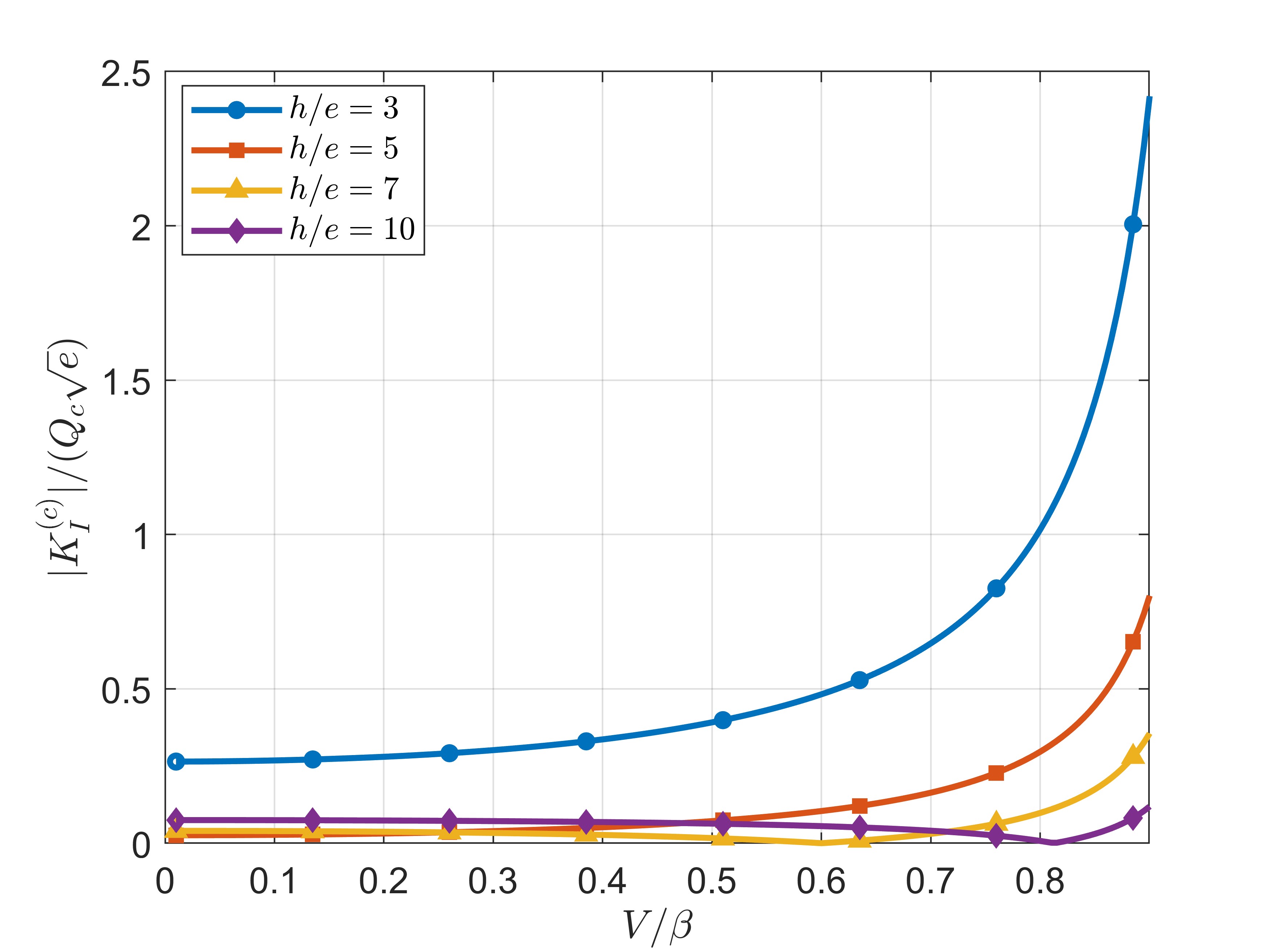}
\caption{Inner crack tip.}
\end{subfigure}

\caption{Variation of the normalized Mode-I stress intensity factors at the outer and inner crack tips with the crack-speed ratio for different values of the normalized strip half-thickness $h/e$.}
\label{fig:thickness}
\end{figure*}

Figure~\ref{fig:thickness} presents the variation of the normalized Mode-I stress intensity factors with the crack-speed ratio for different values of the normalized strip half-thickness $h/e$. An increase in strip thickness leads to a reduction in the normalized Mode-I stress intensity factors at both the inner and outer crack tips over the entire range of crack speeds. The reduction is relatively small at lower crack velocities but becomes increasingly significant as the crack speed approaches the limiting shear-wave velocity.
The observed behavior can be attributed to the influence of the strip boundaries on the crack-tip stress field. For relatively thin strips, stress waves reflected from the upper and lower boundaries interact strongly with the propagating cracks, thereby amplifying stress concentration at the crack tips. As the strip thickness increases, these boundary interactions become progressively weaker, and the medium approaches the response of an unbounded elastic strip. Consequently, the finite-thickness effects diminish, resulting in lower stress intensity factors.

\subsubsection{Effect of loading parameters}

\begin{figure*}[!t]
\centering

\begin{subfigure}{0.48\textwidth}
\includegraphics[width=\linewidth]{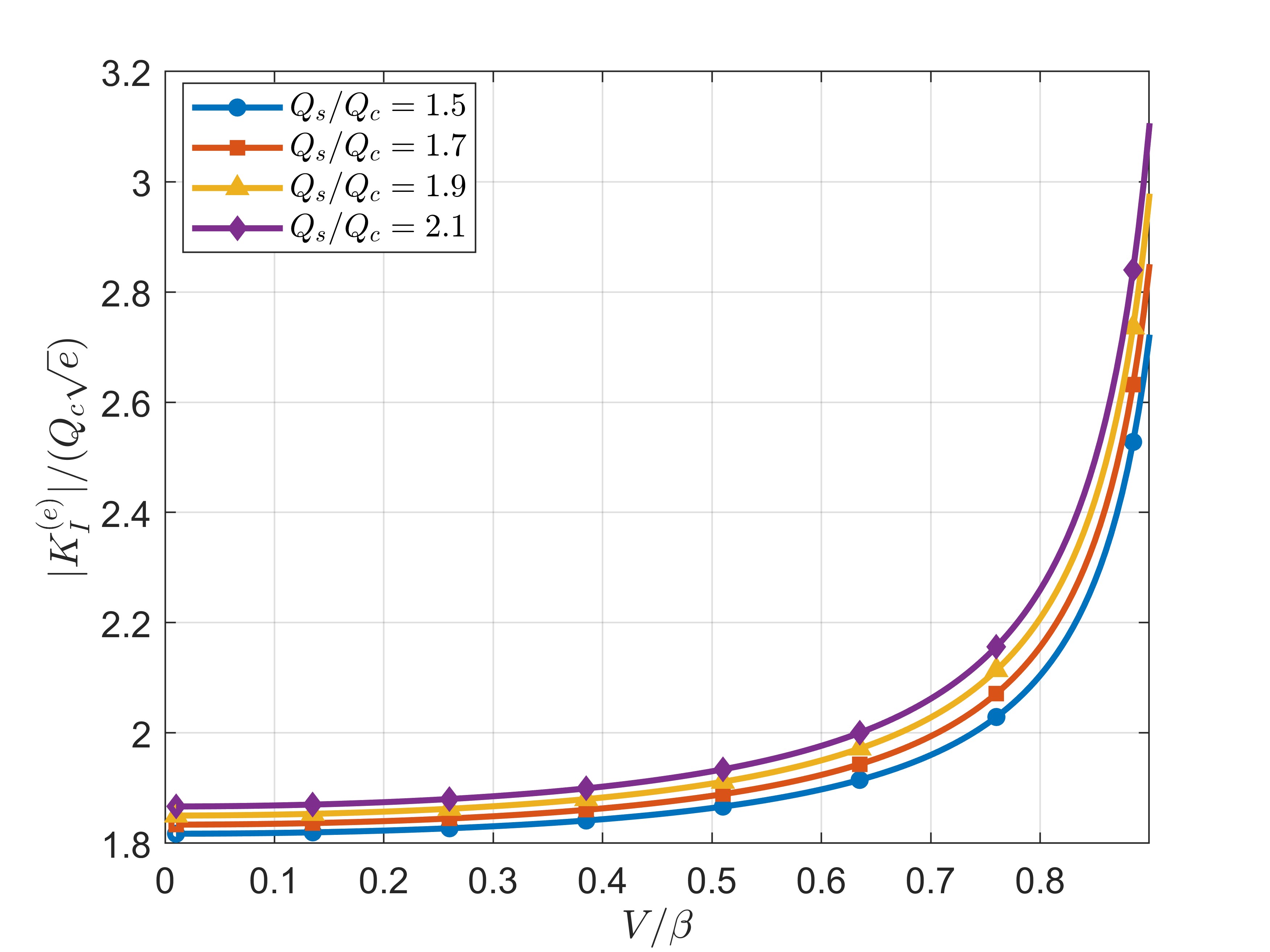}
\caption{Outer crack tip for different tangential loading ratios $Q_s/Q_c$.}
\label{fig:loading(a)}
\end{subfigure}
\hfill
\begin{subfigure}{0.48\textwidth}
\includegraphics[width=\linewidth]{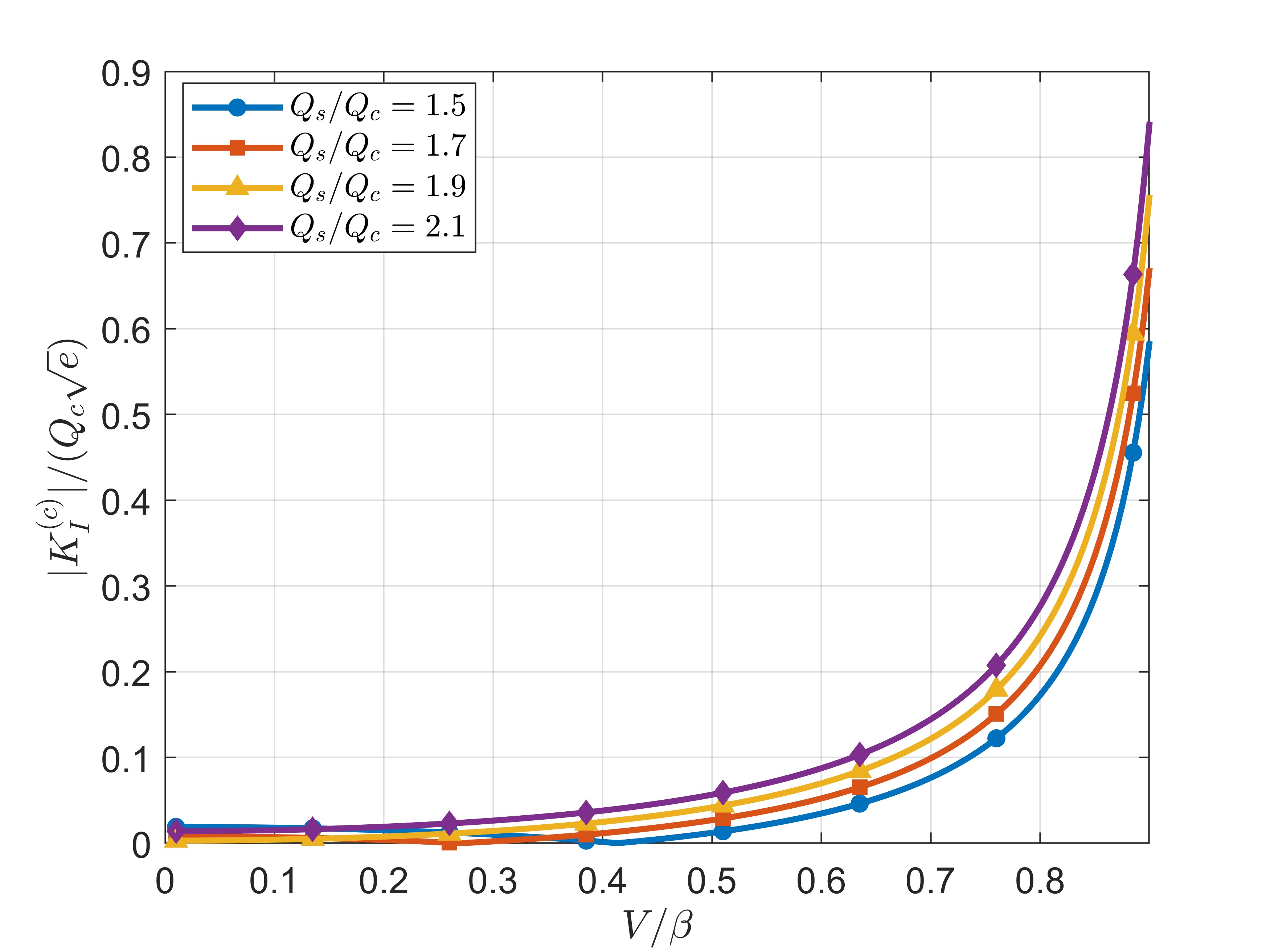}
\caption{Inner crack tip for different tangential loading ratios $Q_s/Q_c$.}
\label{fig:loading(b)}
\end{subfigure}

\vspace{0.3cm}

\begin{subfigure}{0.48\textwidth}
\includegraphics[width=\linewidth]{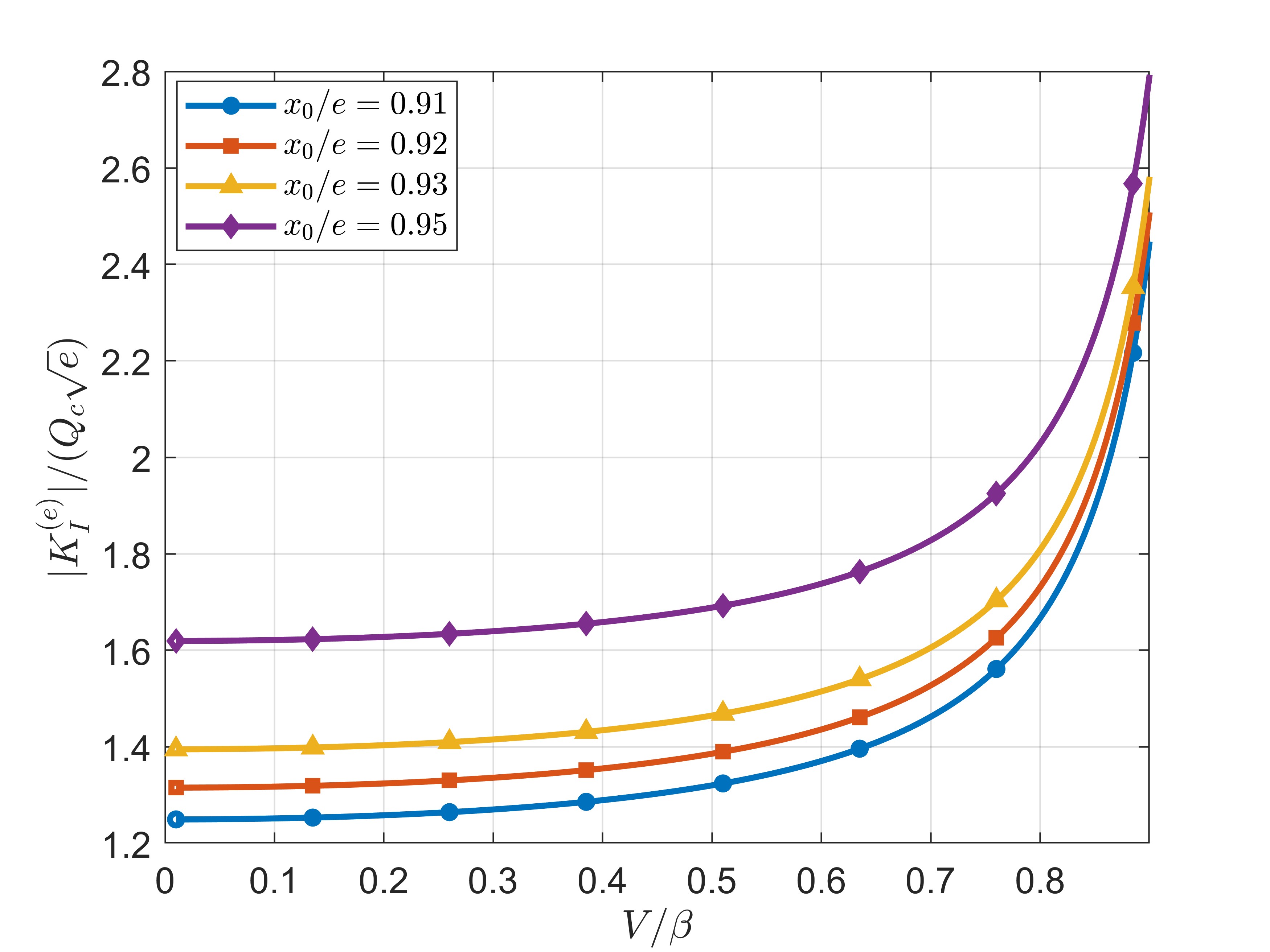}
\caption{Outer crack tip for different normalized load positions $x_0/e$.}
\label{fig:loading(c)}
\end{subfigure}
\hfill
\begin{subfigure}{0.48\textwidth}
\includegraphics[width=\linewidth]{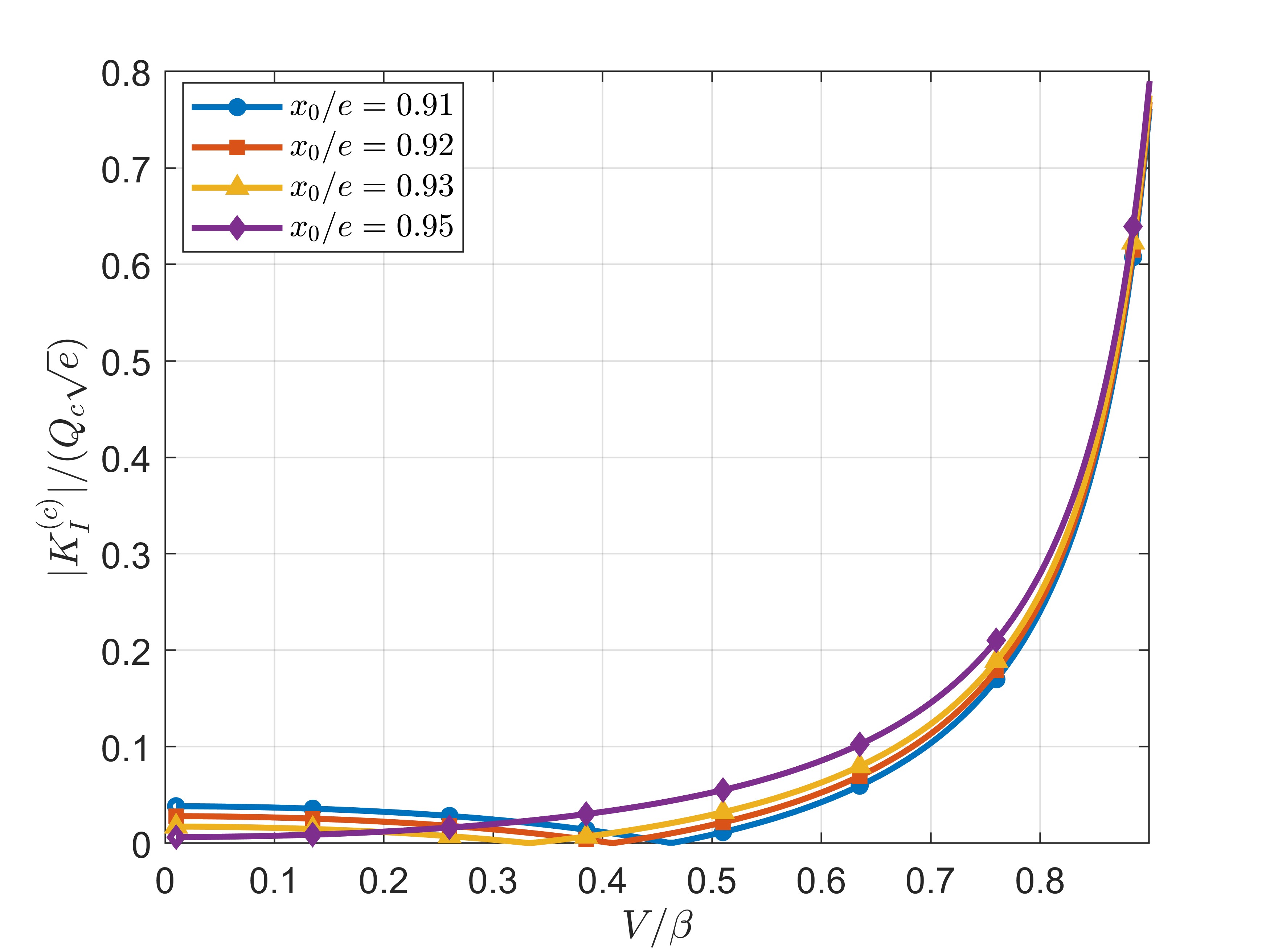}
\caption{Inner crack tip for different normalized load positions $x_0/e$.}
\label{fig:loading(d)}
\end{subfigure}

\caption{Variation of the normalized Mode-I stress intensity factors at the outer and inner crack tips with the crack-speed ratio for different loading parameters.}
\label{fig:loading}
\end{figure*}

Figure~\ref{fig:loading} illustrates the variation of the normalized Mode-I stress intensity factors with the crack-speed ratio for different loading parameters. Figures~\ref{fig:loading(a)} and \ref{fig:loading(b)} present the influence of the tangential-to-normal loading ratio, $Q_s/Q_c$, on the outer and inner crack-tip stress intensity factors, respectively. It is observed that increasing the tangential loading ratio results in a gradual increase in the stress intensity factors throughout the investigated range of crack speeds. The increase becomes more pronounced as the crack speed approaches the limiting shear-wave velocity, indicating that the additional tangential loading intensifies the dynamic stress concentration at crack tips.
The influence of the moving load position is depicted in Figs.~\ref{fig:loading(c)} and \ref{fig:loading(d)}. As the normalized load position $x_0/e$ increases, corresponding to the moving concentrated load approaching the outer crack tip, the stress intensity factors increase significantly. This behavior is attributed to the localization of the applied load in the vicinity of the crack tip, which produces a stronger stress concentration and enhances the dynamic fracture response. The effect is particularly pronounced for the outer crack tip because of its direct proximity to the moving load, whereas the inner crack tip exhibits a comparatively weaker sensitivity to variations in the load position.
The results indicate that both the tangential loading ratio and the load position play significant roles in governing the dynamic fracture behavior. While an increase in the tangential loading component amplifies the crack-tip stress field, positioning the moving load closer to the crack tip further intensifies the stress concentration, thereby increasing the likelihood of crack propagation under dynamic loading.

\subsubsection{Contour representation of the normalized Mode-I stress intensity factors}
\begin{figure*}[!t]
\centering
\begin{subfigure}{0.48\textwidth}
    \centering
    \includegraphics[width=\linewidth]{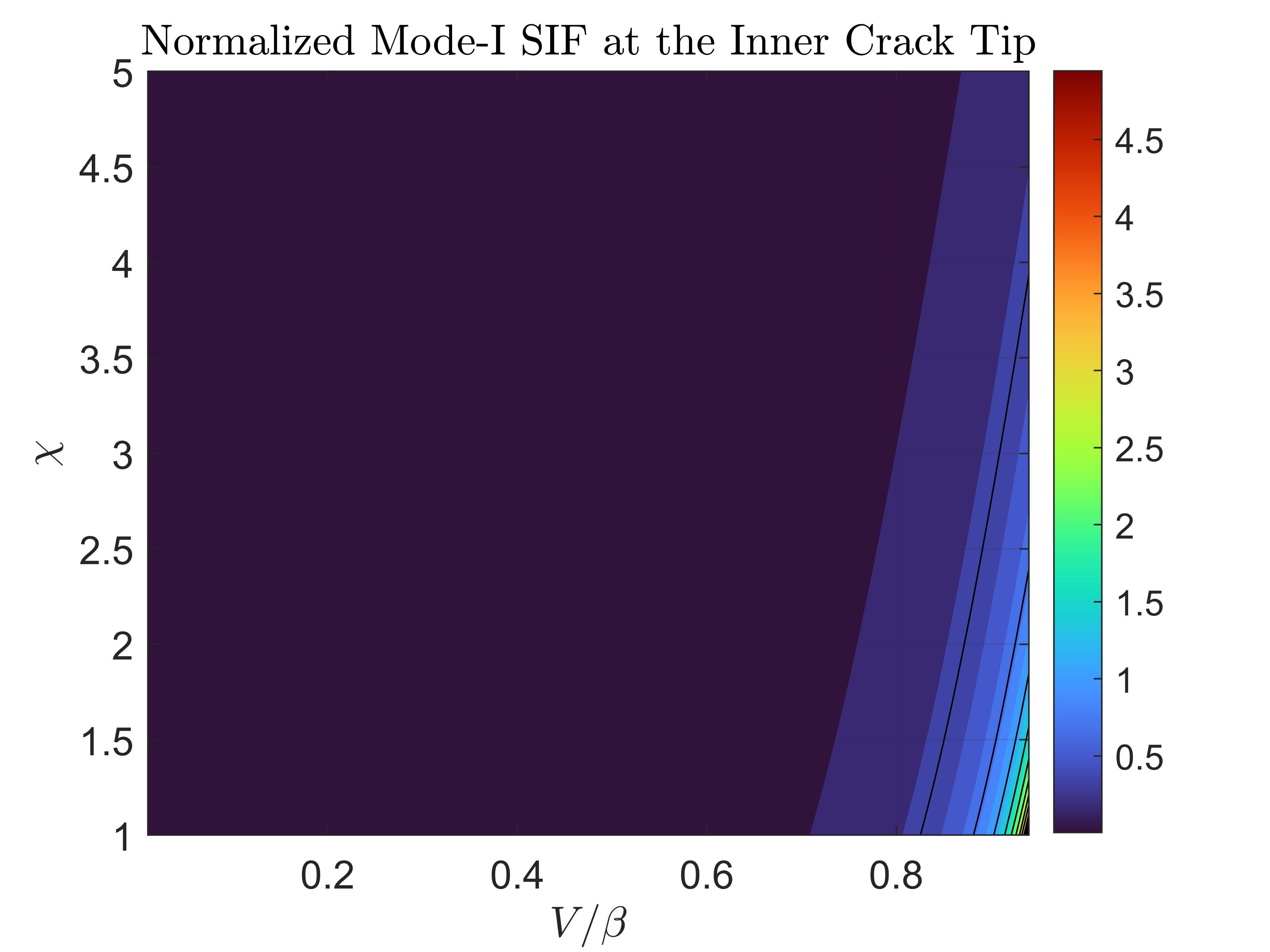}
    \caption{}
\end{subfigure}
\hfill
\begin{subfigure}{0.48\textwidth}
    \centering
    \includegraphics[width=\linewidth]{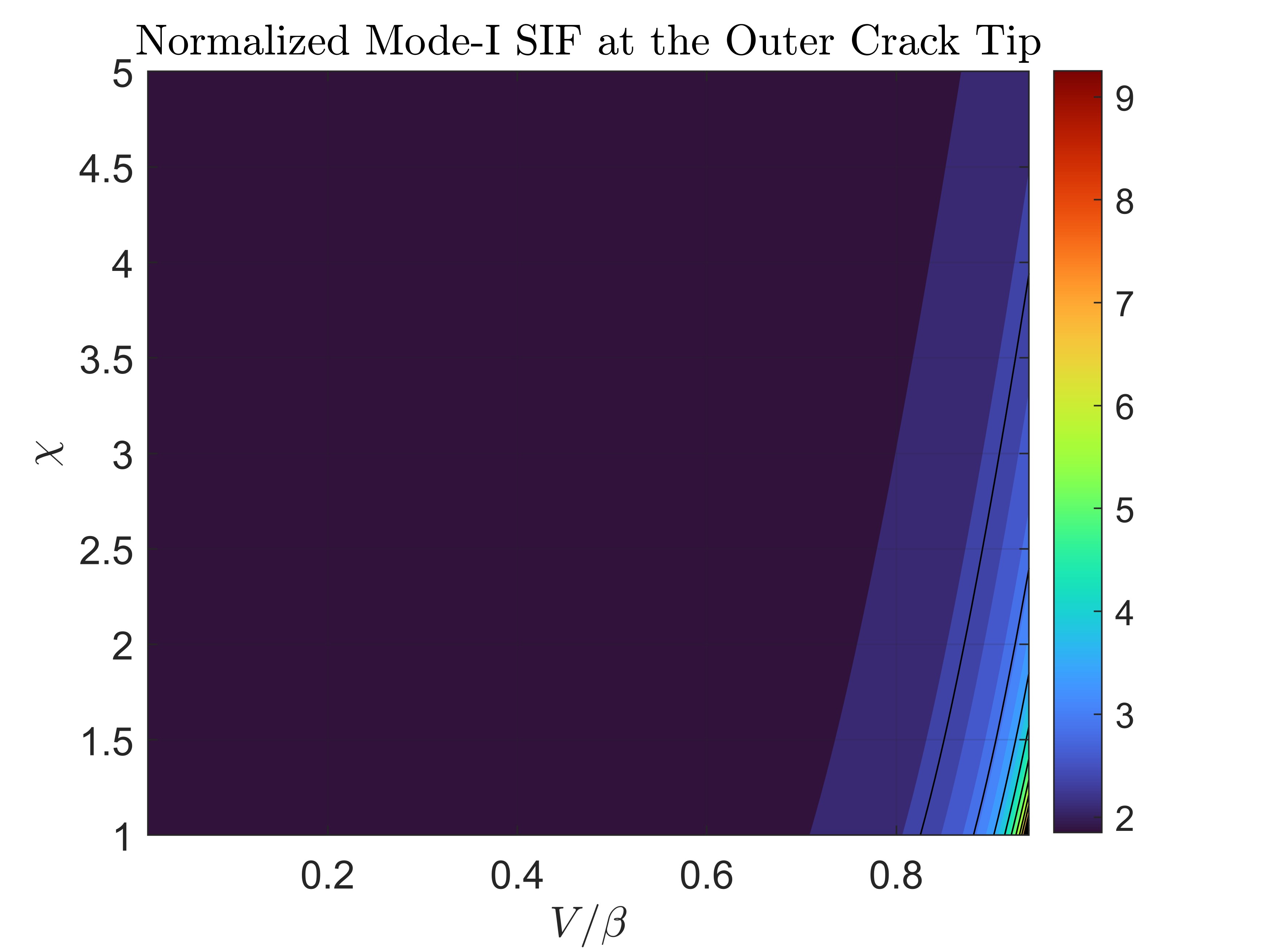}
    \caption{}
\end{subfigure}
\caption{Contour plots of the normalized Mode-I stress intensity factors as functions of the normalized crack-speed ratio $V/\beta$ and the sandiness parameter $\chi$ for (a) the inner crack tip and (b) the outer crack tip.}
\label{fig:contour_sandiness}
\end{figure*}

\begin{figure*}[!t]
\centering
\begin{subfigure}{0.48\textwidth}
    \centering
    \includegraphics[width=\linewidth]{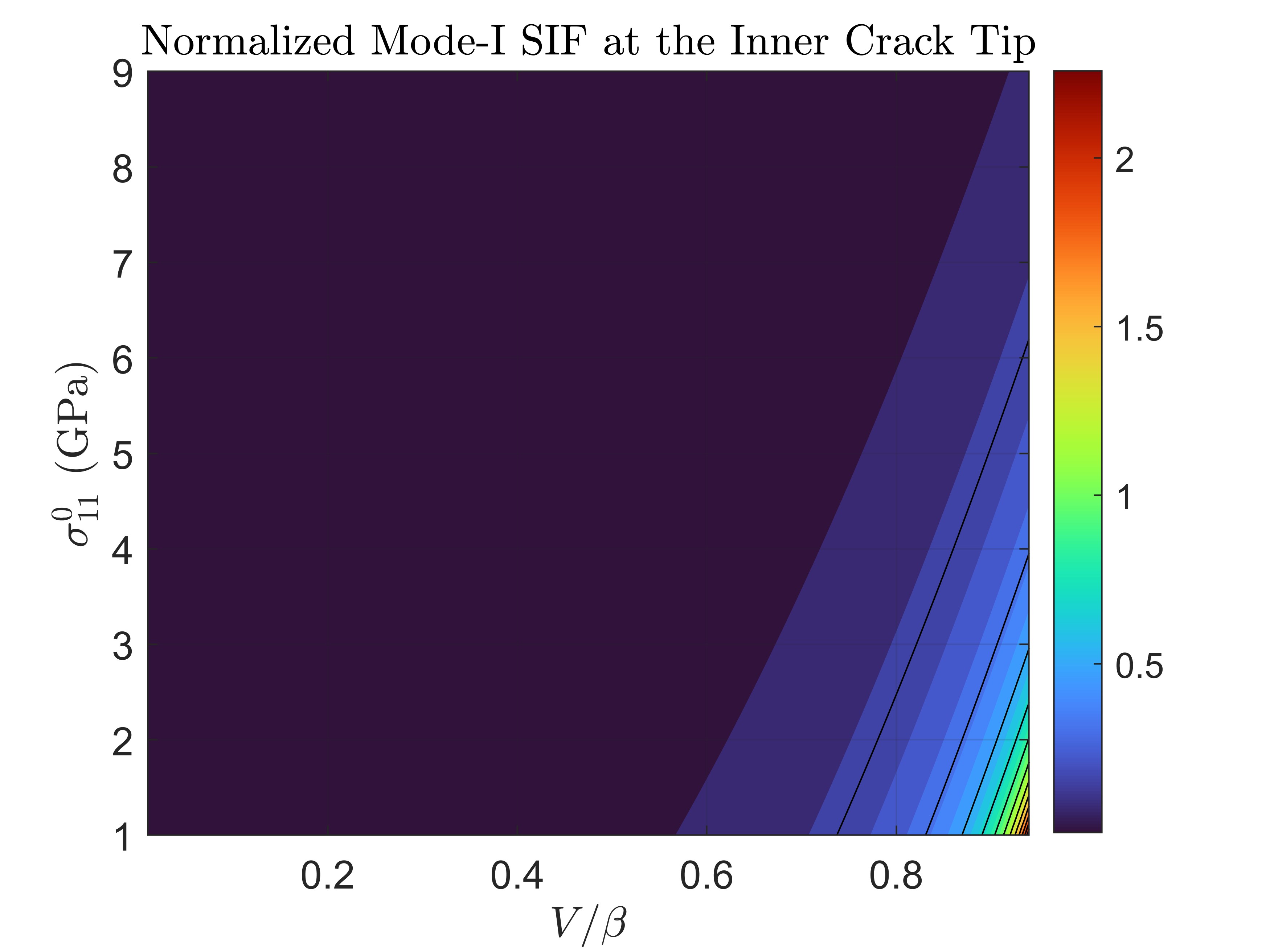}
    \caption{}
\end{subfigure}
\hfill
\begin{subfigure}{0.48\textwidth}
    \centering
    \includegraphics[width=\linewidth]{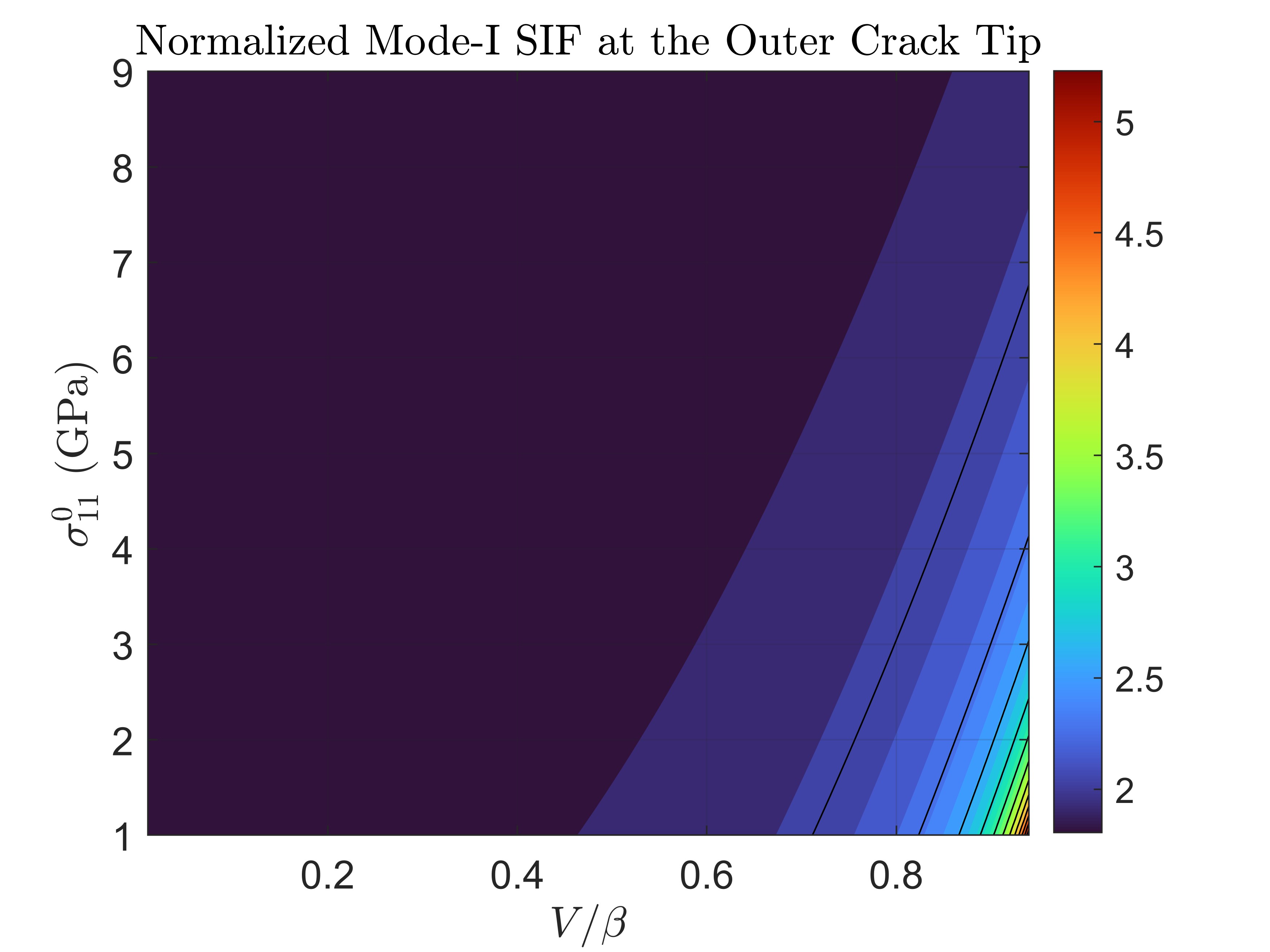}
     \caption{}
\end{subfigure}

\vspace{0.25cm}

\begin{subfigure}{0.48\textwidth}
    \centering
    \includegraphics[width=\linewidth]{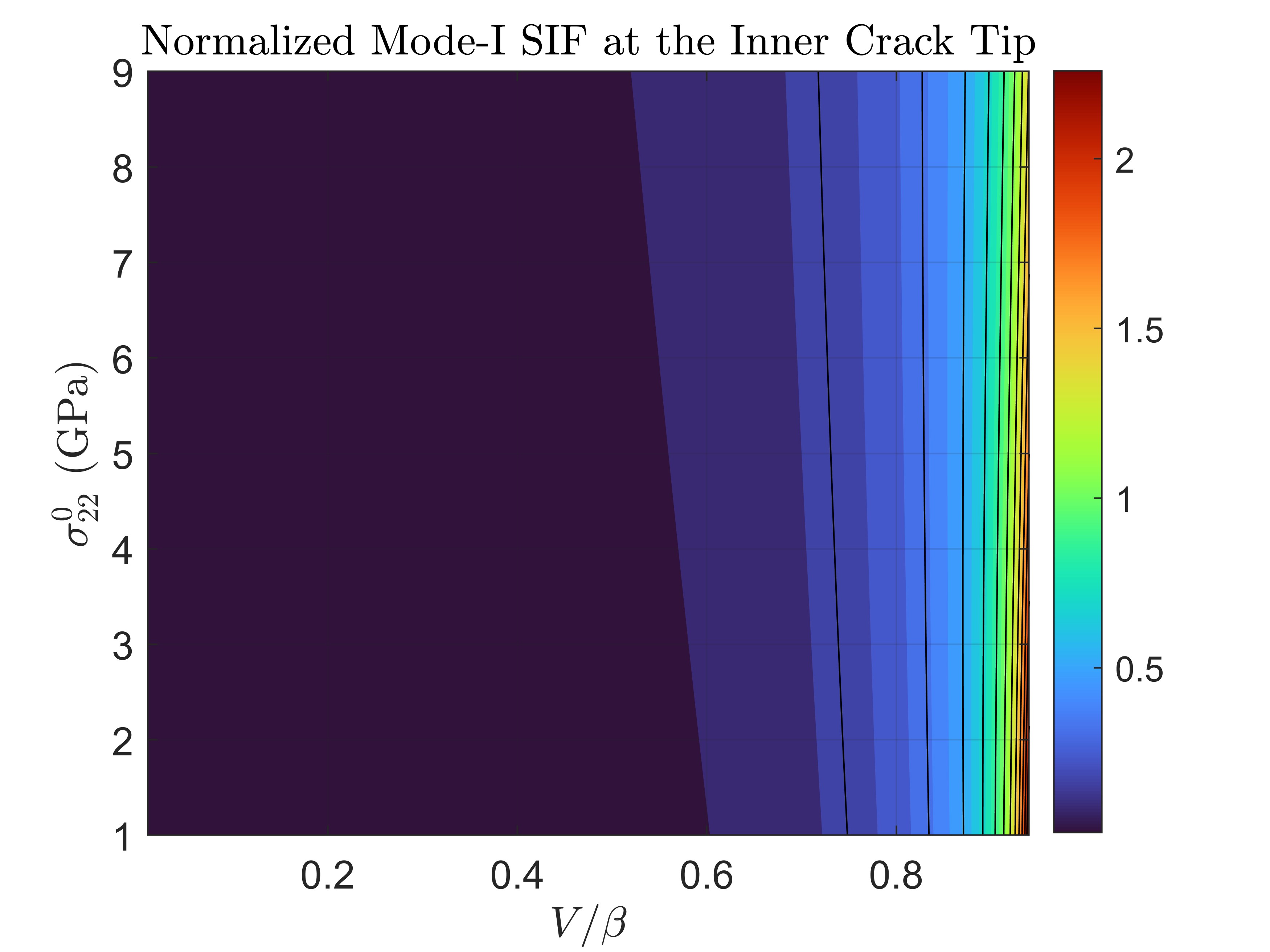}
    \caption{}
\end{subfigure}
\hfill
\begin{subfigure}{0.48\textwidth}
    \centering
    \includegraphics[width=\linewidth]{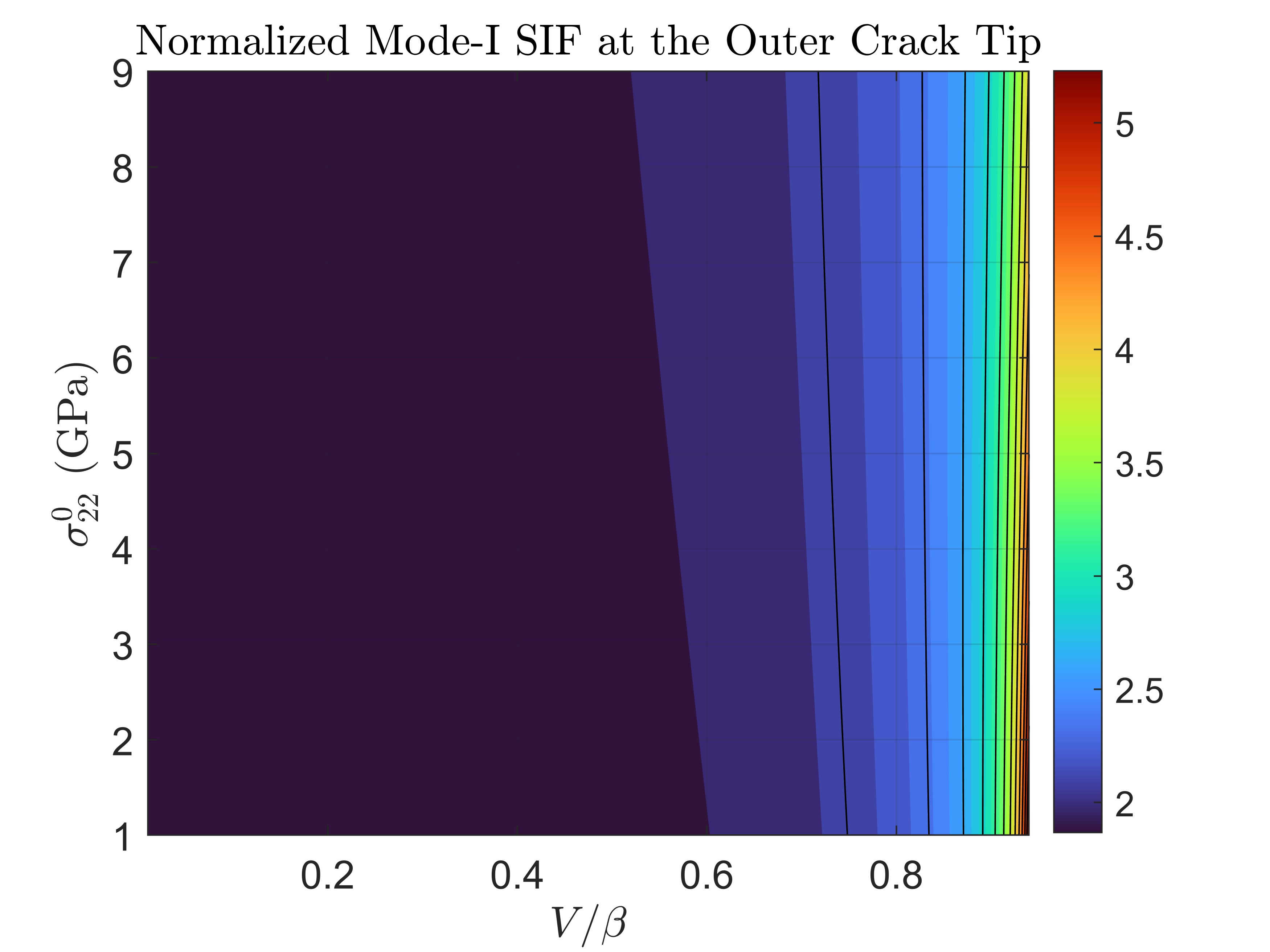}
    \caption{}
\end{subfigure}

\caption{Contour plots of the normalized Mode-I stress intensity factors as functions of the normalized crack-speed ratio $V/\beta$ and the initial stresses: (a)--(b) initial horizontal stress $\sigma_{11}^{0}$ and (c)--(d) initial vertical stress $\sigma_{22}^{0}$ for the inner and outer crack tips.}
\label{fig:contour_prestress}
\end{figure*}

\begin{figure*}[!t]
\centering
\begin{subfigure}{0.48\textwidth}
    \centering
    \includegraphics[width=\linewidth]{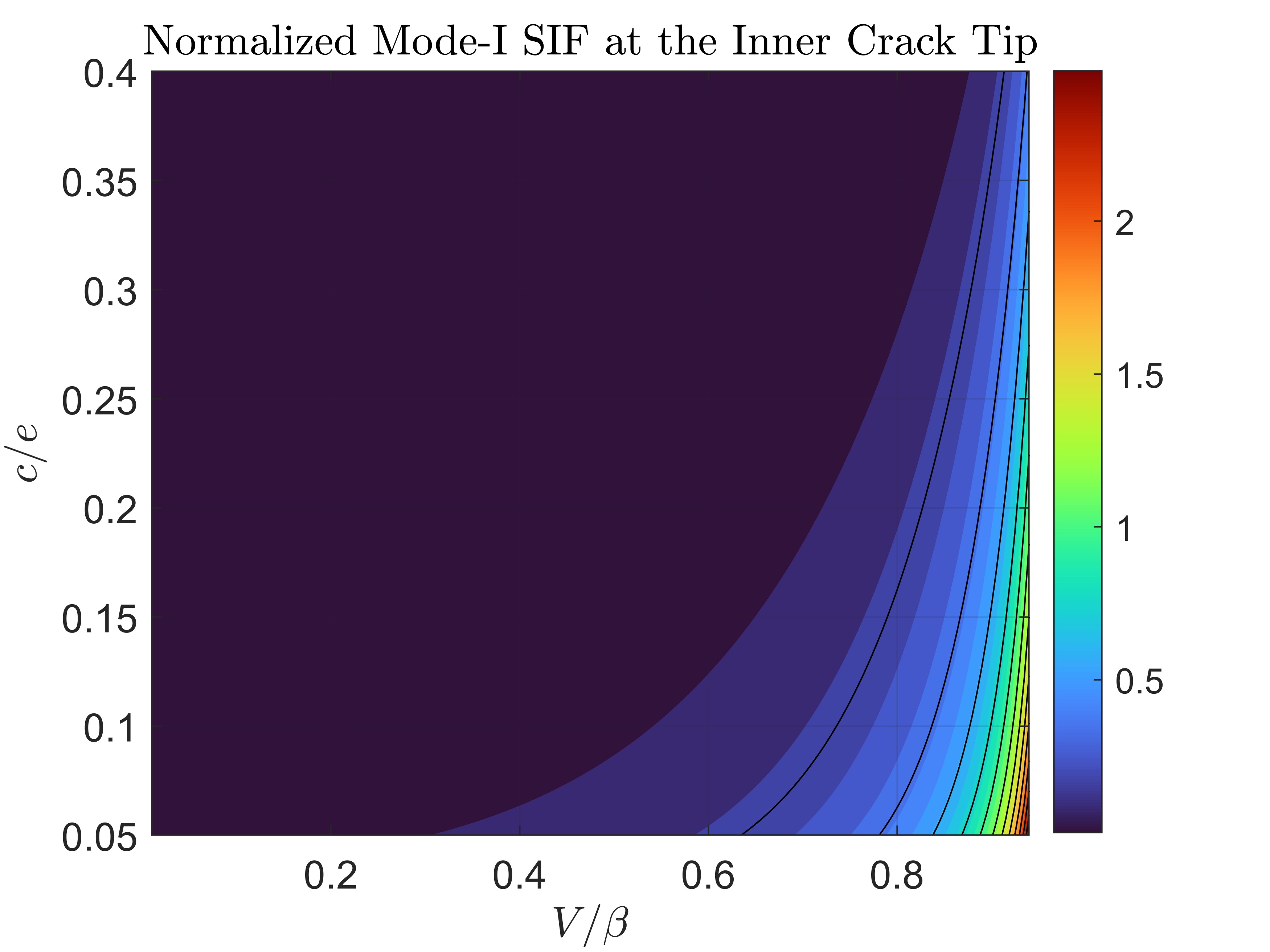}
    \caption{}
\end{subfigure}
\hfill
\begin{subfigure}{0.48\textwidth}
    \centering
    \includegraphics[width=\linewidth]{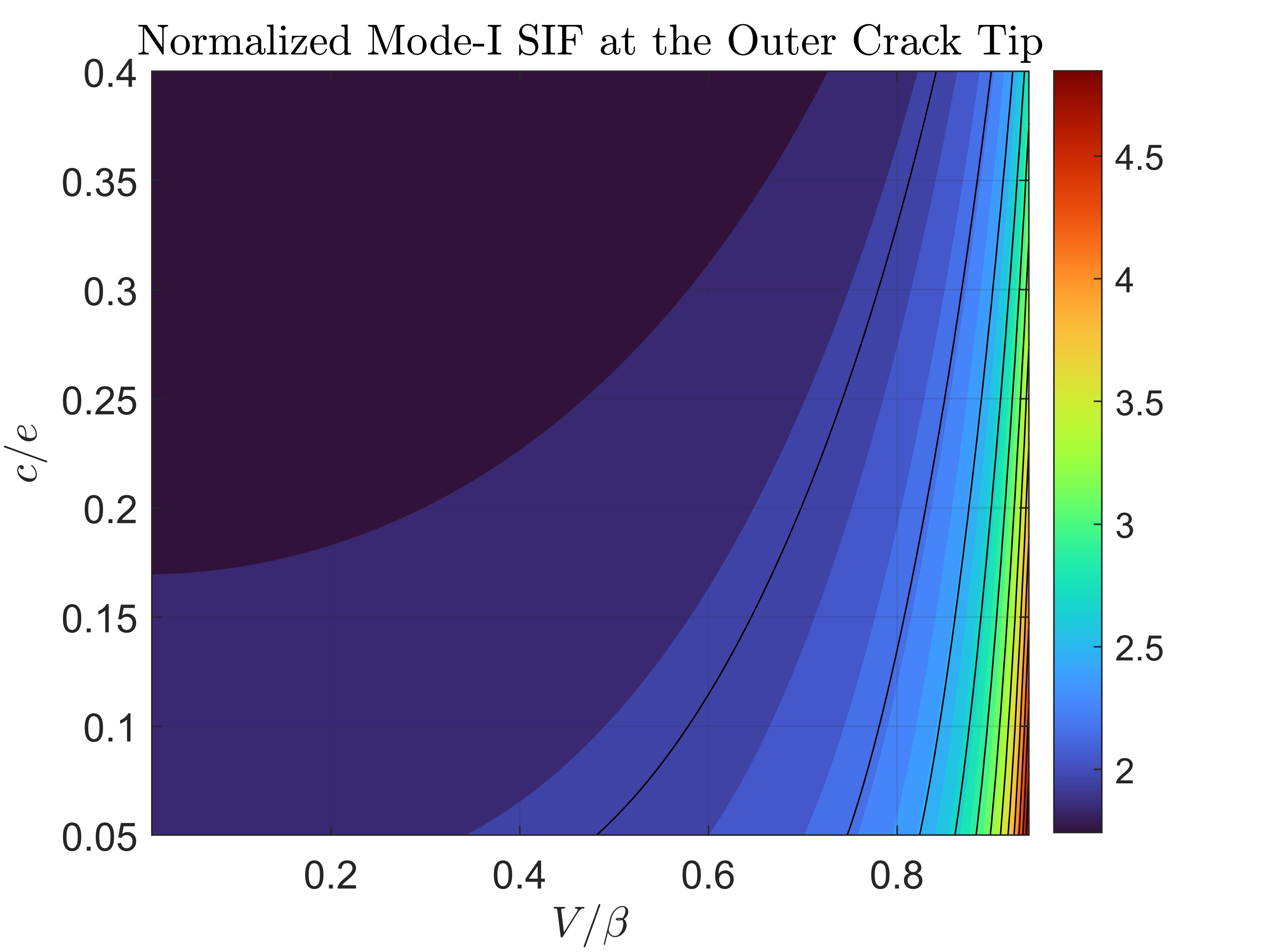}
     \caption{}
\end{subfigure}
\caption{Contour plots of the normalized Mode-I stress intensity factors as functions of the normalized crack-speed ratio $V/\beta$ and the normalized crack geometry parameter $c/e$ for (a) the inner crack tip and (b) the outer crack tip.}
\label{fig:contour_geometry}
\end{figure*}

\begin{figure*}[!t]
\centering
\begin{subfigure}{0.48\textwidth}
    \centering
    \includegraphics[width=\linewidth]{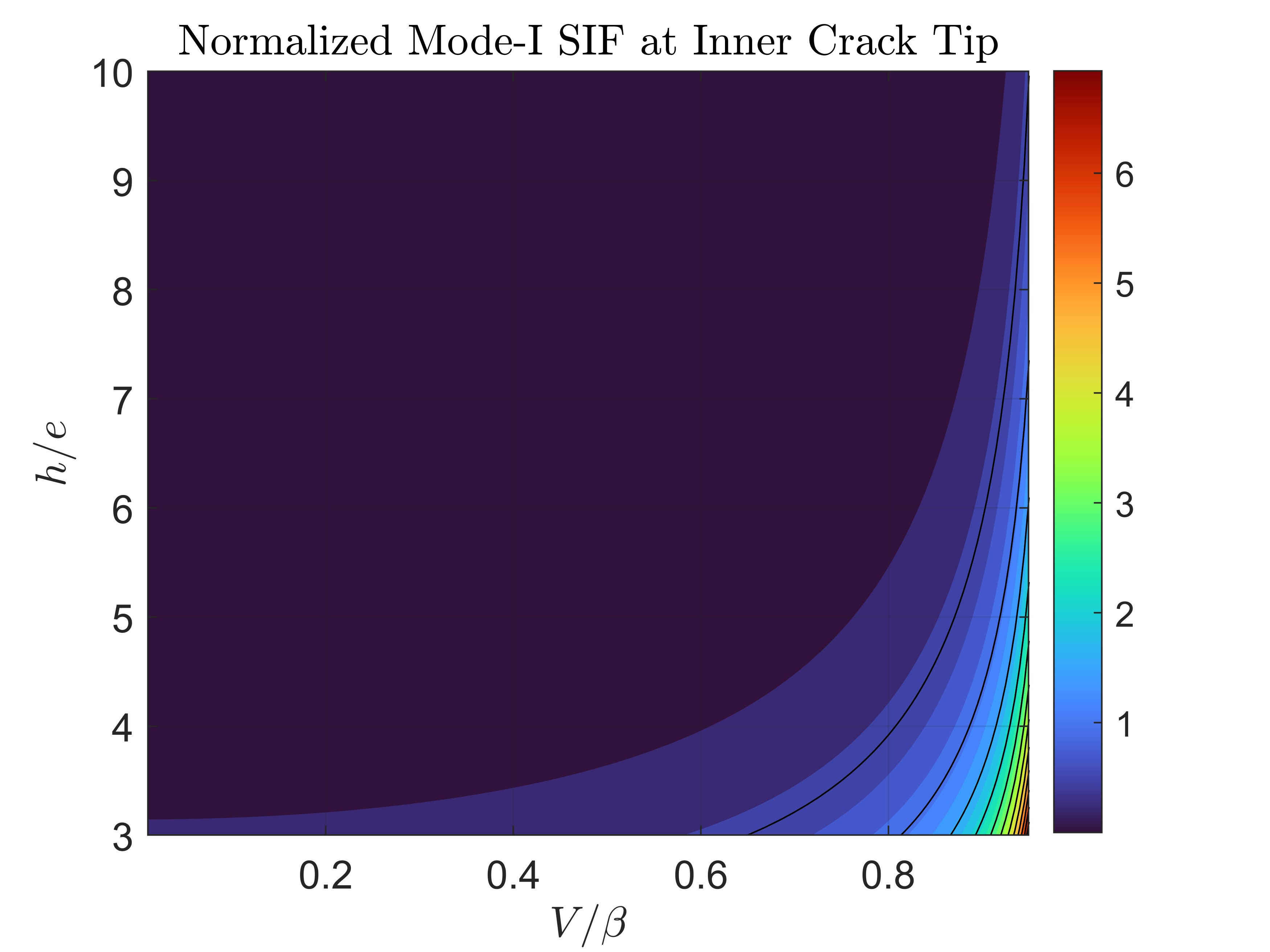}
\end{subfigure}
\hfill
\begin{subfigure}{0.48\textwidth}
    \centering
    \includegraphics[width=\linewidth]{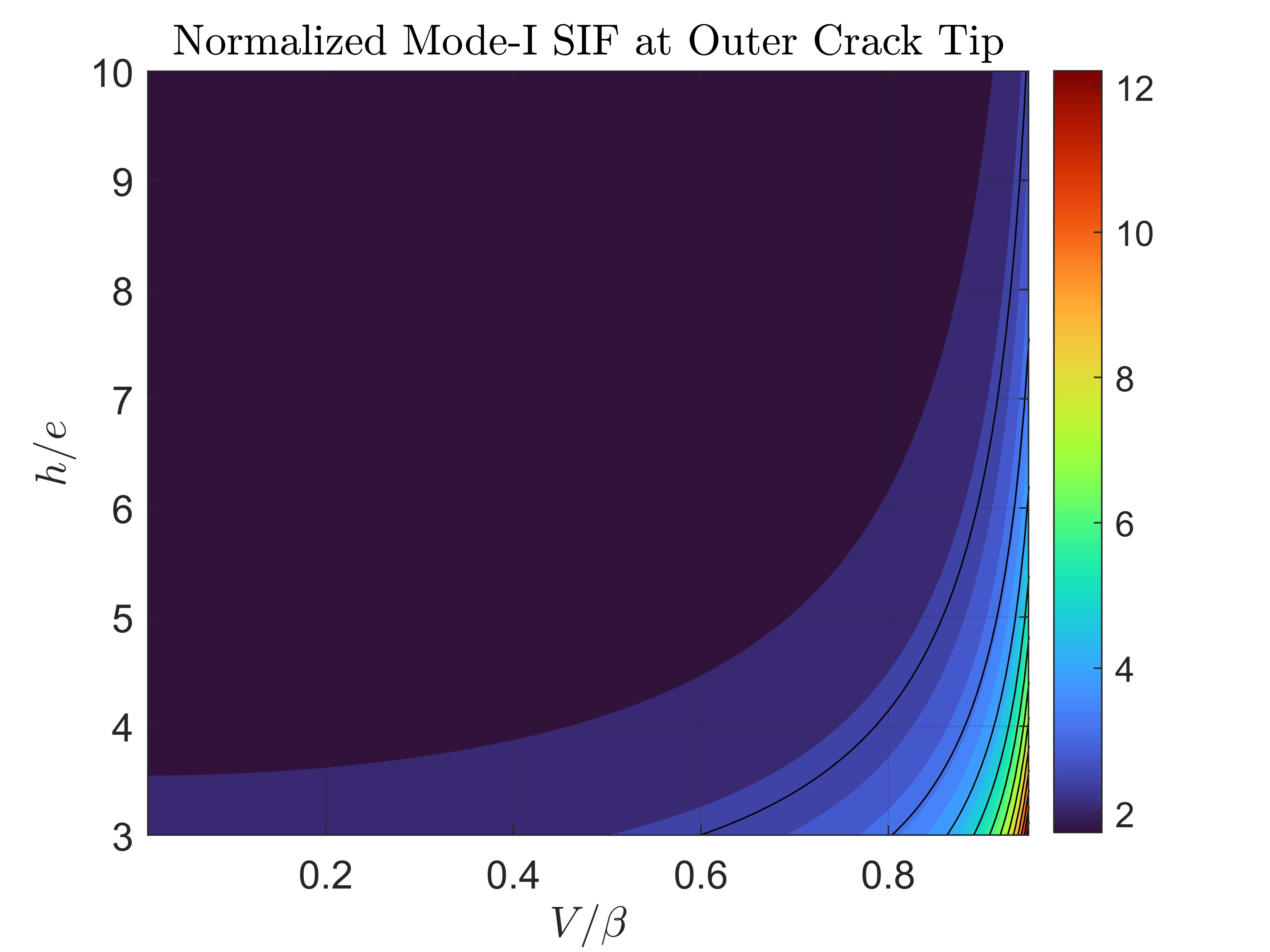}
\end{subfigure}
\caption{Contour plots of the normalized Mode-I stress intensity factors as functions of the normalized crack-speed ratio $V/\beta$ and the normalized strip half-thickness $h/e$ for (a) the inner crack tip and (b) the outer crack tip.}
\label{fig:contour_height}
\end{figure*}

\begin{figure*}[!t]
\centering
\begin{subfigure}{0.48\textwidth}
    \centering
    \includegraphics[width=\linewidth]{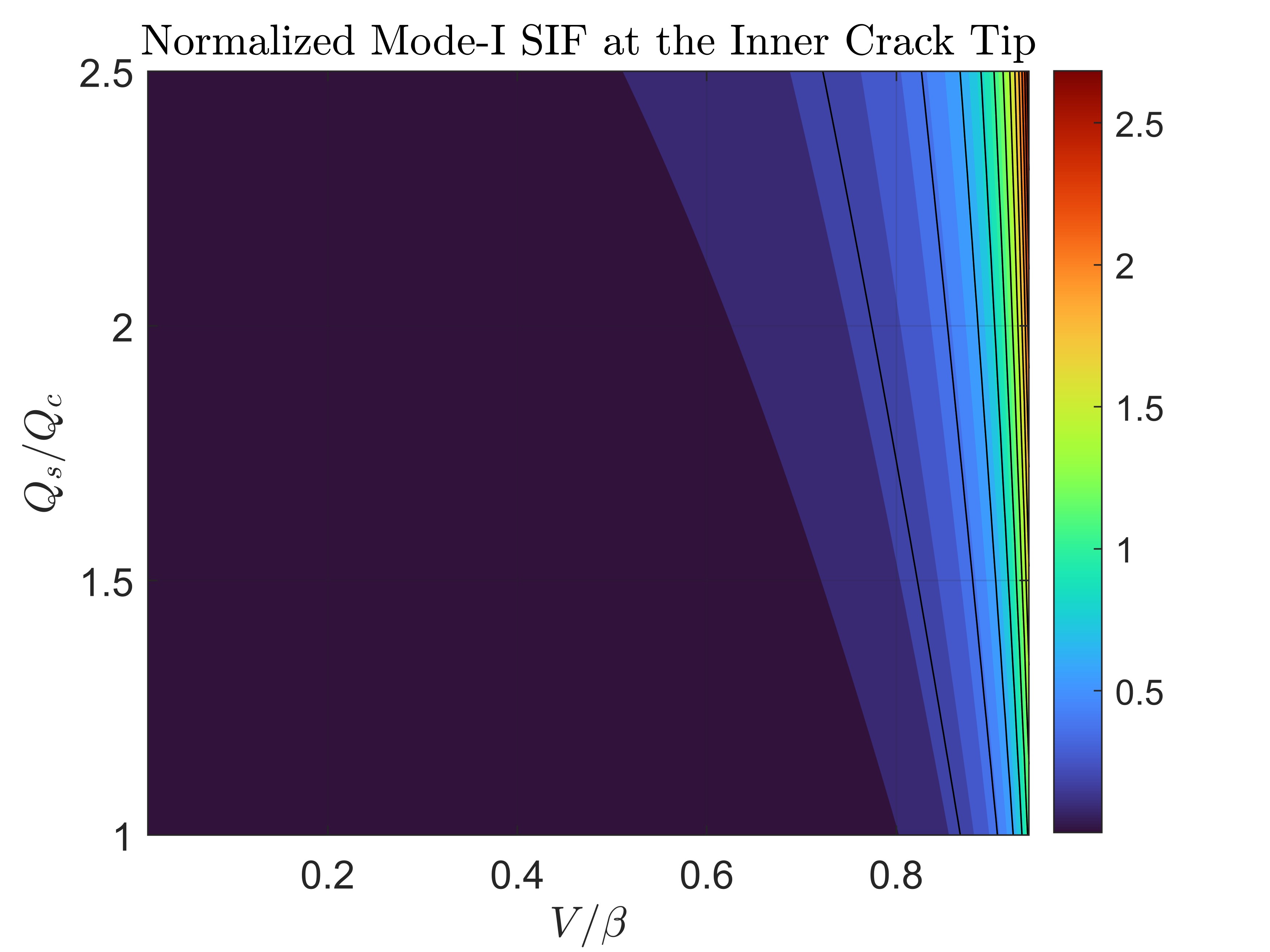}
    \caption{}
\end{subfigure}
\hfill
\begin{subfigure}{0.48\textwidth}
    \centering
    \includegraphics[width=\linewidth]{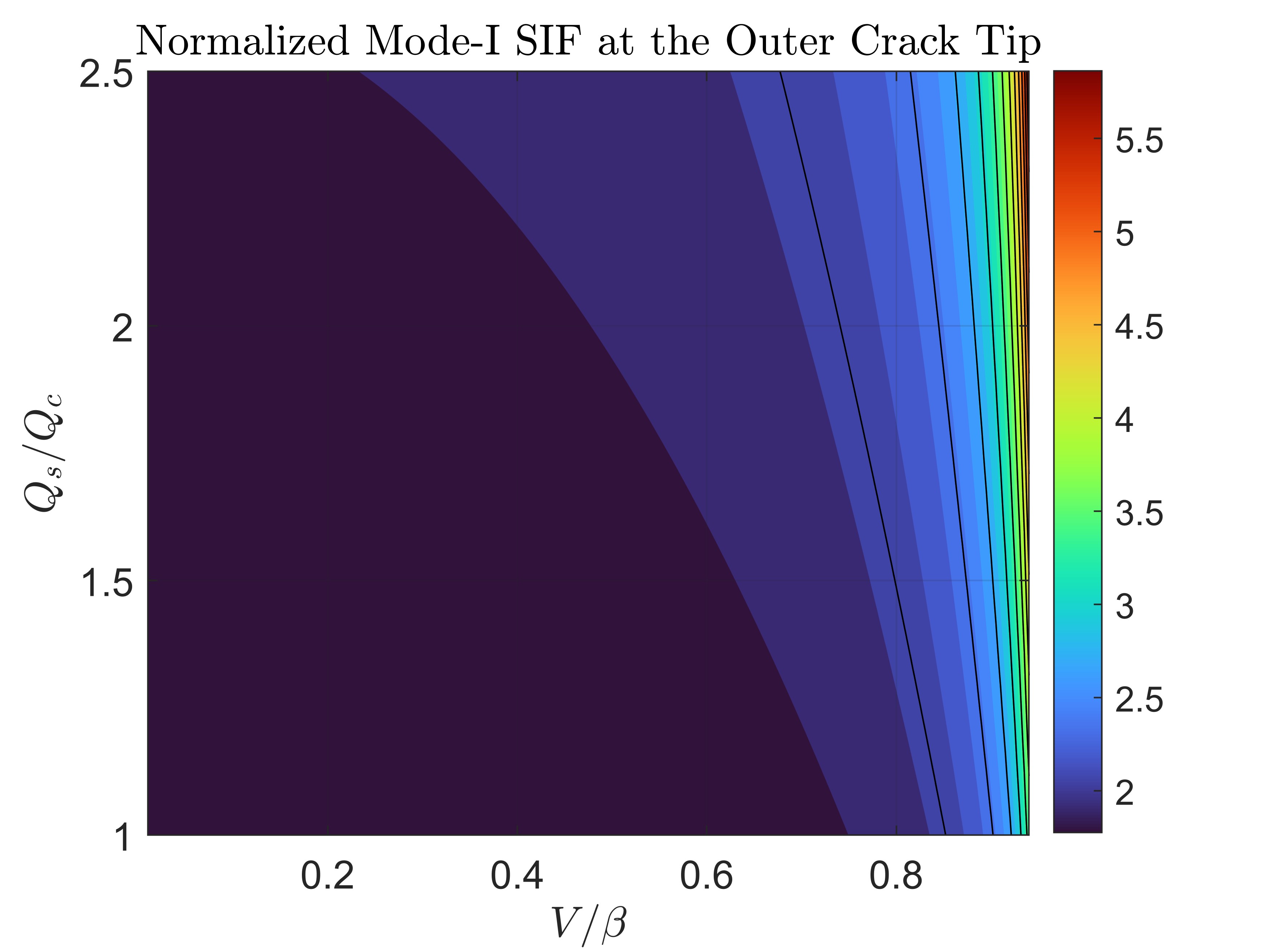}
    \caption{}
\end{subfigure}

\vspace{0.25cm}

\begin{subfigure}{0.48\textwidth}
    \centering
    \includegraphics[width=\linewidth]{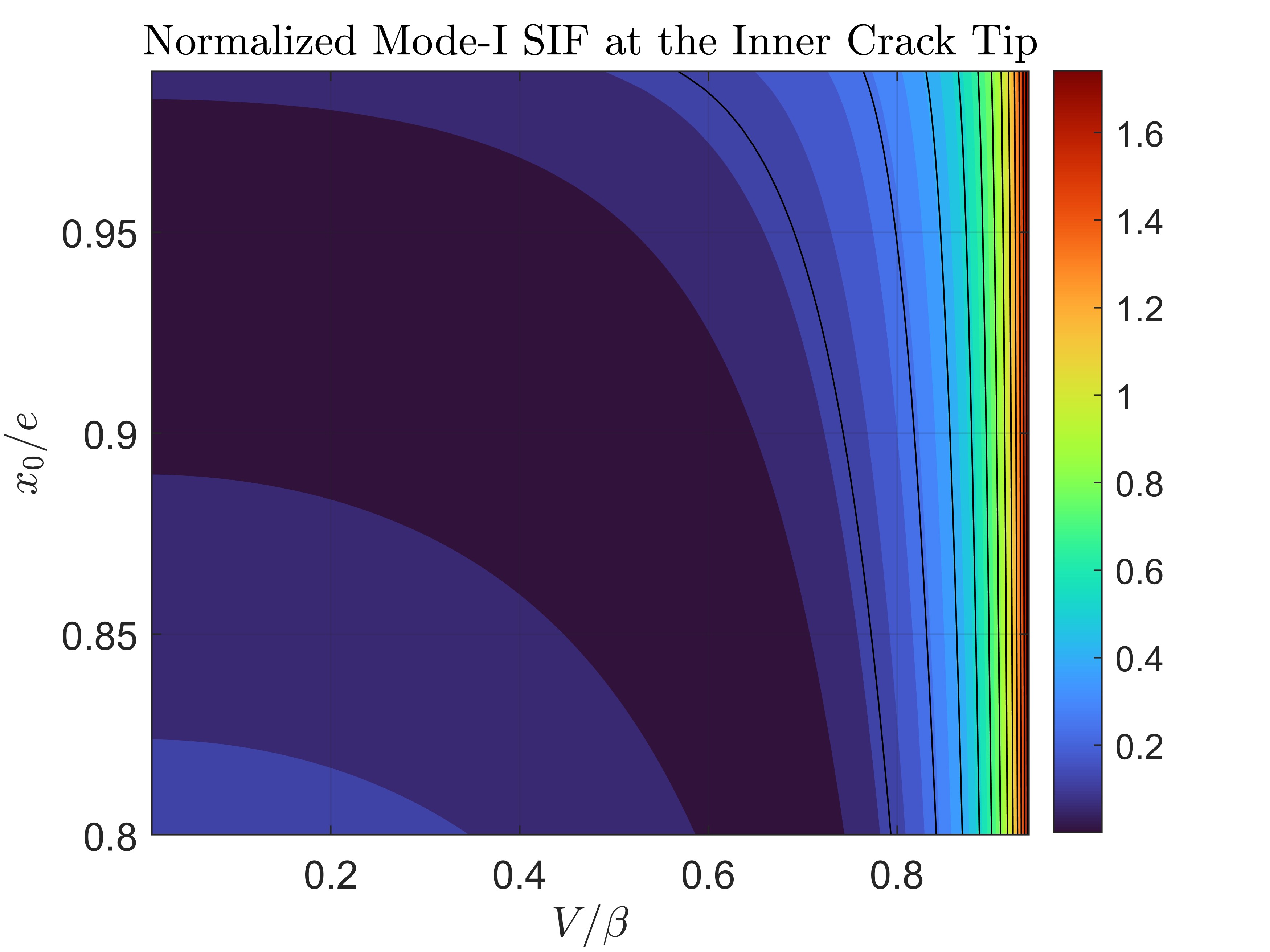}
    \caption{}
\end{subfigure}
\hfill
\begin{subfigure}{0.48\textwidth}
    \centering
    \includegraphics[width=\linewidth]{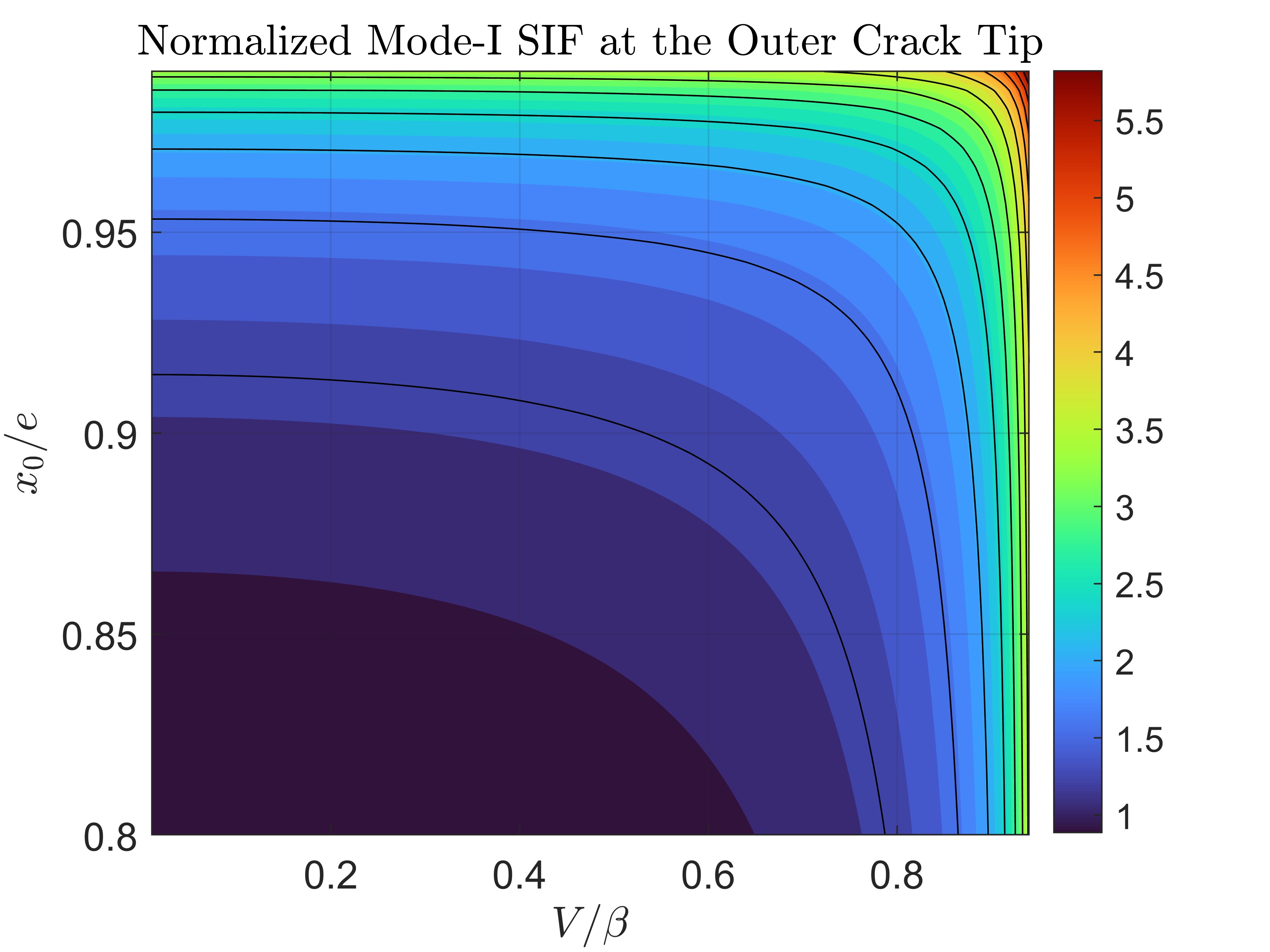}
    \caption{}
\end{subfigure}

\caption{Contour plots of the normalized Mode-I stress intensity factors as functions of the normalized crack-speed ratio $V/\beta$ and the loading parameters: (a)--(b) tangential loading ratio $Q_s/Q_c$ and (c)--(d) normalized load position $x_0/e$ for the inner and outer crack tips.}
\label{fig:contour_loading}
\end{figure*}

The contour plots presented in Figs.~\ref{fig:contour_sandiness}--\ref{fig:contour_loading} provide a comprehensive visualization of the combined influence of the normalized crack-speed ratio and the governing material, initial stress, geometrical, and loading parameters on the normalized Mode-I stress intensity factors at both crack tips. Specifically, Fig.~\ref{fig:contour_sandiness} illustrates the influence of the sandiness parameter, Fig.~\ref{fig:contour_prestress} presents the effects of the initial horizontal and vertical stresses, Fig.~\ref{fig:contour_geometry} depicts the influence of the normalized crack geometry parameter, Fig.~\ref{fig:contour_height} shows the effect of the normalized strip half-thickness, and Fig.~\ref{fig:contour_loading} illustrates the effects of the tangential loading ratio and the normalized load position. In contrast to the one-dimensional line plots presented in the preceding sections, the contour representations illustrate the variation of the fracture response over the entire parameter domain, thereby facilitating a clearer understanding of the coupled interaction between crack speed and the controlling parameters.

Among all the governing variables, the normalized crack-speed ratio $V/\beta$ is observed to be the dominant parameter controlling the fracture behaviour of the strip, as evident from Figs.~\ref{fig:contour_sandiness}--\ref{fig:contour_loading}. In each contour map, the normalized Mode-I stress intensity factors increase continuously with increasing $V/\beta$ and attain their maximum values as $V/\beta$ approaches unity. This behaviour is reflected by the progressively denser contour lines in the high-speed regime, whereas the comparatively wider contour spacing at lower values of $V/\beta$ indicates a more gradual variation of the normalized stress intensity factors.

The contour maps further demonstrate that the sandiness parameter $\chi$ (Fig.~\ref{fig:contour_sandiness}) produces comparatively moderate changes in the normalized Mode-I stress intensity factors throughout the investigated range. In contrast, the initial stresses $\sigma_{11}^{0}$ and $\sigma_{22}^{0}$ (Fig.~\ref{fig:contour_prestress}) exert a more pronounced influence, particularly at higher values of the normalized crack-speed ratio $V/\beta$, with $\sigma_{11}^{0}$ producing a stronger modification of the fracture response than $\sigma_{22}^{0}$. Variations in the normalized crack geometry parameter $c/e$ (Fig.~\ref{fig:contour_geometry}) and the normalized strip half-thickness $h/e$ (Fig.~\ref{fig:contour_height}) alter the interaction of stress waves with the crack boundaries, thereby producing appreciable changes in the normalized Mode-I stress intensity factors, especially in the high-speed regime. Likewise, the loading parameters $Q_s/Q_c$ and $x_0/e$ (Fig.~\ref{fig:contour_loading}) substantially modify the local stress concentration around the crack tips, resulting in corresponding changes in the normalized Mode-I stress intensity factors.

A comparison between the contour plots corresponding to the inner and outer crack tips reveals that the outer crack tip consistently experiences higher normalized Mode-I stress intensity factors throughout the investigated parameter space. This behaviour is attributed to the closer proximity of the outer crack tip to the moving concentrated load, which produces stronger dynamic stress concentration and greater crack-tip amplification. Overall, the contour plots corroborate the trends observed in the corresponding line plots while providing a comprehensive visualization of the coupled influence of crack propagation speed and the governing material, initial stress, geometrical, and loading parameters on the dynamic fracture behaviour of the initially stressed dry sandy strip.


\subsection{Variation of the normalized Mode-I stress intensity factors with the normalized crack-tip position}

Figures~\ref{fig:cracklength_chi}--\ref{fig:cracklength_x0} present the variation of the normalized Mode-I stress intensity factors at the inner and outer crack tips with the normalized crack geometry parameter $c/e$ under different material, geometrical, loading, and initial stress conditions. The corresponding reference values of the parameters used in these parametric studies are listed in Table~\ref{tab:cracklength_parameters}.
Irrespective of the governing parameter, the normalized inner crack-tip position has a pronounced influence on the fracture behaviour of the strip, although the responses at the two crack tips differ significantly.
At the inner crack tip, the normalized Mode-I stress intensity factor exhibits a non-monotonic variation with increasing $c/e$. It decreases rapidly for small values of the normalized crack geometry parameter, attains a minimum around $c/e \approx 0.10$--$0.15$, and subsequently increases as $c/e$ increases further. In contrast, the normalized Mode-I stress intensity factor at the outer crack tip decreases continuously with increasing $c/e$, with a rapid reduction at small crack lengths followed by a more gradual decline over the remaining range. Thus, the normalized inner crack-tip position influences the fracture response differently at the two crack tips.
The specific effects of the sandiness parameter, strip thickness, loading ratio, initial stresses, and load position on these crack-length-dependent variations are discussed separately in the following subsections.

\begin{table}[!ht]
\centering

\rotatebox{90}{%
\begin{minipage}{0.94\textheight}
\centering

\captionof{table}{Reference values of the material, geometrical, loading, and initial stress parameters used to obtain the crack-length-dependent normalized Mode-I stress intensity factors shown in Figs.~\ref{fig:cracklength_chi}--\ref{fig:cracklength_x0}. In each figure, only the parameter indicated as ``Varied'' is changed, while all remaining parameters are maintained at their corresponding reference values.}
\label{tab:cracklength_parameters}

\vspace{0.3cm}

\renewcommand{\arraystretch}{1.2}
\small

\begin{tabular}{lccccc}
\hline
\textbf{Parameter} &
\textbf{Fig.~\ref{fig:cracklength_chi}} &
\textbf{Fig.~\ref{fig:cracklength_height}} &
\textbf{Fig.~\ref{fig:cracklength_loading}} &
\textbf{Fig.~\ref{fig:cracklength_prestress}} &
\textbf{Fig.~\ref{fig:cracklength_x0}} \\
\hline

Lam\'e constant, $\lambda$ (Pa)
& $2.510\times10^{10}$ & $2.510\times10^{10}$ &
$2.510\times10^{10}$ & $2.510\times10^{10}$ &
$2.510\times10^{10}$ \\

Shear modulus, $\mu$ (Pa)
& $1.987\times10^{10}$ & $1.987\times10^{10}$ &
$1.987\times10^{10}$ & $1.987\times10^{10}$ &
$1.987\times10^{10}$ \\

Density, $\rho$ (kg\,m$^{-3}$)
& 4705 & 4705 & 4705 & 4705 & 4705 \\

Sandiness parameter, $\chi$
& Varied & 1.30 & 1.30 & 1.30 & 1.30 \\

Normalized strip half-thickness, $h/e$
& 5 & Varied & 5 & 5 & 5 \\

Tangential loading ratio, $Q_s/Q_c$
& 2.1 & 2.1 & Varied & 2.1 & 2.1 \\

Initial horizontal stress, $\sigma_{11}^{0}$ (Pa)
& $1.0\times10^{9}$ & $1.0\times10^{9}$ &
$1.0\times10^{9}$ & Varied &
$1.0\times10^{9}$ \\

Initial vertical stress, $\sigma_{22}^{0}$ (Pa)
& $1.0\times10^{9}$ & $1.0\times10^{9}$ &
$1.0\times10^{9}$ & Varied &
$1.0\times10^{9}$ \\

Normalized load position, $x_0/e$
& 0.5 & 0.5 & 0.5 & 0.5 & Varied \\

Normalized inner crack-tip position, $c/e$
& Varied & Varied & Varied & Varied & Varied \\

Normal concentrated load, $Q_c$ (N)
& $1.0\times10^{8}$ &
$1.0\times10^{8}$ &
$1.0\times10^{8}$ &
$1.0\times10^{8}$ &
$1.0\times10^{8}$ \\

Outer crack-tip coordinate, $e$
& 1.40 & 1.40 & 1.40 & 1.40 & 1.40 \\

Normalized crack-speed ratio, $V/\beta$
& 0.3 &
 0.3 &
 0.3 &
 0.3 &
 0.3 \\
\hline

\end{tabular}

\end{minipage}
}

\end{table}

\subsubsection{Effect of the sandiness parameter}

\begin{figure*}[!t]
\centering

\begin{subfigure}{0.48\textwidth}
    \centering
    \includegraphics[width=\linewidth]{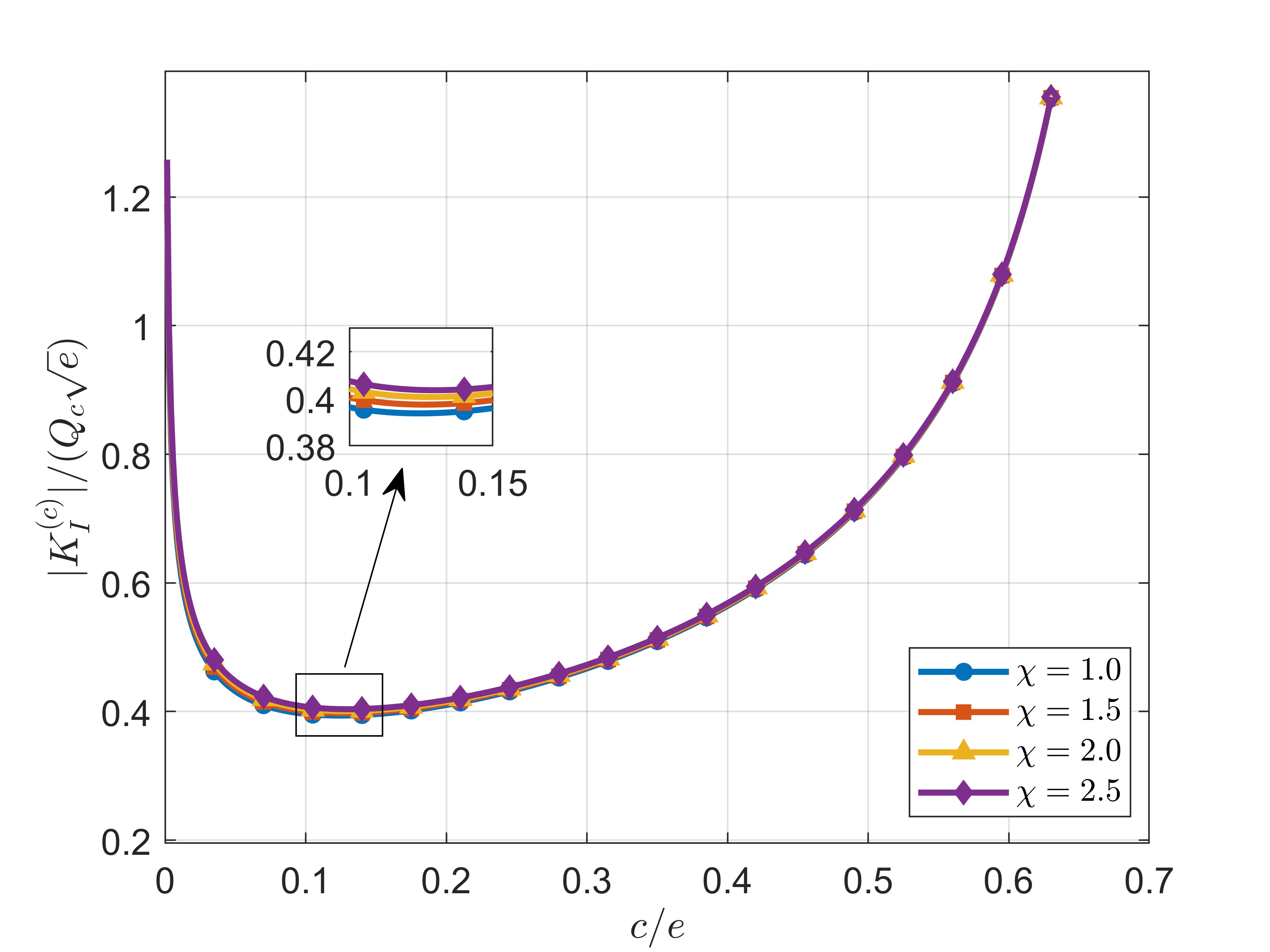}
    \caption{Inner crack tip}
    \label{fig:cracklength_chi(a)}
\end{subfigure}
\hfill
\begin{subfigure}{0.48\textwidth}
    \centering
    \includegraphics[width=\linewidth]{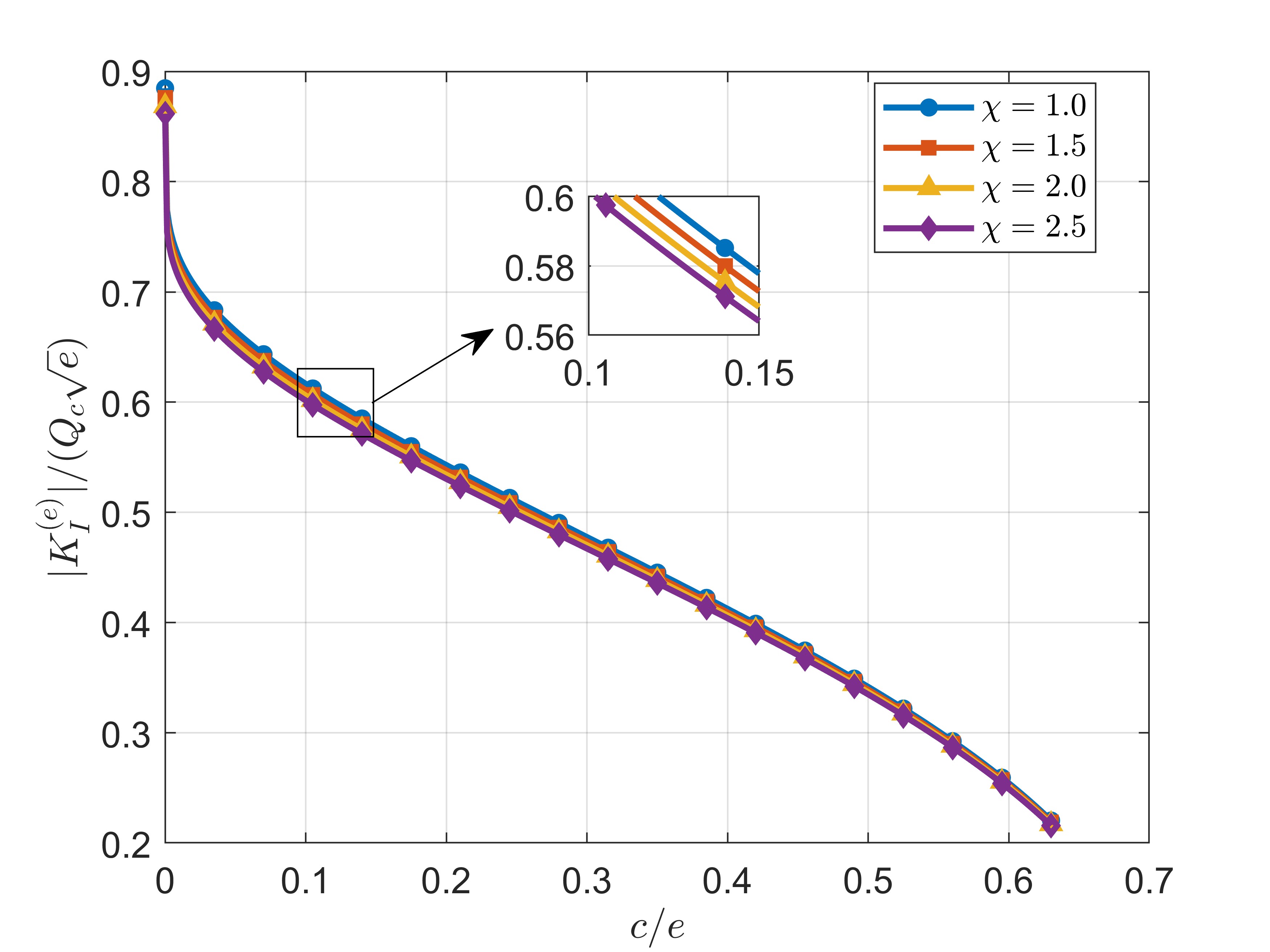}
    \caption{Outer crack tip}
    \label{fig:cracklength_chi(b)}
\end{subfigure}

\caption{Variation of the normalized Mode-I stress intensity factors with the normalized crack geometry parameter $c/e$ for different values of the sandiness parameter $\chi$.}
\label{fig:cracklength_chi}
\end{figure*}

The influence of the sandiness parameter $\chi$ on the normalized Mode-I stress intensity factors at the inner and outer crack tips is presented in Fig.~\ref{fig:cracklength_chi}. It is observed from Fig.~\ref{fig:cracklength_chi(a)} that increasing the sandiness parameter $\chi$ increases the normalized Mode-I stress intensity factor at the inner crack tip over the entire investigated range of the normalized crack geometry parameter $c/e$. In contrast, Fig.~\ref{fig:cracklength_chi(b)} shows that increasing $\chi$ produces a slight reduction in the normalized Mode-I stress intensity factor at the outer crack tip, as can be seen more clearly from the enlarged view. Thus, the sandiness parameter exhibits opposite influences on the fracture response at the two crack tips, enhancing the stress concentration at inner crack tip while slightly reducing it at outer crack tip.

\subsubsection{Effect of the strip thickness}
\begin{figure*}[!t]
\centering

\begin{subfigure}{0.48\textwidth}
    \centering
    \includegraphics[width=\linewidth]{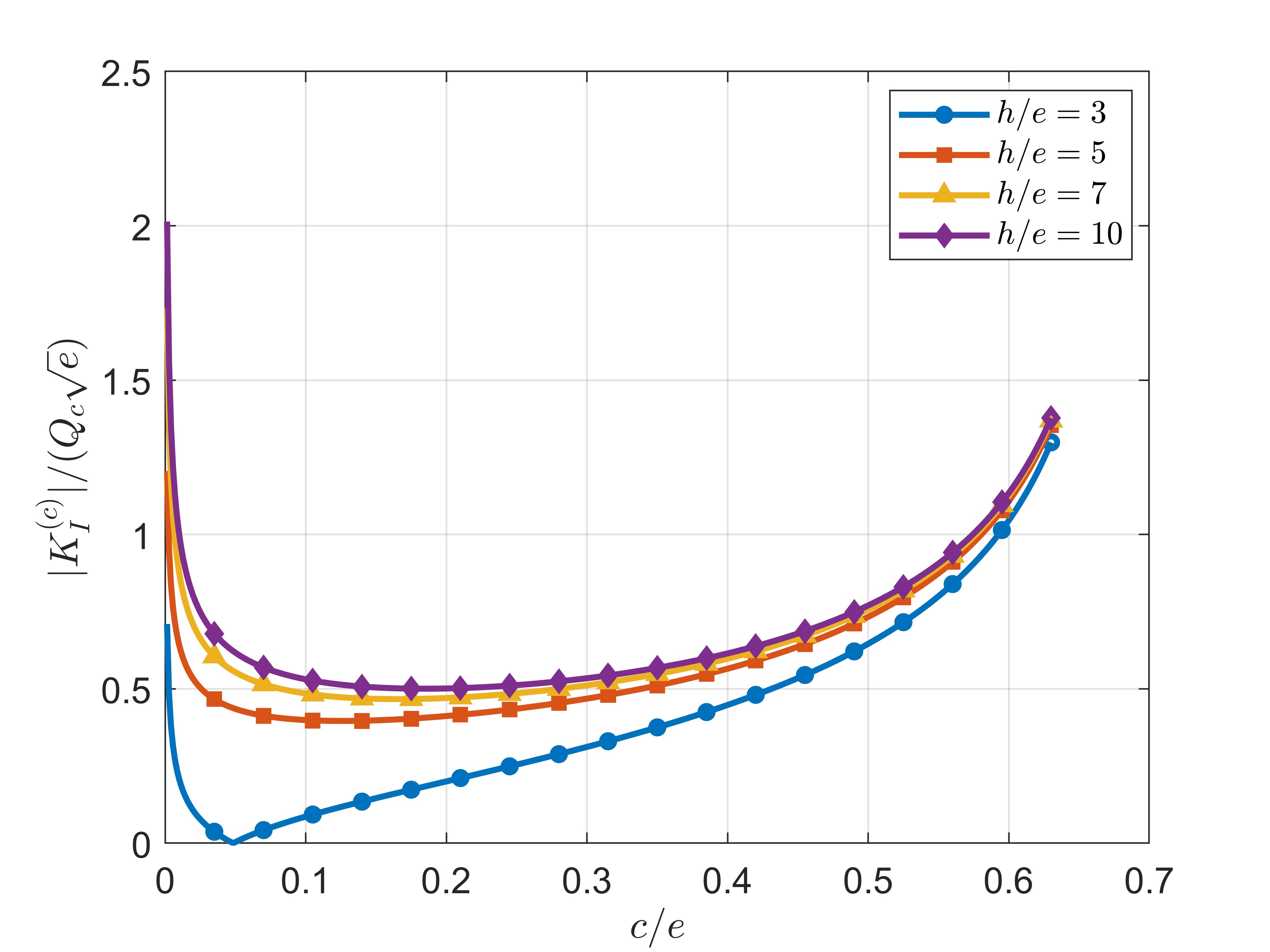}
    \caption{Inner crack tip}
    \label{fig:cracklength_height(a)}
\end{subfigure}
\hfill
\begin{subfigure}{0.48\textwidth}
    \centering
    \includegraphics[width=\linewidth]{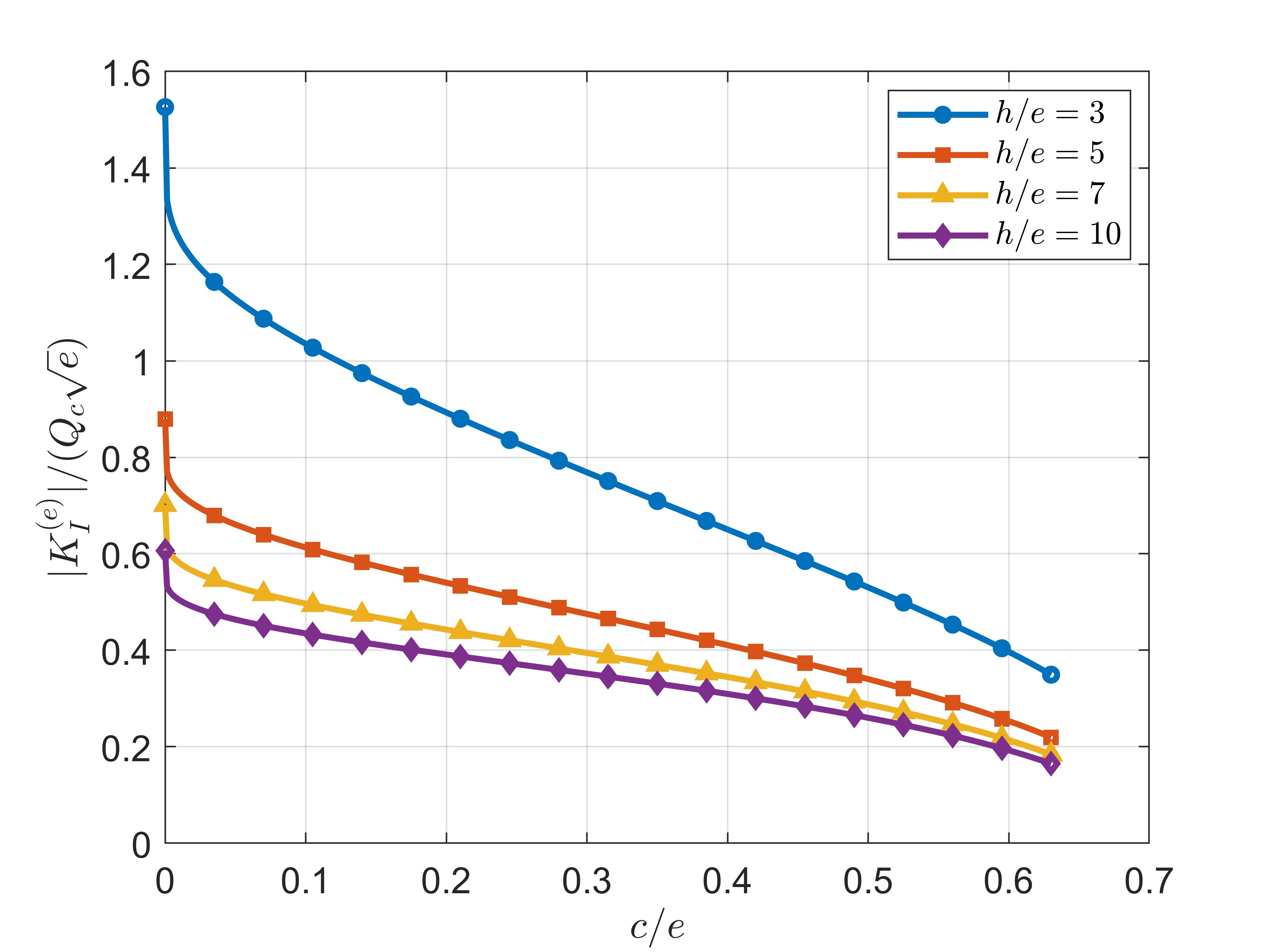}
    \caption{Outer crack tip}
    \label{fig:cracklength_height(b)}
\end{subfigure}

\caption{Variation of the normalized Mode-I stress intensity factors with the normalized crack geometry parameter $c/e$ for different normalized strip half-thicknesses $h/e$.}
\label{fig:cracklength_height}
\end{figure*}

Figure~\ref{fig:cracklength_height} demonstrates the influence of the normalized strip half-thickness $h/e$ on the normalized Mode-I stress intensity factors at the inner and outer crack tips.
It is observed from Fig.~\ref{fig:cracklength_height(a)} that increasing the normalized strip half-thickness increases the normalized Mode-I stress intensity factor at the inner crack tip throughout the investigated range of the normalized crack geometry parameter $c/e$. The increase is more pronounced for relatively small crack lengths, whereas the separation between the curves gradually decreases with increasing $c/e$.
In contrast, Fig.~\ref{fig:cracklength_height(b)} shows that increasing the normalized strip half-thickness decreases the normalized Mode-I stress intensity factor at the outer crack tip. The reduction remains evident over the entire range of $c/e$, indicating that increasing the strip thickness alleviates the stress concentration at the outer crack tip while simultaneously enhancing it at the inner crack tip. Thus, the strip thickness exerts opposite influences on fracture behaviour at two crack tips.

\subsubsection{Effect of the loading parameter}

\begin{figure*}[!t]
\centering

\begin{subfigure}{0.48\textwidth}
    \centering
    \includegraphics[width=\linewidth]{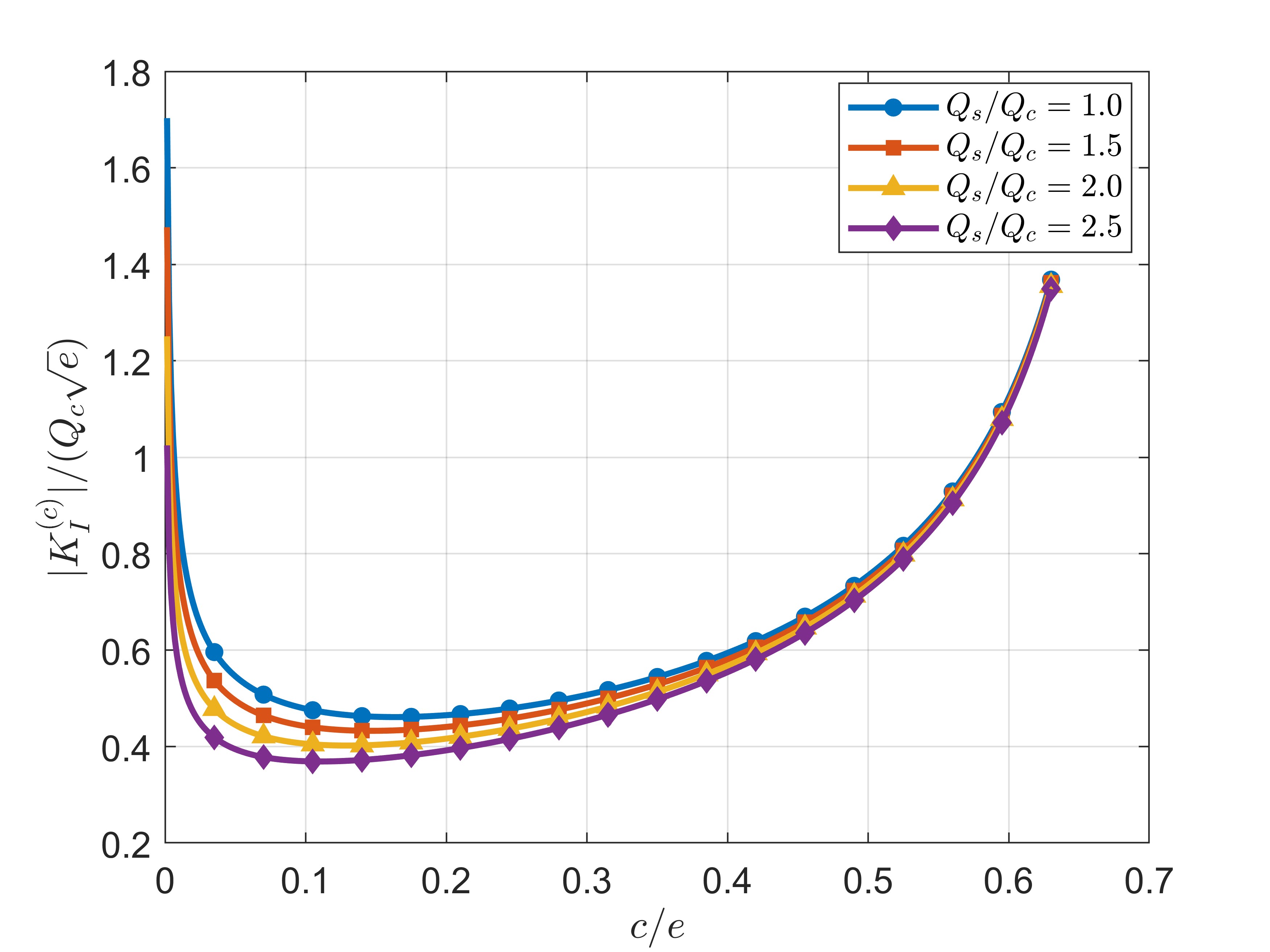}
    \caption{Inner crack tip}
    \label{fig:cracklength_loading(a)}
\end{subfigure}
\hfill
\begin{subfigure}{0.48\textwidth}
    \centering
    \includegraphics[width=\linewidth]{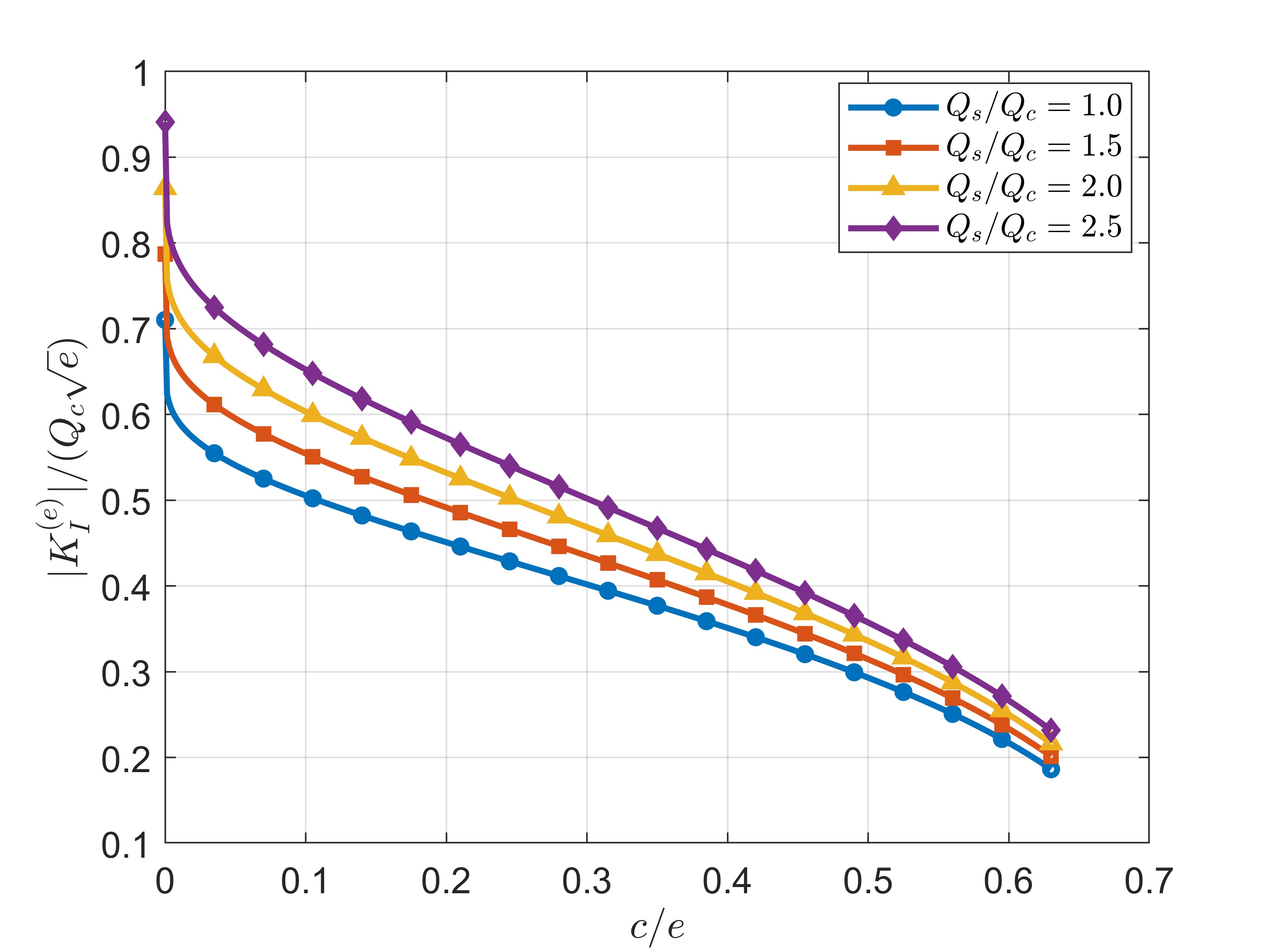}
    \caption{Outer crack tip}
    \label{fig:cracklength_loading(b)}
\end{subfigure}

\caption{Variation of the normalized Mode-I stress intensity factors with the normalized crack geometry parameter $c/e$ for different tangential loading ratios $Q_s/Q_c$.}
\label{fig:cracklength_loading}
\end{figure*}

The effect of the normalized loading ratio $Q_s/Q_c$ on the normalized Mode-I stress intensity factors at the inner and outer crack tips is presented in Fig.~\ref{fig:cracklength_loading}.
It is observed from Fig.~\ref{fig:cracklength_loading(a)} that increasing the loading ratio $Q_s/Q_c$ decreases the normalized Mode-I stress intensity factor at the inner crack tip throughout the investigated range of the normalized crack geometry parameter $c/e$. This behaviour indicates that a larger distributed punch load relative to the concentrated load alleviates the stress concentration at the inner crack tip.
In contrast, Fig.~\ref{fig:cracklength_loading(b)} shows that increasing the loading ratio $Q_s/Q_c$ increases the normalized Mode-I stress intensity factor at the outer crack tip. The upward shift of the curves demonstrates that increasing the distributed punch load intensifies the stress concentration at the outer crack tip. Therefore, the loading parameter exerts opposite influences on the fracture response at the two crack tips by reducing the normalized Mode-I stress intensity factor at inner crack tip while enhancing it at outer crack tip.

\subsubsection{Effect of the initial stresses}
\begin{figure*}[!t]
\centering

\begin{subfigure}{0.48\textwidth}
    \centering
    \includegraphics[width=\linewidth]{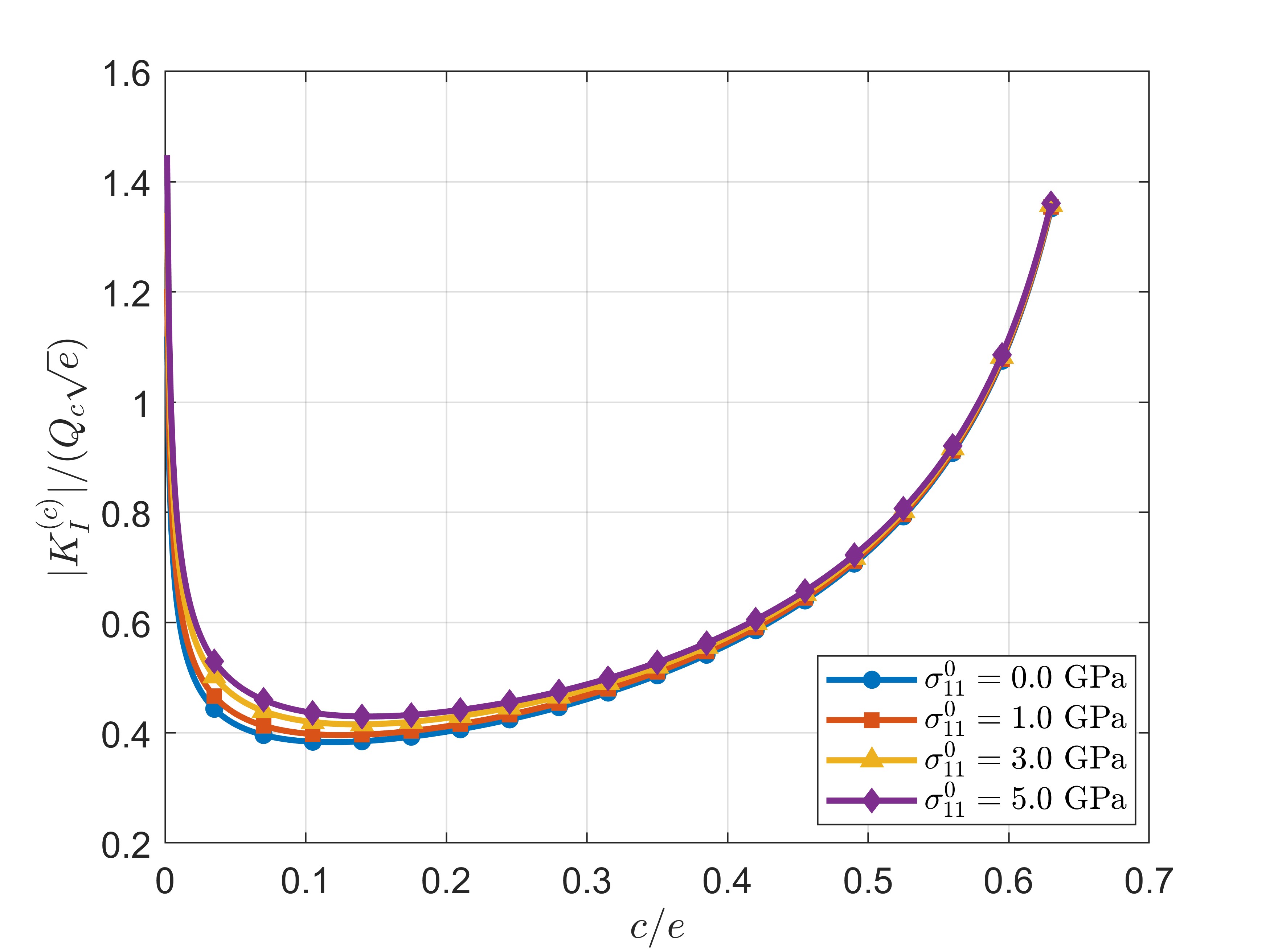}
    \caption{Inner crack tip ($\sigma_{11}^{0}$)}
    \label{fig:cracklength_prestress(a)}
\end{subfigure}
\hfill
\begin{subfigure}{0.48\textwidth}
    \centering
    \includegraphics[width=\linewidth]{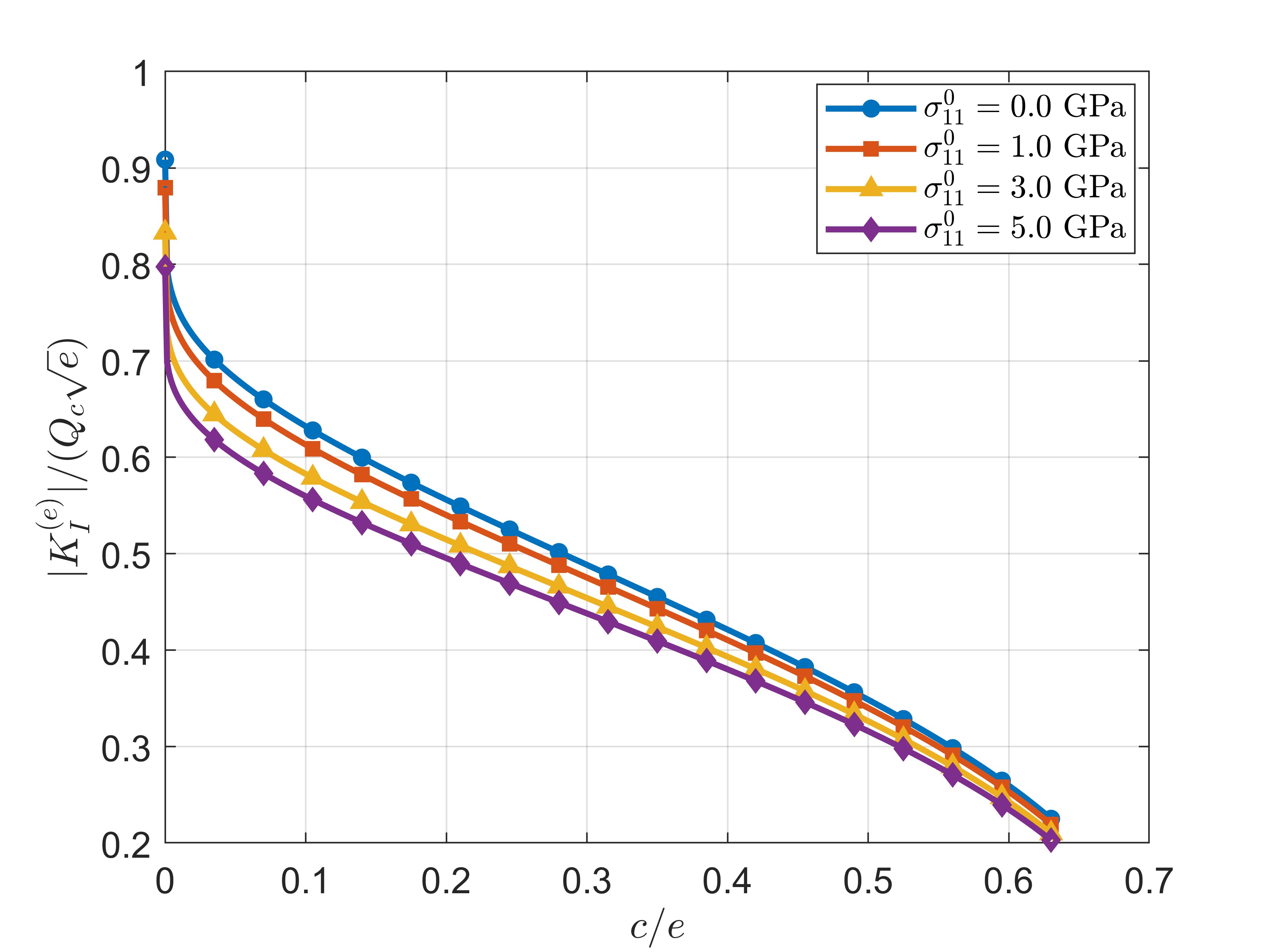}
    \caption{Outer crack tip ($\sigma_{11}^{0}$)}
    \label{fig:cracklength_prestress(b)}
\end{subfigure}

\vspace{0.25cm}

\begin{subfigure}{0.48\textwidth}
    \centering
    \includegraphics[width=\linewidth]{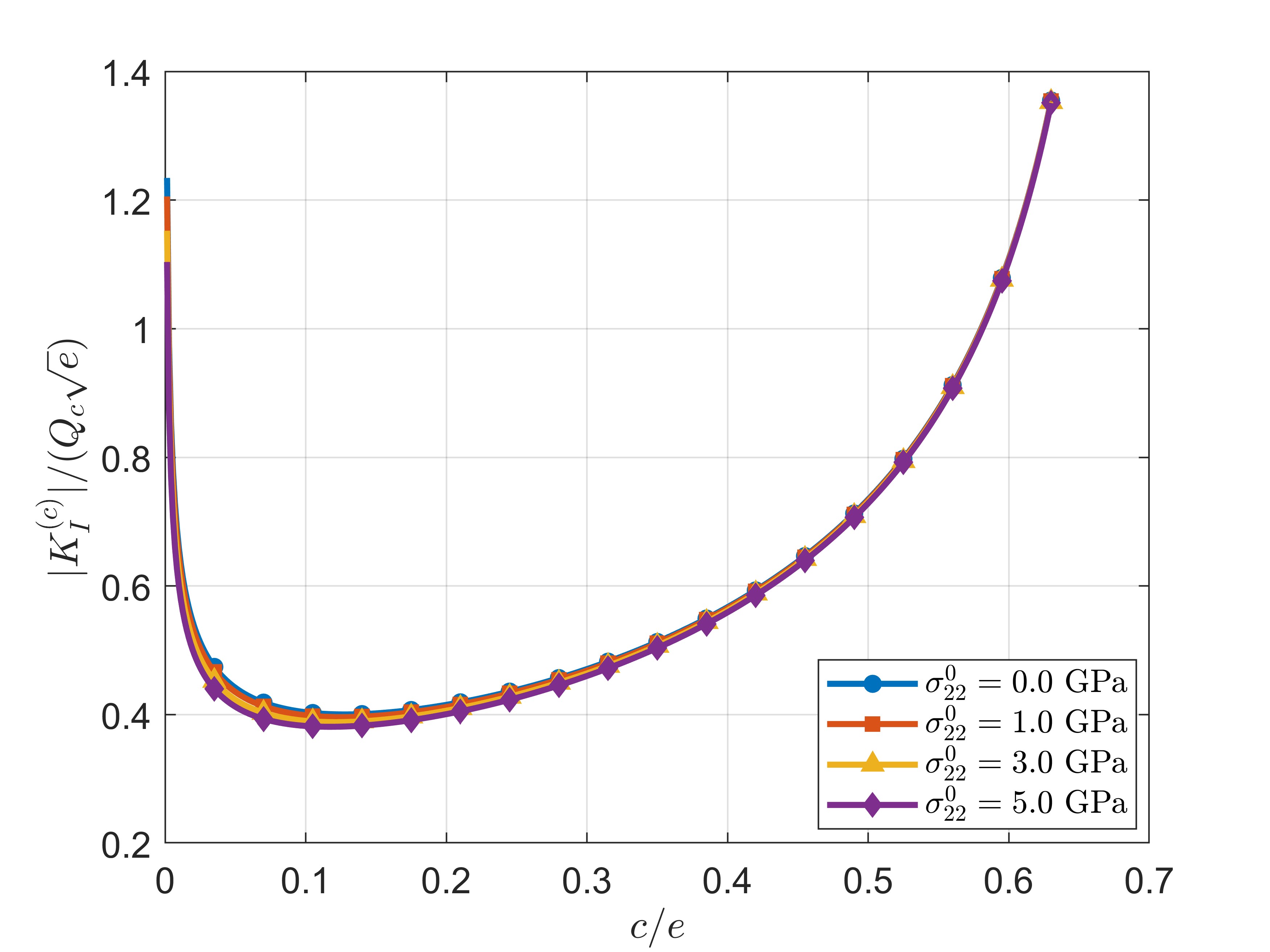}
    \caption{Inner crack tip ($\sigma_{22}^{0}$)}
    \label{fig:cracklength_prestress(c)}
\end{subfigure}
\hfill
\begin{subfigure}{0.48\textwidth}
    \centering
    \includegraphics[width=\linewidth]{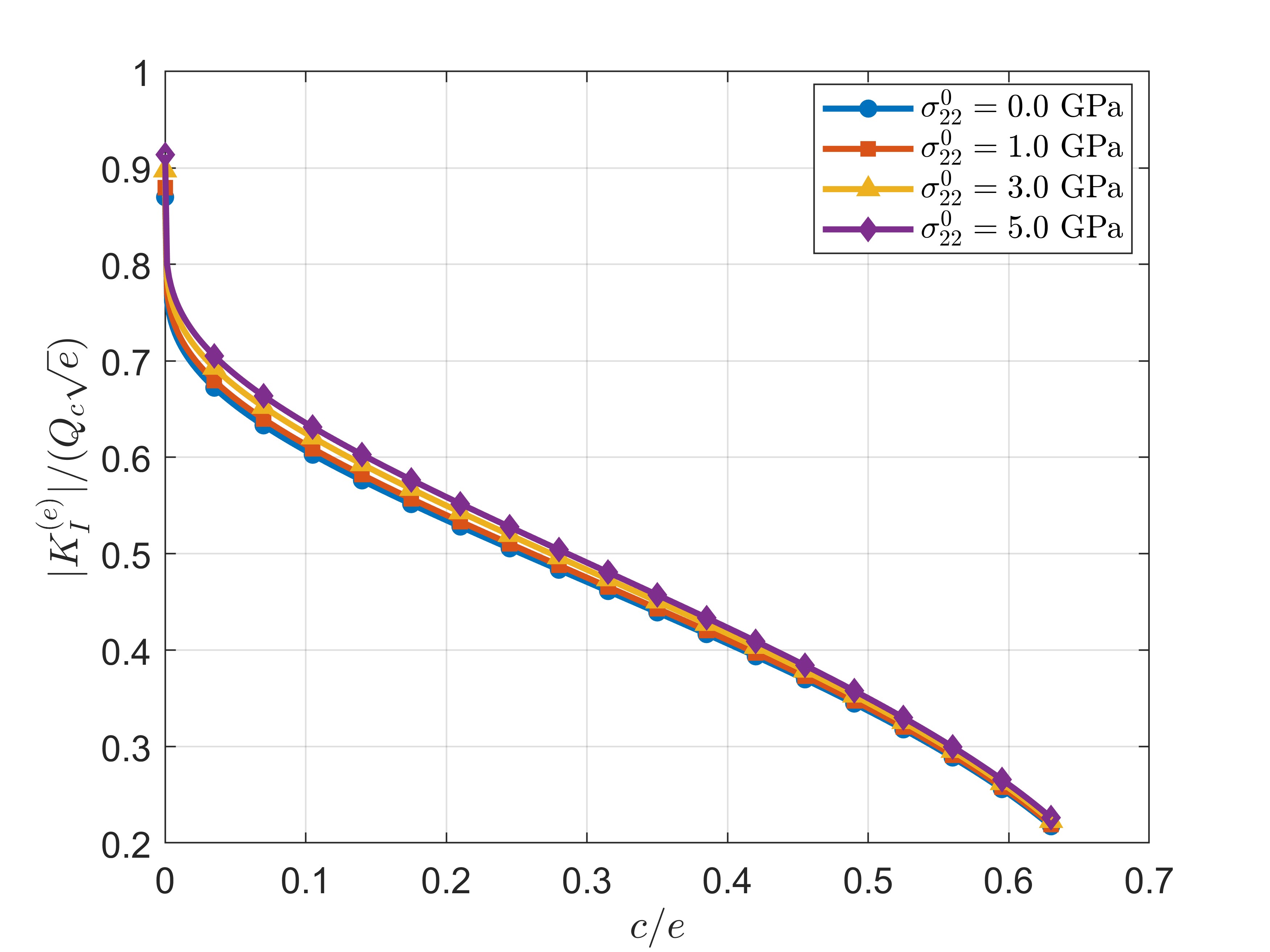}
    \caption{Outer crack tip ($\sigma_{22}^{0}$)}
    \label{fig:cracklength_prestress(d)}
\end{subfigure}

\caption{Variation of the normalized Mode-I stress intensity factors with the normalized crack geometry parameter $c/e$ for different initial horizontal and vertical stresses.}
\label{fig:cracklength_prestress}
\end{figure*}

The fracture response under different initial horizontal and vertical stress levels is presented in Fig.~\ref{fig:cracklength_prestress}, where the corresponding variations in the normalized Mode-I stress intensity factors at the inner and outer crack tips are illustrated.
It is observed from Figs.~\ref{fig:cracklength_prestress(a)} and \ref{fig:cracklength_prestress(b)} that the initial horizontal stress affects the two crack tips differently. Increasing $\sigma_{11}^{0}$ increases the normalized Mode-I stress intensity factor at the inner crack tip, whereas it decreases the normalized Mode-I stress intensity factor at the outer crack tip. This opposite behaviour indicates that the initial horizontal stress redistributes the stress field between the two crack tips, enhancing the stress concentration at the inner crack tip while reducing it at the outer crack tip.
In contrast, Figs.~\ref{fig:cracklength_prestress(c)} and \ref{fig:cracklength_prestress(d)} demonstrate that the initial vertical stress $\sigma_{22}^{0}$ also exerts opposite influences on the normalized Mode-I stress intensity factors at the two crack tips. Increasing $\sigma_{22}^{0}$ decreases the normalized Mode-I stress intensity factor at the inner crack tip, whereas it increases the normalized Mode-I stress intensity factor at the outer crack tip. Nevertheless, the separation between the curves is smaller than that observed for the initial horizontal stress, indicating that the influence of the initial vertical stress on the fracture response is comparatively weaker. Therefore, both initial stress components redistribute the crack-tip stress field in opposite directions, although the effect of the initial horizontal stress is more pronounced than that of the initial vertical stress.

\subsubsection{Effect of the normalized load position}
\begin{figure*}[!t]
\centering

\begin{subfigure}{0.48\textwidth}
    \centering
    \includegraphics[width=\linewidth]{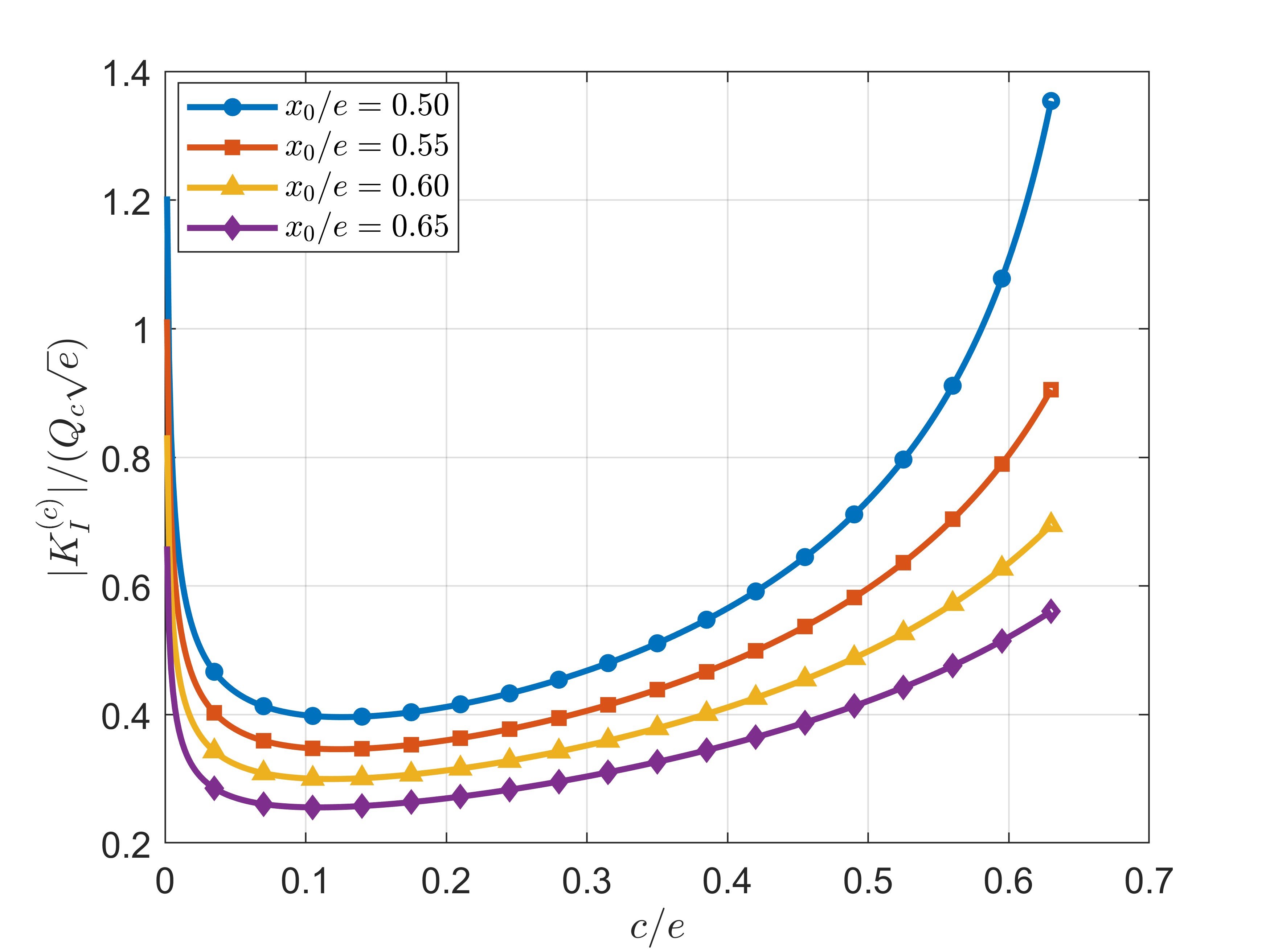}
    \caption{Inner crack tip}
    \label{fig:cracklength_x0(a)}
\end{subfigure}
\hfill
\begin{subfigure}{0.48\textwidth}
    \centering
    \includegraphics[width=\linewidth]{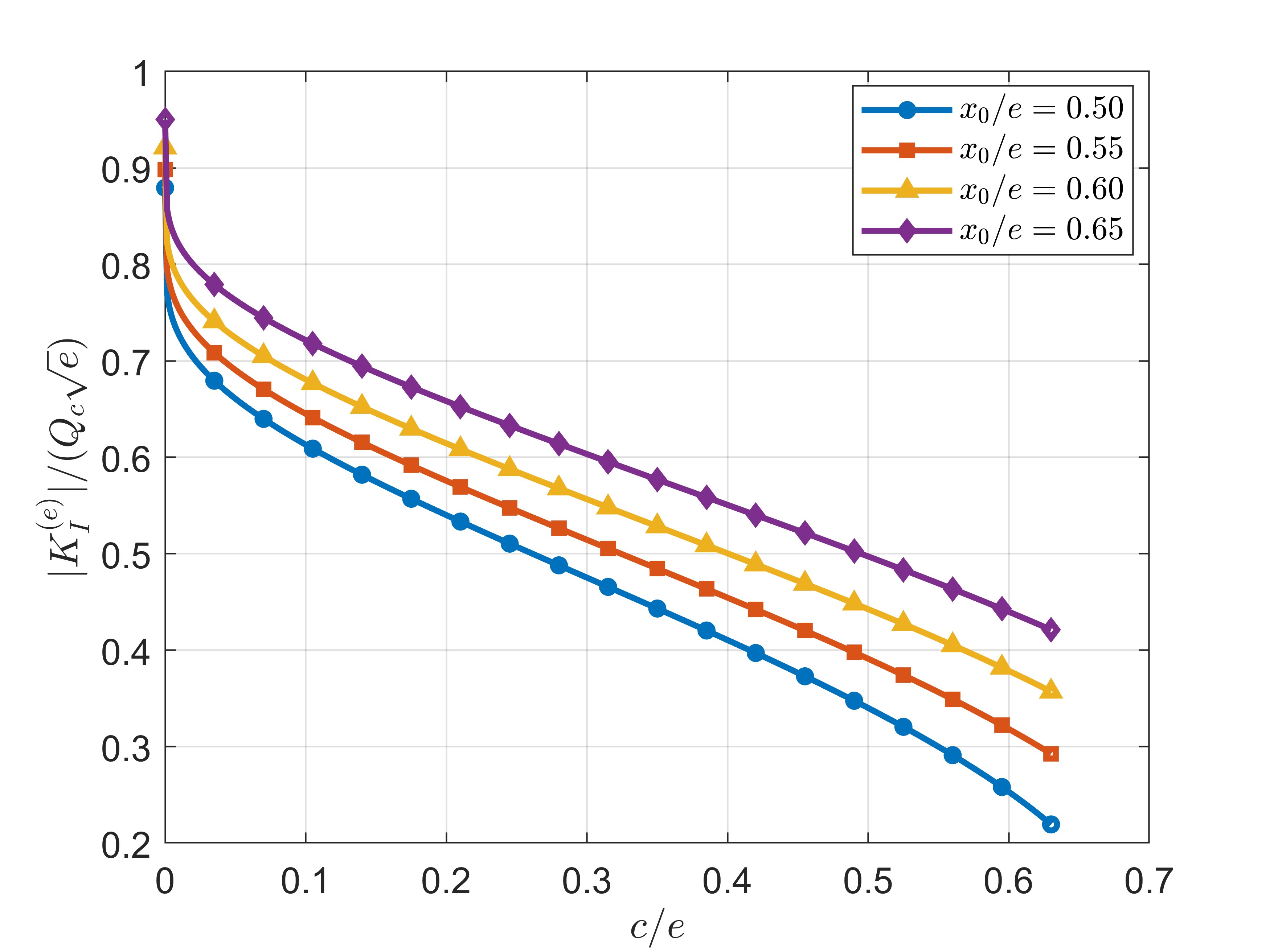}
    \caption{Outer crack tip}
    \label{fig:cracklength_x0(b)}
\end{subfigure}

\caption{Variation of the normalized Mode-I stress intensity factors with the normalized crack geometry parameter $c/e$ for different normalized load positions $x_0/e$.}
\label{fig:cracklength_x0}
\end{figure*}

Figure~\ref{fig:cracklength_x0} illustrates the influence of the normalized load position $x_{0}/e$ on the normalized Mode-I stress intensity factors at the inner and outer crack tips. It is observed from Fig.~\ref{fig:cracklength_x0(a)} that increasing the normalized load position increases the normalized Mode-I stress intensity factor at the inner crack tip throughout the investigated range of the normalized crack geometry parameter $c/e$. This indicates that positioning the concentrated load closer to the outer crack tip enhances the stress concentration at the inner crack tip.
In contrast, Fig.~\ref{fig:cracklength_x0(b)} shows that increasing the normalized load position decreases the normalized Mode-I stress intensity factor at the outer crack tip. The reduction becomes progressively more pronounced with increasing normalized inner crack-tip position, resulting in a substantial separation between the corresponding curves. These opposite trends demonstrate that varying the load position redistributes the stress field between the two crack tips, increasing the fracture intensity at the inner crack tip while simultaneously reducing it at the outer crack tip.

\subsubsection{Contour representation of the normalized Mode-I stress intensity factors}

\begin{figure*}[!ht]
\centering

\begin{subfigure}{0.48\textwidth}
    \centering
    \includegraphics[width=\linewidth]{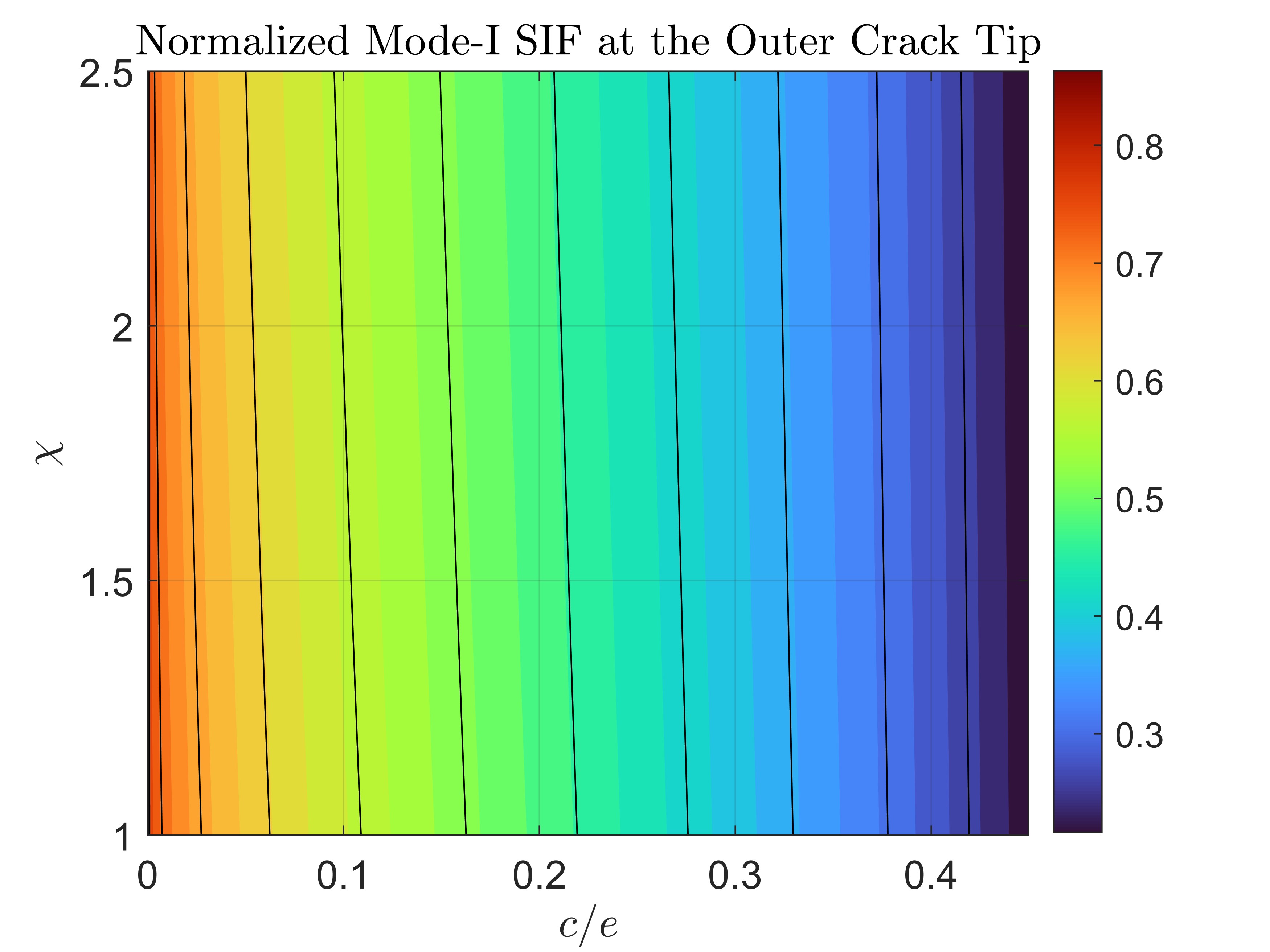}
    \caption{}
    \label{fig:contour_chi_outer}
\end{subfigure}
\hfill
\begin{subfigure}{0.48\textwidth}
    \centering
    \includegraphics[width=\linewidth]{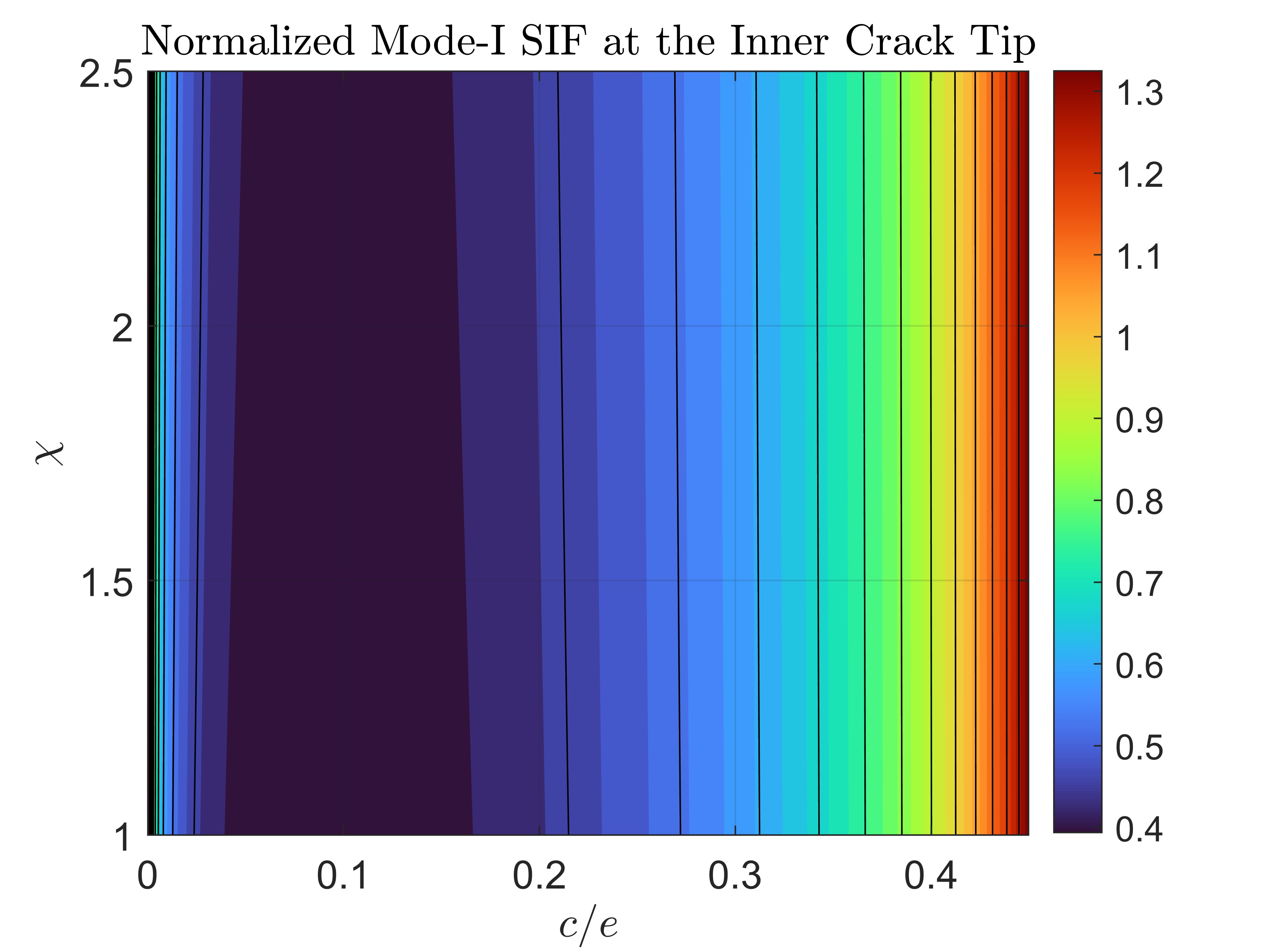}
    \caption{}
    \label{fig:contour_chi_inner}
\end{subfigure}

\caption{Contour representation of the normalized Mode-I stress intensity factors as functions of the normalized inner crack-tip position $(c/e)$ and the sandiness parameter $(\chi)$: (a) outer crack tip and (b) inner crack tip.}
\label{fig:contour_chi}

\end{figure*}

\begin{figure*}[!ht]
\centering

\begin{subfigure}{0.48\textwidth}
    \centering
    \includegraphics[width=\linewidth]{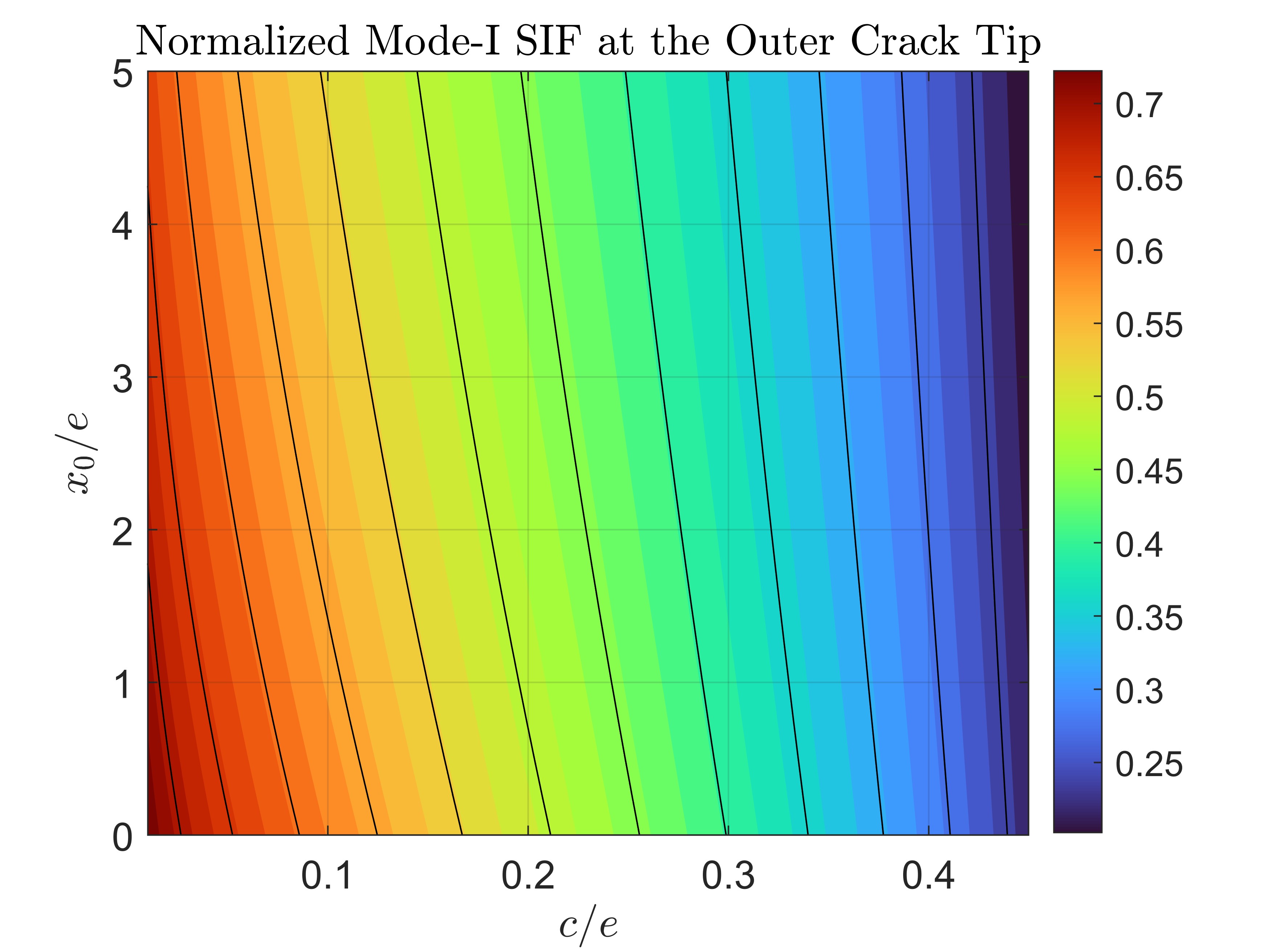}
    \caption{}
    \label{fig:contour_sigma11_outer}
\end{subfigure}
\hfill
\begin{subfigure}{0.48\textwidth}
    \centering
    \includegraphics[width=\linewidth]{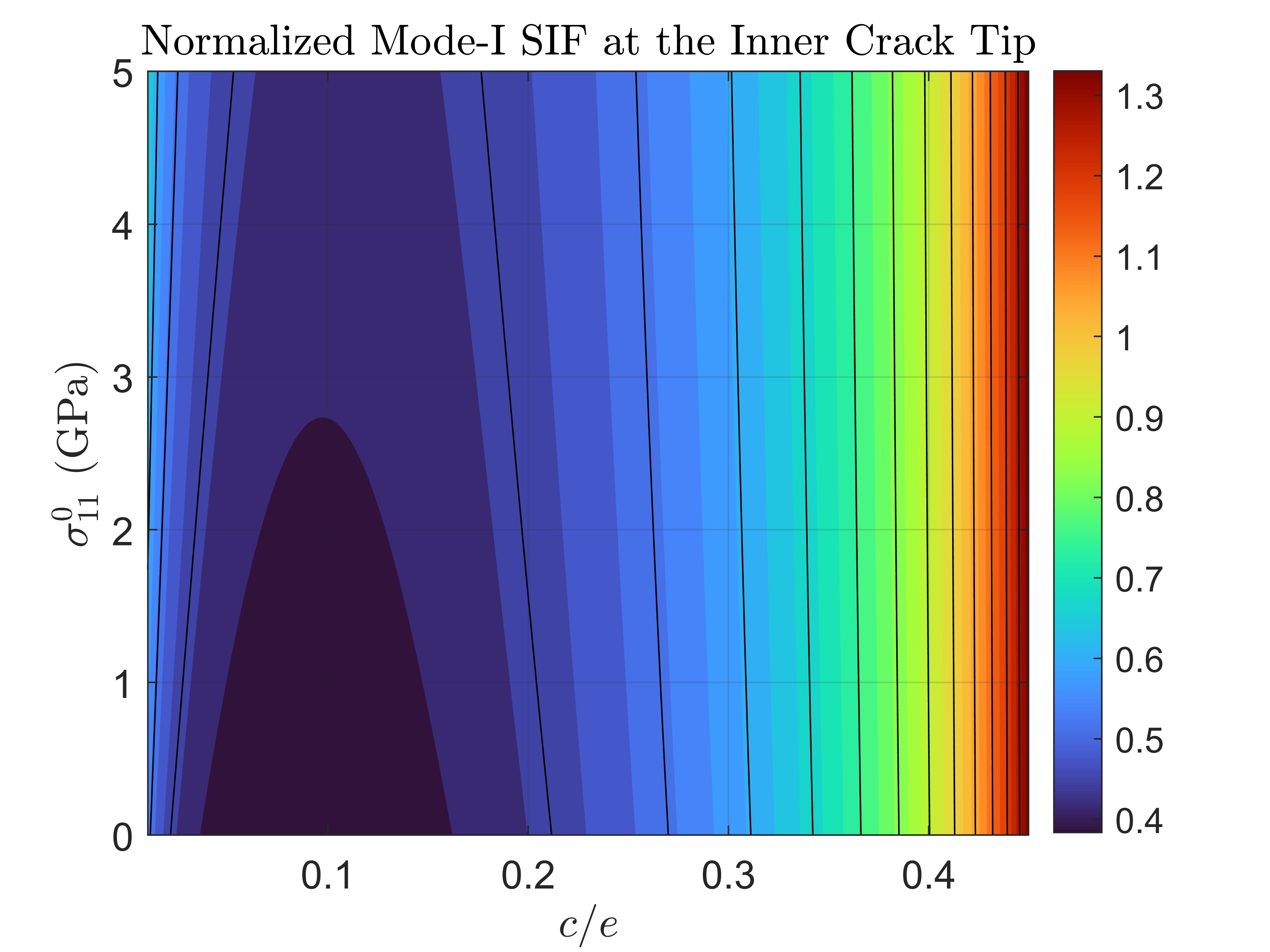}
    \caption{}
    \label{fig:contour_sigma11_inner}
\end{subfigure}

\vspace{0.4cm}

\begin{subfigure}{0.48\textwidth}
    \centering
    \includegraphics[width=\linewidth]{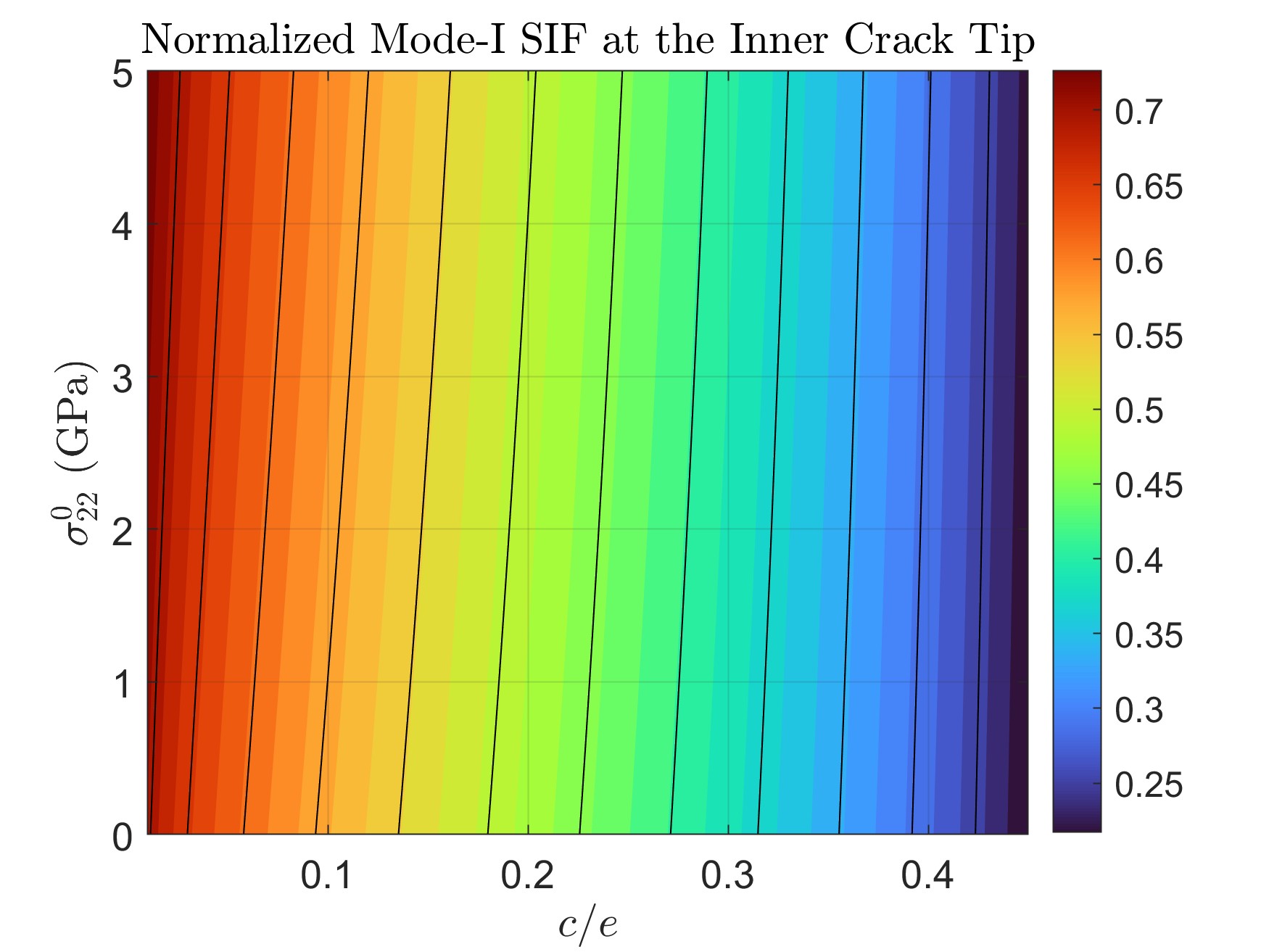}
    \caption{}
    \label{fig:contour_sigma22_outer}
\end{subfigure}
\hfill
\begin{subfigure}{0.48\textwidth}
    \centering
    \includegraphics[width=\linewidth]{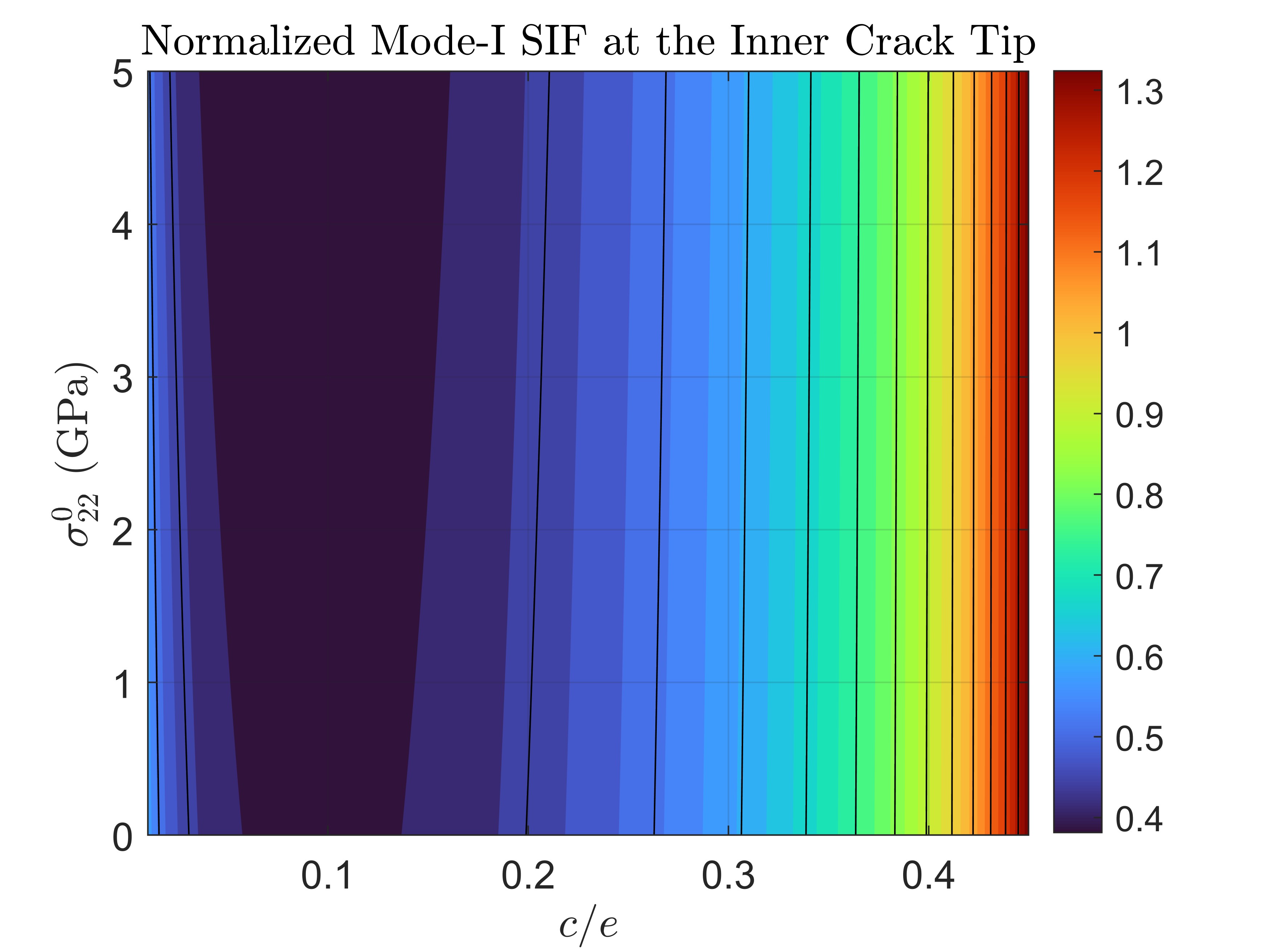}
    \caption{}
    \label{fig:contour_sigma22_inner}
\end{subfigure}

\caption{Contour representation of the normalized Mode-I stress intensity factors as functions of the normalized inner crack-tip position $(c/e)$ and the initial stresses: (a) outer crack tip for the initial horizontal stress $\sigma_{11}^{0}$, (b) inner crack tip for the initial horizontal stress $\sigma_{11}^{0}$, (c) outer crack tip for the initial vertical stress $\sigma_{22}^{0}$, and (d) inner crack tip for the initial vertical stress $\sigma_{22}^{0}$.}
\label{fig:contour_prestress2}

\end{figure*}

\begin{figure*}[!ht]
\centering

\begin{subfigure}{0.48\textwidth}
    \centering
    \includegraphics[width=\linewidth]{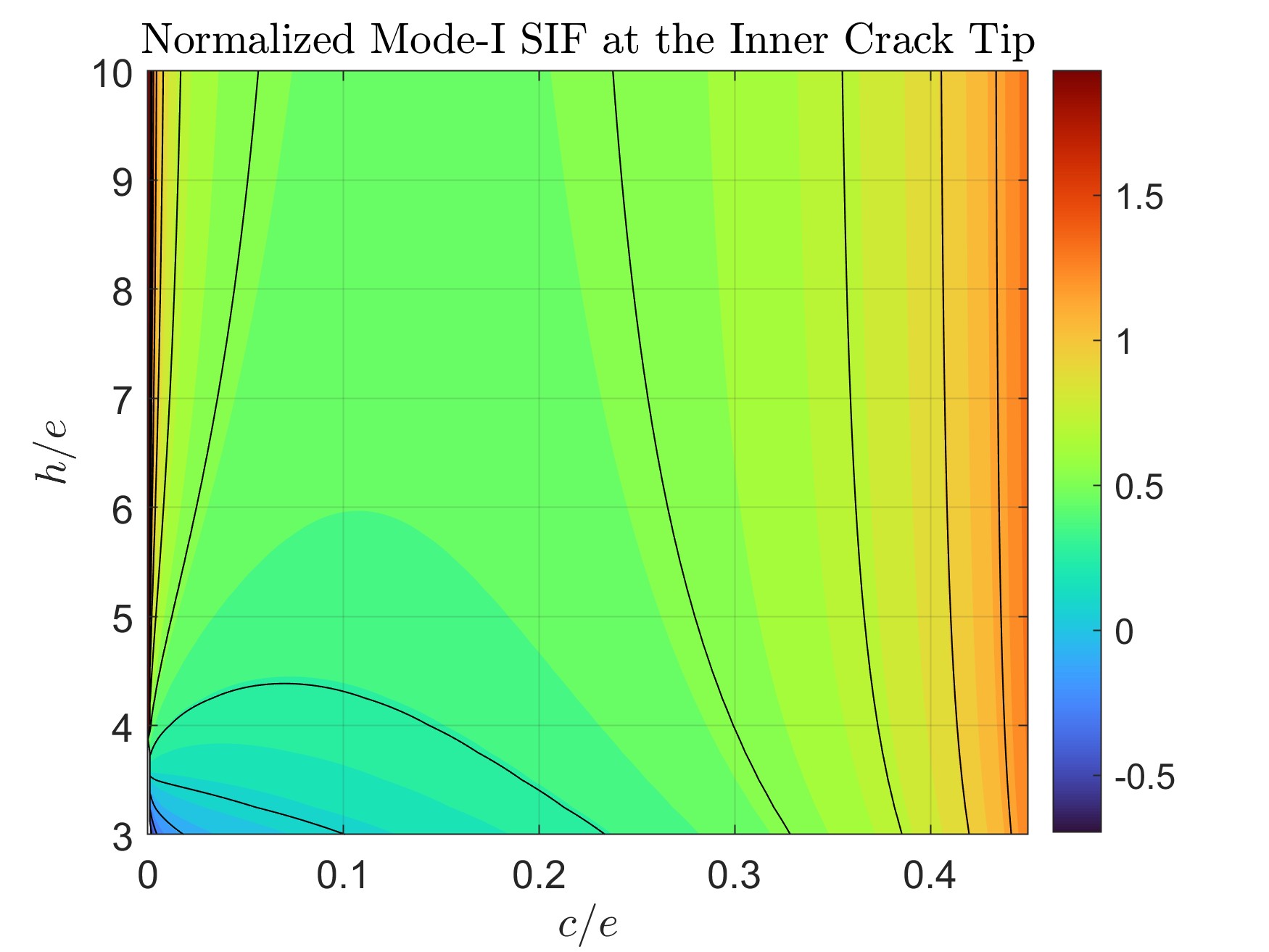}
    \caption{}
    \label{fig:contour_height_outer}
\end{subfigure}
\hfill
\begin{subfigure}{0.48\textwidth}
    \centering
    \includegraphics[width=\linewidth]{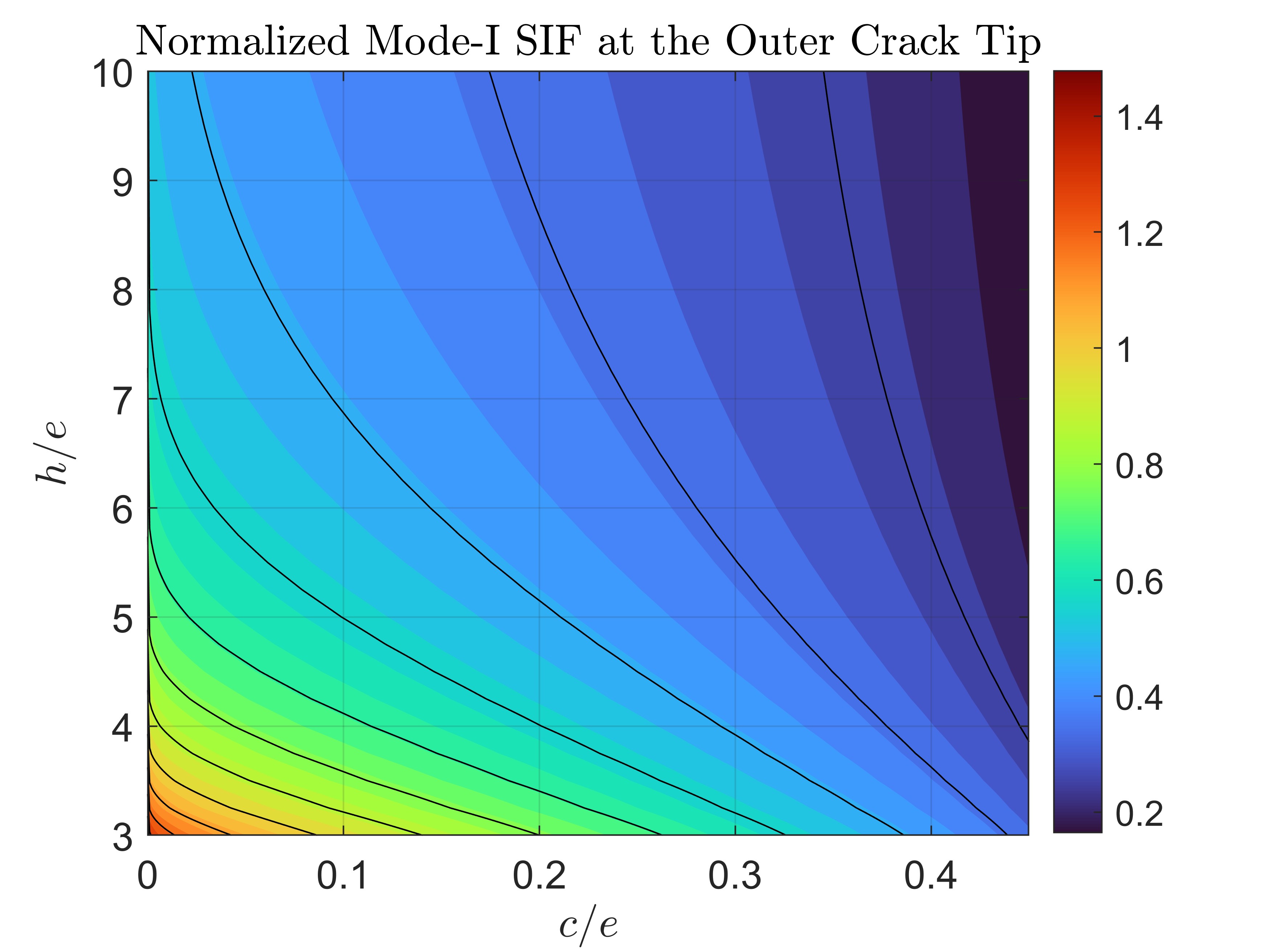}
    \caption{}
    \label{fig:contour_height_inner}
\end{subfigure}

\caption{Contour representation of the normalized Mode-I stress intensity factors as functions of the normalized inner crack-tip position $(c/e)$ and the normalized strip half-thickness $(h/e)$: (a) outer crack tip and (b) inner crack tip.}
\label{fig:contour_height2}

\end{figure*}

\begin{figure*}[!ht]
\centering

\begin{subfigure}{0.48\textwidth}
    \centering
    \includegraphics[width=\linewidth]{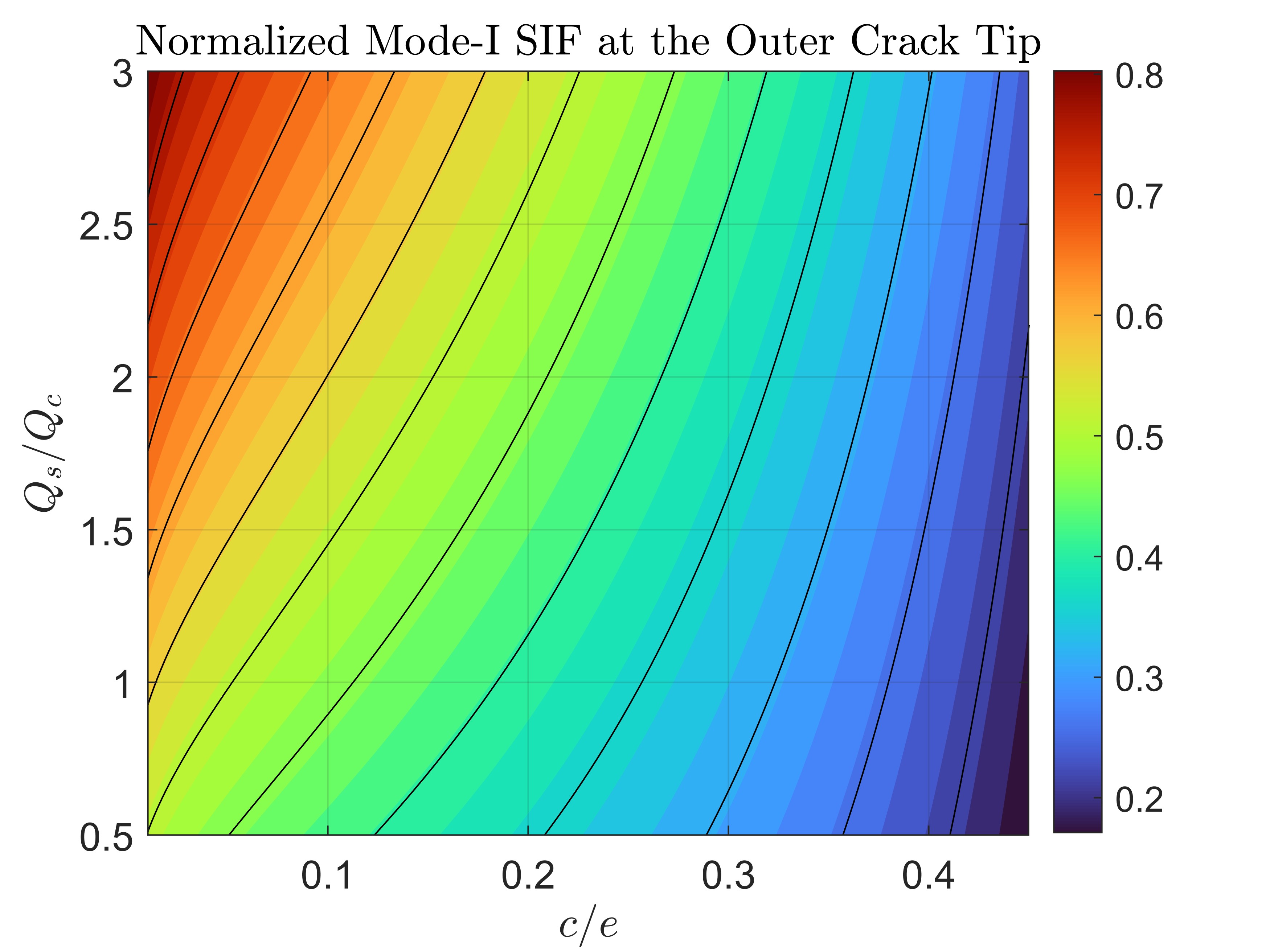}
    \caption{}
    \label{fig:contour_loading_outer}
\end{subfigure}
\hfill
\begin{subfigure}{0.48\textwidth}
    \centering
    \includegraphics[width=\linewidth]{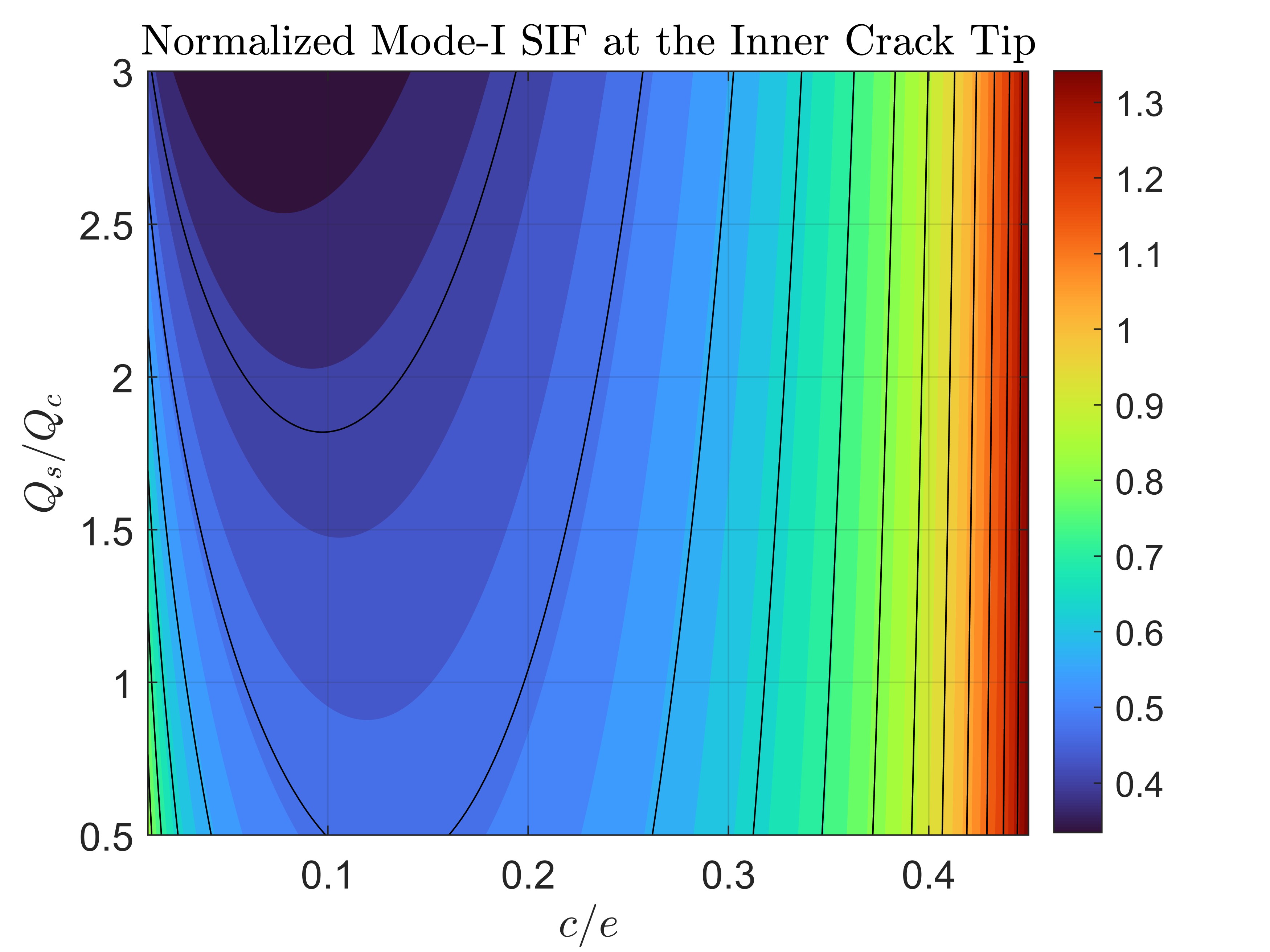}
    \caption{}
    \label{fig:contour_loading_inner}
\end{subfigure}

\caption{Contour representation of the normalized Mode-I stress intensity factors as functions of the normalized inner crack-tip position $(c/e)$ and the loading ratio $(Q_s/Q_c)$: (a) outer crack tip and (b) inner crack tip.}
\label{fig:contour_loading2}

\end{figure*}

\begin{figure*}[!ht]
\centering

\begin{subfigure}{0.48\textwidth}
    \centering
    \includegraphics[width=\linewidth]{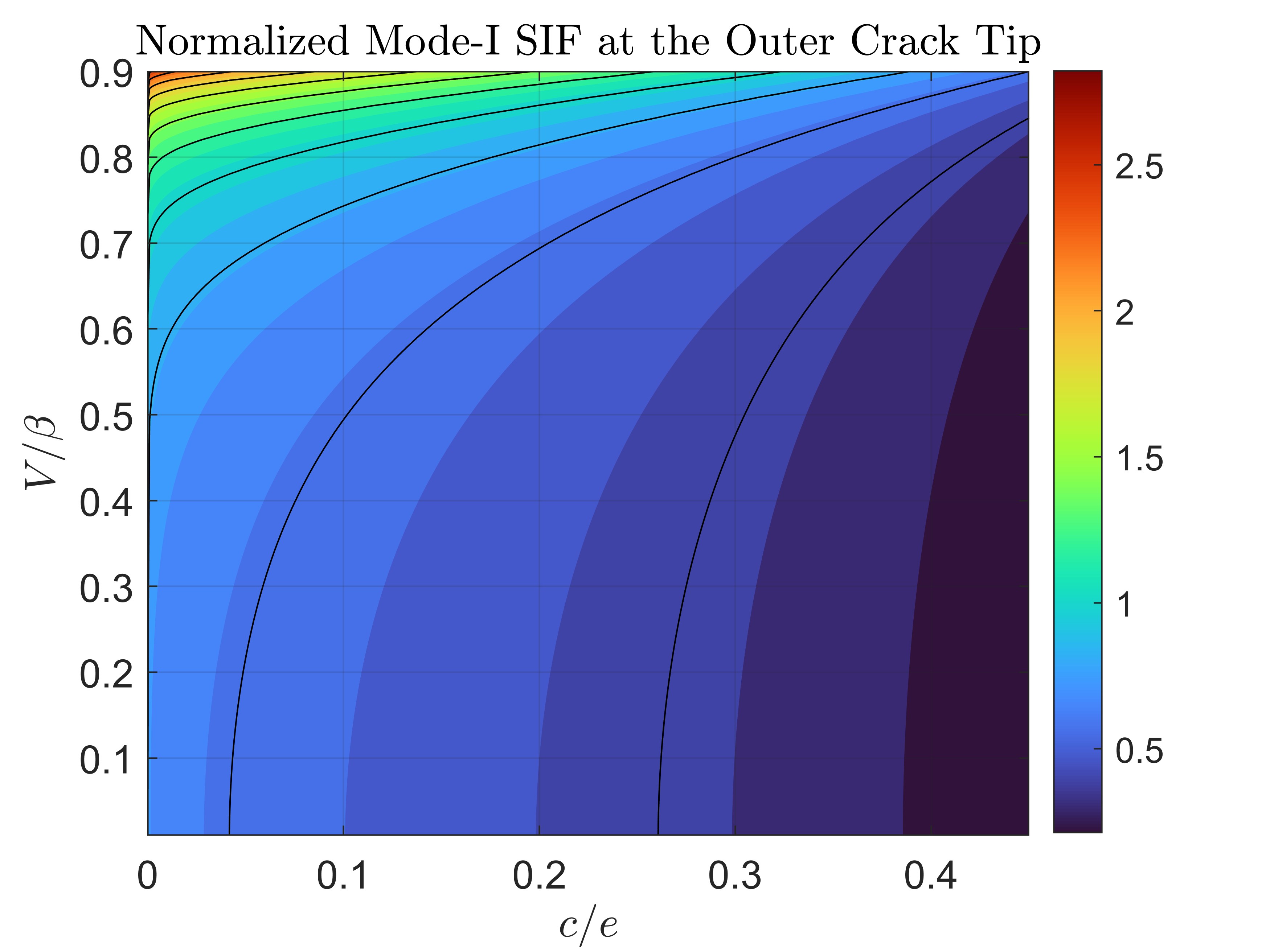}
    \caption{}
    \label{fig:contour_speed_outer}
\end{subfigure}
\hfill
\begin{subfigure}{0.48\textwidth}
    \centering
    \includegraphics[width=\linewidth]{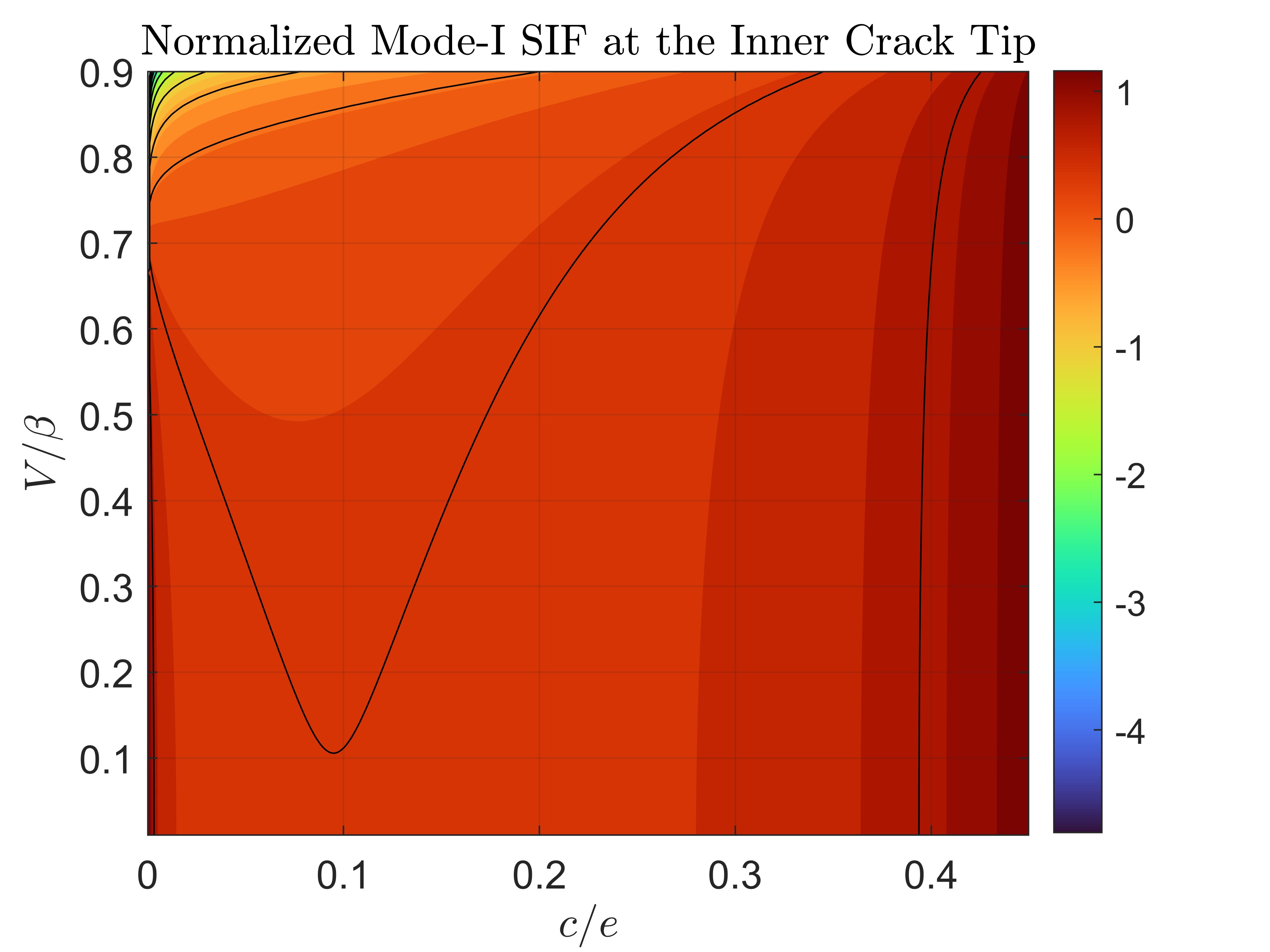}
    \caption{}
    \label{fig:contour_speed_inner}
\end{subfigure}

\caption{Contour representation of the normalized Mode-I stress intensity factors as functions of the normalized inner crack-tip position $(c/e)$ and the crack-speed ratio $(V/\beta)$: (a) outer crack tip and (b) inner crack tip.}
\label{fig:contour_speed}

\end{figure*}

The contour plots presented in Figs.~\ref{fig:contour_chi}--\ref{fig:contour_speed} provide a comprehensive representation of the normalized Mode-I stress intensity factors at the inner and outer crack tips over the entire range of the governing parameters. Specifically, Fig.~\ref{fig:contour_chi} illustrates the influence of the sandiness parameter $\chi$, Fig.~\ref{fig:contour_prestress2} presents the effects of the initial horizontal and vertical stresses $(\sigma_{11}^{0},\sigma_{22}^{0})$, Fig.~\ref{fig:contour_height2} depicts the influence of the normalized strip half-thickness $(h/e)$, Fig.~\ref{fig:contour_loading2} shows the effect of the loading ratio $(Q_s/Q_c)$, and Fig.~\ref{fig:contour_speed} illustrates the influence of the normalized crack-speed ratio $(V/\beta)$. Compared with the corresponding line plots, these contour maps simultaneously display the variation of the normalized stress intensity factors with crack length and the governing parameter, thereby providing a clearer visualization of their coupled effects.

It is observed from all contour maps that the normalized inner crack-tip position $(c/e)$ significantly influences the fracture response of the strip. As the crack length increases, the contour levels undergo noticeable changes for both crack tips, indicating continuous redistribution of the crack-tip stress field caused by the interaction between the two collinear cracks. This behaviour is manifested by the gradual variation of the contour bands along the horizontal direction and reflects the strong dependence of the normalized Mode-I stress intensity factors on crack geometry.

The influence of the individual governing parameters can also be clearly identified from the contour distributions. The sandiness parameter $\chi$ (Fig.~\ref{fig:contour_chi}) produces relatively smooth variations in the contour levels, indicating that changes in the granular characteristics of the medium modify the crack-tip stress field without substantially altering its overall distribution. The initial stresses $\sigma_{11}^{0}$ and $\sigma_{22}^{0}$ (Fig.~\ref{fig:contour_prestress2}) shift the contour levels owing to the superposition of the pre-existing stress field and the dynamically induced stresses around the crack tips. The effect of the normalized strip half-thickness $h/e$ (Fig.~\ref{fig:contour_height2}) is more pronounced, since variations in the strip thickness modify the interaction between stress waves reflected from the strip boundaries and those generated at the crack tips, thereby influencing the intensity of crack-tip stress concentration. Likewise, the loading ratio $Q_s/Q_c$ (Fig.~\ref{fig:contour_loading2}) alters the relative contribution of the concentrated and distributed loads, leading to corresponding changes in the stress field and consequently in the normalized Mode-I stress intensity factors. The contour plots corresponding to the crack-speed ratio $V/\beta$ (Fig.~\ref{fig:contour_speed}) demonstrate that increasing crack speed considerably modifies the dynamic stress field, and the normalized stress intensity factors increase rapidly as the crack velocity approaches the characteristic shear-wave speed, reflecting the well-known amplification associated with dynamic fracture.


\section{Conclusion}
\label{Conclusion}

The present study investigates the dynamic fracture behaviour of two moving collinear Griffith cracks embedded in an initially stressed dry sandy elastic strip subjected to moving concentrated and distributed loading. An analytical formulation based on the Fourier transform technique is developed to determine the crack-tip stress fields and the corresponding normalized Mode-I stress intensity factors at both the inner and outer crack tips.tips. The principal findings of the present investigation are summarized as follows:

\begin{itemize}
    \item The normalized crack-speed ratio $(V/\beta)$ has a significant influence on the dynamic fracture response. The normalized Mode-I stress intensity factors increase continuously with increasing crack speed and exhibit rapid amplification as the crack velocity approaches the characteristic shear-wave speed.

    \item The sandiness parameter $(\chi)$ modifies the crack-tip stress intensity factors by altering the effective mechanical properties of the dry sandy medium. Although its influence is moderate compared with the dynamic effects, it contributes to the overall fracture response.

    \item The initial horizontal and vertical stresses $(\sigma_{11}^{0},\,\sigma_{22}^{0})$ affect the crack-tip stress field through stress superposition, demonstrating that the initial state of stress plays an important role in the fracture behaviour of the strip.

    \item The normalized strip half-thickness $(h/e)$ substantially influences the stress intensity factors. Variations in strip thickness modify the interaction between reflected stress waves and the crack-tip stress field, thereby changing the level of stress concentration.

    \item The loading ratio $(Q_s/Q_c)$ provides an effective means of controlling the fracture response. Changes in the relative magnitudes of the moving concentrated and distributed loads produce noticeable variations in the normalized Mode-I stress intensity factors.

    \item The outer crack tip generally experiences larger normalized Mode-I stress intensity factors than the inner crack tip, indicating the combined influence of crack interaction, asymmetric stress redistribution, and the proximity of the moving loads.

    \item The contour representations complement the one-dimensional parametric studies by providing a comprehensive visualization of the coupled influence of crack length and the governing material, initial stress, geometrical, loading, and dynamic parameters on the fracture behaviour of the strip.

    \end{itemize}
The analytical formulation presented in this study provides a comprehensive framework for investigating the dynamic fracture behaviour of initially stressed dry sandy media containing moving collinear cracks under moving concentrated and distributed loading. The analytical solutions obtained herein may serve as benchmark results for validating numerical and computational techniques, including finite element, boundary element, meshfree, and phase-field methods developed for dynamic fracture analysis. The findings also provide useful insight into the fracture assessment and structural integrity evaluation of engineering systems involving sandy or granular materials subjected to moving loads, such as railway and highway pavements, airport runways, geotechnical foundations, underground tunnels, and earth-retaining structures. Furthermore, the present formulation establishes a basis for future extensions to more realistic material models, including layered, functionally graded, anisotropic, poroelastic, and thermoelastic media, as well as more general crack configurations and dynamic loading conditions.

\section*{Acknowledgements}
The first author gratefully acknowledges the University Grants Commission (UGC), Government of India, for providing financial support through the UGC Junior Research Fellowship (JRF). The authors also express their sincere gratitude to the National Institute of Technology Hamirpur, Himachal Pradesh, India, for providing the necessary research facilities to carry out this work.

%
\section*{Conflict of interest}
 The authors declare that they have no conflict of interest.

\bibliographystyle{spmpsci}      
\bibliography{refrences}   

\end{document}